\pdfoutput=1
\documentclass[11pt]{article}
\usepackage[right=1in,left=1in,top=1in,bottom=1in]{geometry}
\usepackage{hyperref}
\hypersetup{colorlinks, citecolor=blue, filecolor=blue, linkcolor=blue, urlcolor=blue}
\usepackage{graphicx}
\usepackage{url}
\usepackage{adjustbox}
\usepackage[round]{natbib}
\usepackage{amsmath,amsthm}
\usepackage{engord}
\usepackage{float}
\usepackage{subfig}
\usepackage{pdflscape}
\usepackage{booktabs}
\usepackage{endnotes}
\usepackage{pgfplots}
\usepackage{algorithm}
\usepackage{algorithmic}
\usepackage{enumitem}
\usepackage{subfig}
\pgfplotsset{compat=1.14}
\pgfplotsset{every axis label/.append style={font=\tiny}}
\usepackage[labelsep=period]{caption} %% This switches "Table 1: Title" to "Table 1. Title"
\usepackage{authblk}
\usepackage{amssymb} %% Necessary, just for the \checkmark command  in tables.
\usepackage{multirow} %% Necessary if we are doing tables in LaTeX

\usepackage{xr}

\newtheorem{assumption}{Assumption}

\newtheorem{proposition}{Proposition}
\newtheorem{corollary}{Corollary}
\newtheorem{lemma}{Lemma}

\newtheorem{theorem}{Theorem}

\usepackage{setspace}
\usepackage{sectsty}
\sectionfont{\large}
\subsectionfont{\normalsize}
\subsubsectionfont{\normalsize}

\title{ \vspace*{-2.5cm} \hspace*{-0.5cm}Digital Platform Consolidation and Offline Expansion: Strategic Convergence and Market Welfare in China's Second-hand Real Estate Market\footnote{
A preliminary version of this paper, entitled \emph{Platform-Mediated Consolidation and Offline Store Expansion: Evidence from Real Estate Brokerages in Major Chinese Cities}, was presented at the 1st Summer Meeting in Urban Economics, China. We are grateful for the comments and suggestions made at the meeting. The code to replicate the paper can be accessed at \href{https://github.com/sergiozxy/RealEstateBrokerage}{https://github.com/sergiozxy/RealEstateBrokerage} The authors gratefully acknowledge financial support from the Fundamental Research Funds for the Central Universities of Sichuan University (SKSYL2023-01)
}}

\date{ \vspace*{0.5cm} First Version: September 2. 2024; This Version: \today}

\begin{document}

\author[1]{Guoying Deng} % dengguoying@scu.edu.cn
% \author[2,3]{Li Gan} % ganli@tamu.edu 
\author[2]{Xuyuan Zhang \thanks{Email Address: \href{mailto:zxuyuan@umich.edu}{zxuyuan@umich.edu}}}
\affil[1]{School of Economics, Sichuan University, Chengdu, China}
% \affil[2]{Department of Economics, Texas A\&M University, College Station, TX, USA}
% \affil[3]{Survey and Research Center for China Household Finance, Southwestern University of Finance and Economics, Chengdu, China}
\affil[2]{Department of Economics, University of Michigan, Ann Arbor, MI, USA}

\bgroup
\let\footnoterule\relax

\begin{singlespace}
\maketitle

% Police brutality, law enforcement, and crime: Evidence from Chicago
\begin{abstract}
    This study analyzes the impact of offline expansion and online platform consolidation in China's second-hand real estate market. Using micro-level transaction data and difference-in-differences estimations, we find offline store entry significantly boosts transaction volumes (9-10\%) and reduces price concessions (1\%) initially, though effects diminish over time. Platform consolidation via Lianjia's Agent Cooperation Network yields delayed yet persistent transaction volume increases (5-6\%), particularly in less concentrated markets, and consistently lowers price concessions. These strategies sustainably enhance brokerage competitiveness, bargaining power, and market welfare, despite increased market concentration, ultimately benefiting sellers and improving overall efficiency. 
  \end{abstract}
  
  
  \textbf{JEL} Classification Codes: R30, R31, L85, L86
\end{singlespace}
\thispagestyle{empty}

\clearpage
\egroup
\setcounter{page}{1}

%% Temporary tool to track how this paper is structured. Feel free to comment in or out. 
% \tableofcontents
% \bigskip

%%%%%%%%%%%%%%%%%%%%%%%%%%%%%%%%%%%%%%%%%%%%%%%%%%%%%%%%%%%%%
%%%%\section{Introduction\label{sec:introduction}}

\section{Introduction \label{sec:introduction}}

\noindent

This paper investigates the emerging phenomenon of traditional offline intermediates strategically integrating digital platforms to enhance their competitive position in modern markets. Historically, such intermediaries--whether retail distributors, service agents, or brokerage firms--operated primarily through extensive physical networks and relied heavily on localized market knowledge and personal interactions. With ongoing digital transformation, these intermediaries increasingly consolidate fragmented informational resources onto unified digital platforms, thereby transforming their competitive landscape, altering strategic behaviors, and reshaping market outcomes in terms of efficiency, welfare distribution, and market concentration. This paper tends to answer the following questions: How do offline store expansions and platform-mediated consolidation affect transaction? What are the implications of these changes for market performance, particularly in terms of price concessions and transaction volumes? Do these strategies improve total welfare, and how are the resulting benefits distributed between buyers and sellers?

Our theoretical framework rigorously formalizes the strategic interactions between an intermediary's offline presence, represented by its physical branch networks, and its pricing strategy, specifically concerning price concessions. We assume a segmented market characterized by heterogeneous consumer preferences and geographic constraints, aligning with typical empirical settings. Central to our analysis is the notion of an exogenous enhancement in offline productivity, clearly defined as improvements resulting from advanced technological adoption, superior employee training, or optimized operational processes. We model offline branch-network expansion as a strategic decision by each intermediary, influenced by its exogenously given level of offline productivity. Additionally, we explicitly model how online consolidation—integrating fragmented informational resources onto unified digital platforms—enhances intermediaries' market visibility and bargaining leverage, independently amplifying transaction efficiency and reducing the necessity of aggressive price concessions.

We analyze how offline productivity enhancements expand intermediaries' effective market coverage, consequently increasing transaction volumes and allowing them to recalibrate pricing optimally. Simultaneously, we examine the distinct effect of online consolidation, which significantly reduces transaction costs and enhances market matching efficiency through unified informational resources and increased visibility. The framework particularly examines conditions under which enhanced offline presence and price concessions act as strategic substitutes, explicitly showing that intermediaries leverage expanded offline networks and digital consolidation synergistically to reduce price concessions strategically, thereby maximizing profits. The integration of online consolidation effects highlights how digital strategies reinforce the competitive positioning and profitability of intermediaries beyond offline productivity enhancements alone.

Moreover, our model incorporates heterogeneous buyer search behavior—local searchers, who focus exclusively on properties within their immediate geographic segment, and global searchers, who explore properties across adjacent segments.\footnote{Inherited from \citep{10.1257/aer.20141772}.} Proposition \ref{prop:searchers} reveals that intermediaries respond differently depending on searcher composition: an increase in global searchers prompts a more substantial expansion of branch networks than an equivalent increase in local searchers, particularly benefiting larger intermediaries with higher operational efficiency. Proposition \ref{prop:coexistence} further highlights that differences in intermediaries' intrinsic productivity and operating costs allow both large and small intermediates to coexist, explaining persistent market segmentation observed in practice.

We first establish the theoretical foundations by demonstrating monotonic relationships between intermediaries' offline expansion and pricing adjustments, ensuring the existence of well-defined best responses and Nash equilibrium in these strategic interactions. Proposition \ref{prop:offline_expansion} demonstrates that increased offline productivity compels intermediaries to expand their physical market presence while strategically reducing price concessions, clearly articulating the underlying economic rationale of substitutability between these strategies. Extending this analysis, Theorem \ref{theorem:simul_efficiency} illustrates that simultaneous productivity improvements among multiple intermediaries result in substantial expansions of offline networks and increased transaction volumes, albeit with diminishing marginal returns due to intensified competition.  Building on these offline results, Proposition \ref{prop:platform_consolidation} shows that when intermediaries consolidate their listings onto a single platform, the combined increase in market visibility leads directly to higher transaction volumes and reduced pressure to offer price concessions. This means intermediaries jointly benefit from enhanced market power through improved matching between buyers and sellers, even without immediate offline expansion.

In terms of the welfare implications, Theorem \ref{theorem:simul_welfare} delineates how simultaneous efficiency gains redistribute market shares toward more productive intermediaries, thus intensifying competitive pressures and reshaping welfare outcomes. Furthermore, Proposition \ref{prop:platform_welfare} identifies the distinctive welfare effects of platform consolidation, revealing persistent reductions in price concessions due to improved bargaining power and enhanced matching efficiency. These effects uniquely emerge from integrated informational resources and unified digital presence, independent of competitors' simultaneous offline productivity enhancements. Importantly, the result recognizes varying welfare implications across different market structures, noting that while consolidation generally increases total welfare and seller surplus under the scenario where the largest intermediate has dominant position, competitive market structures yield nuanced outcomes where consumer surplus might increase without guaranteeing total welfare enhancement.

Empirically, we validate our theoretical predictions using comprehensive transaction-level data from China's real estate brokerage market, providing an empirical setting characterized by clearly defined market segmentation, established physical branch networks, and explicit pricing practices. We measure offline presence by the year Lianjia opens its first physical store in each business area. Online consolidation is proxied by Lianjia's rollout of its Agent Cooperation Network (ACN) on the Beke platform. Treating both entry and consolidation as plausibly exogenous shocks, we then estimate their effects on log-transaction volumes and price concessions using two-way fixed-effects and dynamic event-study DID models. Our findings explicitly reveal a substantial initial increase in transaction volumes, ranging approximately from 9\% to 10\%, accompanied by significant reductions in price concessions, aligning closely with theoretical expectations of improved market efficiency driven by offline productivity enhancements. Detailed analysis further uncovers that these initial positive effects gradually diminish over time as competing brokerages similarly enhance their offline presence, prompting dynamic equilibrium adjustments and market share redistribution. Robustness checks employing market concentration indices (Herfindahl-Hirschman Index, HHI) explicitly confirm the reliability and generalizability of these outcomes, particularly highlighting the role of intensified competitive pressures in moderating the marginal benefits of offline productivity improvements.

Complementarily, we delve into the distinct effects of platform consolidation strategies by analyzing the prominent case of China's brokerage platform, Lianjia's Agent Cooperation Network (ACN), through a dynamic DID approach. Our nuanced empirical investigation explicitly shows minimal immediate effects on transaction volumes following digital consolidation. However, subsequent periods exhibit significant and sustained increases in transaction activity, with volume gains of approximately 5\% to 6\%, particularly pronounced in markets characterized by lower initial concentration levels. Further, we uncover consistent and persistent reductions in price concessions averaging around 1\%, which explicitly demonstrates enhanced bargaining leverage and improved buyer-seller matching efficiency derived uniquely from platform-driven consolidation efforts. Critically, these consolidation effects remain robust and independent of concurrent offline productivity gains, underscoring their distinctive strategic value. The empirical insights not only strongly corroborate theoretical predictions but also highlight how strategic digital consolidation fundamentally reshapes competitive advantages and bargaining dynamics in intermediary markets.

Overall, our findings explicitly elucidate critical insights into the interplay between digital integration and offline operational efficiency, offering broader theoretical and empirical implications beyond the specific context of real estate brokerage. Our analysis underscores how traditional offline intermediaries strategically leveraging digital platforms can fundamentally enhance competitive positioning, reshape bargaining dynamics, and significantly influence market welfare distributions, thereby contributing valuable insights to the broader economic literature on digital transformations and competitive market structures.

The structure of the paper is organized as follows. Section \ref{sec:literature_review} provides a comprehensive review of the relevant literature. Section \ref{sec:theoretical_framework} models the theoretical framework and outlines the research hypotheses. Section \ref{sec:data} outlines the study's background, including a statistical summary and stylized facts. Section \ref{sec:mechanism_design} examines the impact of Lianjia's offline store expansion and platform-mediated consolidation on transaction. Finally, Section \ref{sec:conclusion} discusses the findings and offers concluding remarks.

\section{Literature Review} \label{sec:literature_review}

The integration of digital platforms by traditional intermediaries, complemented by offline market expansion, represents an emerging phenomenon reshaping competitive dynamics across various industries. While classical economic theory extensively covers market competition and pricing mechanisms, the strategic interplay between online consolidation and offline expansion has received less systematic attention.

In the literature on offline-online channel integration, \citet{10.1287/mnsc.1090.1062} show that the geographic distribution and product assortments of brick-and-mortar stores shape the competitive niche of Internet retailers by altering search costs and consumer reservation utilities across channels. Building on this, \citet{10.1509/jmr.14.0518} exploit the staggered openings of physical outlets to document both substitution and complementarity effects on online sales, thereby highlighting how store presence conveys valuable demand signals and mitigates information frictions. \citet{10.1287/mnsc.2017.2787} develop a formal operations model in which in-store kiosks and mobile ordering apps integrate with online systems to reduce customer wait times, optimize staffing levels, and enhance service profitability, underscoring the procedural synergies of self-order technologies across channels. Finally, \citet{10.1016/j.jom.2018.06.004} provide empirical evidence that ship-to-store services significantly boost in-store traffic and cross-channel sales—despite higher return rates—revealing the trade-offs in omni-channel fulfillment and the operational value of leveraging online information for offline expansion. These literatures remain unsystematic and do not offer a comprehensive framework of integrated offline-online operations, and we now turn to the literature on real estate brokerage, positing that it serves as a salient example of the integration between offline and online channels.

We first consider the background of the real estate market with the foundational work of \citep{Rosen_hedonic} who introduced the hedonic pricing model. This model breaks down property prices by analyzing internal and external factors. However, it's worth noting that this model does not adequately account for market asymmetries and information disparities, leading to potential inaccuracies in pricing. A fundamental study by \citep{Akerlof_1970} highlights the significant impact of asymmetric information on market dynamics, using the market for used cars as an example. Here, the prevalence of low-quality goods, known as `lemons', often leads to market inefficiencies, a problem that is mirrored in the real estate sector. The challenge of asymmetric information in real markets was further emphasized by \citet{grossman_impossibility_1980}, who questioned the feasibility of the effective market assumption, particularly under conditions of information disparity. 

Subsequent studies have expanded on these foundational theories, exploring dynamics specific to real estate pricing and strategic behavior. \citet{550a6ccf-cde2-3dd1-979f-1a8db2b8ceb9} documents that apart from price competition in the market, there is a lot of market inefficiency that stems from non-price competition, which suggests that as the degree of competition in the market increases, the market becomes progressively less efficient, indicating that the entry dividend begins to fall and aggregate social welfare begins to decline. Moreover, \citet{hendel_relative_2009} analyzes two types of listings in the second-hand housing market and finds that For-Sale-By-Owner type of platforms are less effective in terms of time and probability of sale while operating better compared to listing homes for sale as a broker (see also \citep{GENESOVE201231}). In addition, \citet{bailey_economic_2018} uses data from the social media site Facebook to show that social interactions can influence people's economic decisions. Their results show that people who have friends who are geographically distant in real life and who have a hunch that house prices are about to rise are more likely to buy a house than rent one. Other relevant areas of research include \citep{SIRMANS1991207, NIEUWERBURGH_information, salz_intermediation_2022, YANG2021102359}.

The strategic behavior of real estate brokerages has been documented to leverage informational advantages. \citet{AGARWAL2019715} confirms that brokerages, as market intermediaries, possess nuanced knowledge of market conditions, enabling them to negotiate discounts effectively. Furthermore, \citet{HAN2015813} discusses the varying bargaining power of brokerages across unidirectional and bidirectional markets, influencing their operational strategies. This is corroborated by evidence suggesting that properties listed with lower commission rates not only sell less frequently but also take longer to sell \citep{10.1257/app.20160214}. Additionally, studies have documented that brokerages may adopt discriminatory strategies, steering minorities into neighborhoods with lower economic opportunities and higher exposures to crime and pollution, thereby contributing to persistent social and economic inequalities in the United States \citep{RePEc:ucp:jpolec:doi:10.1086/720140}.\footnote{See also \citep{gilbukh_goldsmith-pinkham_2019}.} The advent of online platforms has significantly altered the landscape of real estate transactions. \citet{ZUMPANO2003134} notes that while the duration of property searches has not changed markedly, the scope of searches has broadened to encompass more online listings. Moreover, \citet{ZHANG2021101104} associates the rise of online platforms with a reduction in existing house prices and an increase in sales volumes, a dynamic influenced by factors such as new house prices and household demographics. However, a detailed analysis of the impact of the presence of these platforms on market performance of offline stores remains scant.

Moreover, the overall market influence of real estate intermediaries is multifaceted. Utilizing a model predicated on perfect competition, \citet{williams_agency_1998} illustrates that excessive entry of brokers into the market can surpass the optimal allocation, thereby reducing social welfare. This is corroborated by studies indicating that an increase in the number of brokers can depress house prices and shorten transaction cycles \citep{https://doi.org/10.1002/jae.2891, RePEc:kap:jrefec:v:67:y:2023:i:3:d:10.1007_s11146-021-09858-w}. Additionally, \citet{qu_identifying_2021} highlights the moderating role of broker commissions in disseminating information during transactions, facilitating more efficient home sales. Lastly, \citet{AGARWAL2024103668} utilized second-hand real estate transaction data from Beijing to demonstrate that real estate agents may significantly contribute to the formation of Yin-Yang contracts. They quantified the magnitude of the resulting tax evasion, attributing it to the learning-by-doing effect and peer influence among agents. \citet{hsieh_can_2003} also consider the general welfare implication of the real estate brokerage on the market and finds that the market is more inefficient when more brokerages enter the market. Other related literature includes \citep{doi:10.1080/10527001.2021.2016340, d082a2db-5cce-33f2-87c4-9cb020cc6666, doi:10.1080/10835547.1996.12090852}.

Finally, the concept of the platform as a ``two-sided market'' or kind of intermediates that connects the buyer side and seller side digitally, \citep{10.1162/154247603322493212, Langley_Leyshon_2017, 10.1257/aer.100.4.1642, Armstrong2006}. \citet{https://doi.org/10.1111/j.1756-2171.2006.tb00036.x} characterizes real estate brokerages as a two-sided market. In this model, brokerages must effectively communicate and mediate between sellers and buyers, facilitating transactions and ensuring efficient market operations. Despite the significant attention given to online platforms and real estate brokerage's two-sided market \citep{10.1257/jep.23.3.125}, there is a lack of systematic analysis on the impact of offline stores of online platforms on market performance or on real estate brokerage. Therefore, this paper aims to fill this gap by examining the influence of offline stores associated with online platforms on the real estate market.

Based on the study of existing literature, this paper represents the pioneering empirical investigation into the effects of offline store expansion and platform-mediated consolidation within the real estate brokerage market. Additionally, this paper systematically examines how informational advantages can enhance brokerage revenue. By delving into these dynamics, we aim to contribute to the broader economics literature by providing a nuanced understanding of how offline and online consolidation influences market behavior and outcomes. Our findings offer meaningful insights into the strategic decisions of real estate brokerages, highlighting the significance of informational advantages in shaping competitive advantage and market performance. This research not only fills critical gaps in the literature but also provides a comprehensive framework for future studies to explore the intersection of offline expansion and digital consolidation in various economic contexts. By systematically examining these phenomena, we enhance the understanding of how technological and infrastructural developments can drive revenue generation and market consolidation in the real estate industry and beyond.

\section{Theoretical Framework \label{sec:theoretical_framework}}

\noindent

\subsection{Model Setup}

We consider a general model of traditional offline intermediates. There is a finite set $\mathcal{I} = \{1, \ldots, N\}$ of intermediates. Each intermediate $i \in \mathcal{I}$ can choose one of three options for platform membership:

\begin{equation*}
  \ell_i \in \{P_L, P_S, 0\}, \quad i \in \mathcal{I}
\end{equation*}

where $P_L$ denotes a platform dominated by large intermediates, $P_S$ a platform of small intermediates, and $0$ the ``outside option'' of remaining independent.

There is an exogenous number of potential buyers distributed over a geographic region partitioned into segments $\mathcal{M} = \{1,\ldots,M\}$. Each buyer is initially indifferent among intermediates; however, the number of listings available in the relevant segment affects their choice. Each intermediate $i$ chooses an offline store (branch) configuration 

\begin{equation*}
  \mathbf{n}_i = (n_{i1}, \ldots, n_{iM}) \in \mathbb{Z}_+^M
\end{equation*}

where $n_{im}$ is the number of branches in segment $m$, and a vector of price concession offers

\begin{equation*}
  \mathbf{c}_i = (c_{i1}, \ldots, c_{iM}) \in \mathbb{R}_+^M
\end{equation*}

In each segment $m$, there is an exogenous pool of listings $L_m > 0$. Sellers are indifferent among intermediates except for the offered price concession $c_{im} \geq 0$, which reduces the final transaction price. In segment $m$, the base listing price is $\mu_m$, so that a listing associated with intermediate $i$ will eventually sell at the net price:

\begin{equation*}
  \mu_m - c_{im}.
\end{equation*}

intermediate $i$ earns a commission rate $\beta > 0$, so that its per-transaction profit in segment $m$ is:\footnote{In China's real estate market, the intermediate commission rate is uniformly set across the industry, meaning all intermediates charge the same rate. This standardized commission structure is a common and widely accepted practice in many countries, although specific rates may vary across different markets.}

\begin{equation*}
  p_{im} = \beta (\mu_m - c_{im})
\end{equation*}

\subsection{Listings Capture and Offline Presence}

A key element is the role of offline branches in capturing listings. For each segment $m$, we assume that buyers search for intermediates whose effective presence is given by:

\begin{equation*}
  f(n_{im}, n_{i, \mathcal{N}(m)}) = h(n_{im}) + \gamma \sum_{m^\prime \in \mathcal{N}(m)} \omega_{m, m^\prime} h(n_{im^\prime})
\end{equation*}

where $h: \mathbb{Z}_+ \mapsto \mathbb{R}_+$ is non-decreasing and captures the direct effect of local presence, $\gamma \geq 0$ reflects the strength of cross-segment spillovers, $\omega_{m, m^\prime} \geq 0$ is a weight capturing the proximity between segments $m$ and $m^\prime$, and $\mathcal{N}(m) \subseteq \mathcal{M}$ denotes the set of segments adjacent to $m$ (with $m \notin \mathcal{N}(m)$). We assume that for $\Delta n_{im} > 0$ we have: $\Delta f(n_{im}) = f(n_{im} + \Delta n_{im}) - f(n_{im}) > 0$ has increasing differences. % this is the key part to prove that the intermediate's offline presence is increasing in the transaction efficiency.

Each intermediate has an intrinsic efficiency parameter $\alpha_i > 0$, so that its \emph{effective attractiveness} in segment $m$ is:

\begin{equation*}
  R_{im} \equiv \alpha_i f(n_{im}, n_{i, \mathcal{N}(m)}) \psi(c_{im}, |P(i)|),
\end{equation*}

where $\psi: \mathbb{R}_+ \times \mathbb{N} \mapsto \mathbb{R}_+$ is twice differentiable in its first argument and captures the effect of price concessions. The second argument, $|P(i)|$, denotes the number of intermediates that are members of the same platform as intermediate $i$ (if $\ell_i \neq 0$); for an independent intermediate, (i.e., $\ell_i = 0$), we take $|P(i)| = 1$. The function $\psi$ satisfies:

\begin{equation*}
  \begin{aligned}
  \frac{\partial \psi}{\partial c_{im}}(c_{im}, |P(i)|) > 0 & \text{ for all $c_{im} \geq 0$ and for each fixed $|P(i)|$} \\
  \frac{\partial^2 \psi}{\partial c_{im}^2}(c_{im}, |P(i)|) < 0 & \text{ for all $c_{im} \geq 0$ and for each fixed $|P(i)|$} \\
  \psi(c_{im}, |P(i)|) > \psi(c_{im}, |P(i)^\prime |) & \text{ for each fixed $c_{im} \geq 0$ and $|P(i)| > |P(i)^\prime |$} % because this is a number, you cannot differentiate it
  \end{aligned}
\end{equation*}

If intermediate $i$ chooses to enter segment $m$, its captured share of listings in that segment is defined by:

\begin{equation*}
  \sigma_{im} = \dfrac{R_{im}}{\sum_{k \in \mathcal{I}} R_{km}} \cdot L_{m}
\end{equation*}

If $n_{im} = 0$, then we set $\sigma_{im} = 0$. Notice that by construction, we have the following monotonicity property:

\begin{lemma}[Monotonicity of Coverage] \label{lemma:monotonicity_coverage}
  For any intermediate $i$ and segment $m$,

  \begin{enumerate}[noitemsep,nolistsep]
    \item For fixed $\mathbf{c}_{i}$ and $|P(i)|$, if $n_{im}$ (or any $n_{im^\prime}$ with $m^\prime \in \{m\} \cup \mathcal{N}(m)$) weakly increases, then $f(n_{im}, n_{i, \mathcal{N}(m)})$ and hence $R_{im}$ and $\sigma_{im}$ weakly increase.
    \item For fixed $\mathbf{n}_i$ and $|P(i)|$, if $c_{im}$ weakly increases, then $R_{im}$ and $\sigma_{im}$ weakly increase.
    \item For fixed $\mathbf{n}_i$ and $\mathbf{c}_{i}$, if $|P(i)|$ weakly increases, then $R_{im}$ and $\sigma_{im}$ weakly increase.
  \end{enumerate}
\end{lemma}

intermediates can join the platform to pool and list their listings. Let $P \subseteq \mathcal{I}$ be the set of intermediates that have chosen a given platform (An independent intermediate is simply represented by $P = \{i\})$. In segment $m$, the total coverage of platform $P$ is:

\begin{equation*}
  \Sigma_m(P) = \sum_{i \in P} \sigma_{im}
\end{equation*}

In line with the framework outlined by \citep{10.1257/aer.20141772}, market participants can be categorized into two distinct types of searchers. The first group comprises \emph{local searchers}, denoted as $\lambda_m^L > 0$, who are potential buyers restricted to searching exclusively within their designated home segment $m$. The second group consists of \emph{global searchers}, denoted as $\lambda_m^G > 0$, who are potential buyers considering multiple segments during their search process. Notably, buyers are indifferent to the choice of platforms or intermediates, with their primary focus being the number of listings they encounter.

Formally, each global searcher in segment $m$ considers listings available in both their home segment $m$ and the neighboring segments, i.e., in $\{m\} \cup \mathcal{N}(m)$ and we also require that $m \notin \mathcal{N}(m)$. The equilibrium outcome is determined through a matching process between buyers and listings. In each segment $m$:

\begin{enumerate}[noitemsep,nolistsep]
  \item Local searchers choose a platform in proportion to its coverage in segment $m$, denoted $\nu_{im}^L = \left[ \lambda_m^L \dfrac{\Sigma_m (P(i))}{\sum_{Q \in \mathcal{P}} \Sigma_m (Q)}\right]$
  \item Global searchers choose a platform in proportion to its aggregate coverage over $\{m\} \cup \mathcal{N}(m)$, denoted $\nu_{im}^G = \left[ \lambda_m^G \dfrac{\Sigma_m(P(i)) + \sum_{m^\prime \in \mathcal{N}(m)} \Sigma_{m^\prime}(P(i)) }{ \sum_{Q \in \mathcal{P}} (\Sigma_m(Q)) + \sum_{m^\prime \in \mathcal{N}(m)} \Sigma_{m^\prime}(Q) }  \right]$
  \item After searchers decide which platform to choose, the transaction is achieved with probability $\tau_{im}(\nu_{im}^L + \nu_{im}^G)$ which is proportional to the number of searchers, we assumed that $\tau_{im}(\cdot)$ is twice differentiable, increasing, and has increasing returns to scale, which is a standard assumption of the network property.
\end{enumerate}

Consequently, let $\mathcal{P}$ be set of active platforms, and  the equilibrium transactions for intermediate $i$ in segment $m$ are given by

\begin{equation*}
  Q_{im} = (\nu_{im}^L + \nu_{im}^G) \cdot \tau_{im}(\nu_{im}^L + \nu_{im}^G)  \cdot \dfrac{\sigma_{im}}{\Sigma_m(P(i))}
\end{equation*}

where $P(i)$ denotes the platform to which intermediate $i$ belongs and $\mathcal{P}$ is the set of all active platforms.

The total equilibrium transactions for intermediate $i$ across all segments is

\begin{equation*}
  Q_i = \sum_{m=1}^M Q_{im}
\end{equation*}

In terms of the cost side, intermediate $i$ incurs a fixed cost $FC_{im}$ for entering the segment $m$ and variable cost $VC_{im}(n_{im})$ which is proportional to the number of branches it has in segment $m$. The total cost for intermediate $i$ is:

\begin{equation*}
  C_{im}(n_{im}) = FC_{im} + VC_{im}(n_{im})
\end{equation*}

Finally, each intermediate $i$ chooses its strategy

\begin{equation*}
  s_i = (\mathbf{n}_i, \mathbf{c}_i) \in \mathbb{Z}_+^M \times \mathbb{R}_+^M
\end{equation*}

to maximize its profit

\begin{equation*}
  \pi_{s_i, s_{-i}} = \sum_{m=1}^M p_{im} Q_{im} - \sum_{m=1}^M C_{im}(n_{im}) 
\end{equation*}

where $s_{-i}$ denotes the strategies of all other intermediates. A Nash equilibrium of the game is a profile $\{s_i^\star\}_{i \in \mathcal{I}}$ such that no intermediate can increase its profit by unilaterally deviating.

\begin{lemma}[Existence of Best Responses] \label{lemma:existence_best_response}
  For each intermediate $i$, the strategy space

  \begin{equation*}
    S_i = \{ \mathbf{n}_i \in \prod_{m=1}^M \{0,1,\ldots,\overline{n}_{im}\} \} \times \{ \mathbf{c}_i \in \prod_{m=1}^M [0,\overline{c}_{im}] \}
  \end{equation*}

  is compact and the profit function $\pi_{i} (s_i, s_{-i})$ is continuous in $\mathbf{c}_i$. Therefore, for any fixed strategies $s_{-i}$ of the other intermediates, there exists at least one best response $s_i^\star$. Consequently, the game has at least one Nash equilibrium $\mathbf{s}^\star \in \prod_{i \in \mathcal{I}} S_i.$
\end{lemma}

Now we consider the functional form of the real estate's intermediates. We assume that the intermediate's profit function $\pi_{im}$ satisfies the following properties:

\begin{assumption}[supermodularity and submodularity condition] \label{claim:modularity}
  We say that the intermediate's profits satisfies the \textbf{supermodularity condition} if for any $c_1 < c_2$ and $\Delta n_{im} > 0$:

  \begin{equation*}
    \left[ \pi_{im}(n_{im} + \Delta n_{im}, c_2) - \pi_{im}(n_{im}, c_2)\right] > \left[ \pi_{im}(n_{im} + \Delta n_{im}, c_1) - \pi_{im}(n_{im}, c_1)\right] 
  \end{equation*}

  implying that the intermediate has an incentive to set higher price concessions to attract more transactions. Conversely, we have \textbf{submodularity condition} if for all $c_1 < c_2$ and $\Delta n_{im} > 0$:
  
  \begin{equation*}
      \left[ \pi_{im}(n_{im} + \Delta n_{im}, c_2) - \pi_{im}(n_{im}, c_2)\right] < \left[ \pi_{im}(n_{im} + \Delta n_{im}, c_1) - \pi_{im}(n_{im}, c_1)\right] 
  \end{equation*}
  
  the two strategic variables are substitutes, and higher offline presence diminishes the marginal benefit of concessions.
\end{assumption}

Under these assumptions, equilibrium conditions can be characterized by optimally with respect to each intermediate's strategic variables. Specifically, intermediate $i$ solves:

\begin{equation*}
  \frac{\partial \pi_{im}}{\partial c_{im}} = 0 \quad \text{ and } \quad \pi_{im}(n_{im} + 1, c_{im}) \leq \pi_{im}(n_{im}, c_{im}) \quad \text{ and } \quad \pi_{im}(n_{im} - 1, c_{im}) \leq \pi_{im}(n_{im}, c_{im})
\end{equation*}

reflecting continuous optimality for concessions and discrete optimality for branch presence. Moreover, based on the empirical finding, we can find that the submodularity condition holds, which then implies that the intermediate has an incentive to set higher price concessions to attract more transactions.

\begin{lemma}[Comparative Statics with Respect to intermediate Efficiency] \label{lemma:exogenous_c_alpha}
  Suppose intermediate $i$'s transaction efficiency $\alpha_i$ exogenously increases, and intermediate profits satisfy the the submodularity condition. Then we have intermediate's price concessions increase $\frac{\partial c_{im}}{\partial \alpha_i} < 0$ while other competitor's price concessions decreases: $\frac{\partial c_{km}}{\partial \alpha_i} > 0$ for $k \neq i$.
\end{lemma}

Lemma \ref{lemma:exogenous_c_alpha} intuitively shows that when a intermediate becomes more efficient at converting listings into transactions, it finds that lowering price concessions to gaining more profits is more profitable. In response, rival intermediates face intensified competition, and consequently, competitors react by increasing their own price concessions to attracting more buyers and increasing the number of transactions. Hence, improvements in one intermediate's transaction efficiency lead to strategic price adjustments that intensify market competition.

\begin{assumption}[Dominance Condition for coverage against concession] \label{claim:dominance}
  Under the submodularity condition, consider an exogenous parameter $\theta$, independent of $\mathbf{n}_i$ and $\mathbf{c}_i$, which directly affects intermediate $i$'s effective attractiveness $R_{im}$. Suppose an exogenous increase in $\theta$ occurs, and in equilibrium, intermediates optimally adjust their strategies as follows:

  \begin{itemize}[noitemsep,nolistsep]
    \item intermediate $k \neq i$ changes from $(n_{km}, c_{km}) \to (n_{km} + \Delta n_{km}, c_{km} + \Delta c_{km})$, where $\Delta n_{km} < 0$ and $\Delta c_{km} > 0$ 
    \item intermediate $i$ changes from $(n_{im}, c_{im}) \to (n_{im} + \Delta n_{im}, c_{im} + \Delta c_{im})$, where $\Delta n_{im} > 0$ and $\Delta c_{im} < 0$.
  \end{itemize}
  
  We say the \textbf{dominance condition} holds for intermediate $k$ and $i$ if:

  \begin{equation*}
    \psi(c_{km}, |P(k)|) \Delta f(n_{km}) + f(n_{km}) \Delta \psi(c_{km}) < 0 \text{ and }
    \psi(c_{im}, |P(i)|) \Delta f(n_{im}) + f(n_{im}) \Delta \psi(c_{im}) > 0 
  \end{equation*}

  where $\Delta f(n_{km}) \equiv f(n_{km} + \Delta n_{km}) - f(n_{km})$ and $\Delta \psi(c_{km}) = \psi(c_{km} + \Delta c_{km}, |P(k)|) - \psi(c_{km}, |P(k)|)$ and respectively for $i$.
\end{assumption}

This assumption essentially means that offline presence is critically important and for competitors, losing branches hurts intermediate attractiveness significantly because it reduces visibility, customer trust, and direct interactions, outweighing the benefit of higher price concessions. The condition ensures that intermediates cannot easily substitute between offline presence and price concessions without significant strategic consequences, highlighting the essential role of physical presence in market competition.

\subsection{Results} \label{subsec:results}

\begin{proposition}[Offline Expansion: Higher Coverage and Increased Concessions] \label{prop:offline_expansion}
  Suppose that intermediate $i$ transaction efficiency $\alpha_i$ exogenously increases, and intermediate profits satisfy the the submodularity condition. Then, in equilibrium:

  \begin{enumerate}[noitemsep,nolistsep]
    \item intermediate $i$ weakly increases its branch configuration $\mathbf{n}_i$.
    \item Then, under the Assumption \ref{claim:dominance}, the intermediate $i$'s effective attractiveness $R_{im}$ increases, i.e., $\frac{\partial R_{im}}{\partial \alpha_i} > 0$ and the  captured share of listings $\sigma_{im}$ increases, i.e., $\frac{\sigma_{im}}{\partial \alpha_i}$.
    \item intermediate $i$ can set up higher optimal price concessions $\mathbf{c}_i^\star$ when the profit function $\pi_{im}$ satisfies the supermodularity condition with respect to $n_{im}$ and $c_{im}$. Conversely, when the intermediate satisfies the submodularity condition, intermediate $i$ has an incentive to reduce price concessions. In either case, the total profit $\pi_{im}$ increases as a result of the optimization.
  \end{enumerate}
\end{proposition}

This proposition states that the efficiency of physical operations, such as by better training its staff or by implementing new technology in its offices would prompt an expansion in the intermediate's physical network of branches. A more extensive branch configuration increases the intermediate's effective attractiveness, thereby boosting both its market coverage and transaction volumes. With a larger network of branches, the firm naturally captures a greater share of the available listings.

Lastly, if the branch network ($n_{im}$) and price concession ($c_{im}$) are \emph{strategic complements} in the profit function, the expansion in the offline network raises the marginal benefit of offering higher concessions. As a result, the intermediate sets a higher concession level to attract even more transactions, offsetting the lower per-transaction margin through a sufficiently large volume increase. Consequently, the firm enjoys strictly higher total profit.

On the other hand, if the two strategic variables were \emph{strategic substitutes}, an offline expansion would reduce the marginal benefit of concessions, and the intermediate might respond by lowering its price concessions. In either case, the firm's total profit rises because the expansion in offline capacity increases coverage and the adjusted concession policy aligns to maximize overall profitability.

\begin{theorem}[Simultaneous Increase in Efficiency] \label{theorem:simul_efficiency}
  Consider a set of intermediates partitioned into two groups: an efficiency-increasing set $\mathcal{E} \subseteq \mathcal{I}$ and the remaining set of competitors $\mathcal{C} = \mathcal{I} \setminus \mathcal{E}$. Suppose there is an exogenous and simultaneous increase in the intrinsic efficiency parameters $\{\alpha_i\}_{i \in \mathcal{E}}$:

  \begin{equation*}
    \alpha_i \to \alpha_i + \Delta \alpha_i, \text{ with } \Delta \alpha_i > 0 \ \forall i \in \mathcal{E}
  \end{equation*}

  Then, under the submodularity condition and Assumption \ref{claim:dominance}, in equilibrium:

  \begin{enumerate}[noitemsep,nolistsep]
    \item Each intermediate $i \in \mathcal{E}$ weakly increases its offline branch presence $n_{im}$, and transaction volumes $Q_{im}$ strictly increase. Moreover, price concessions $c_{im}$ strictly decrease, but the magnitude of these reductions is moderated by competitive interactions within $\mathcal{E}$.
    \item Each intermediate $k \in \mathcal{C}$ weakly reduces its offline branch presence $n_{km}$ and transaction volumes $Q_{km}$ weakly decrease.
  \end{enumerate}
\end{theorem}

The theorem shows that when multiple intermediates simultaneously improve their efficiency, there is a distinct equilibrium adjustment pattern. intermediates improving efficiency jointly expand their offline presence, but competitive pressures moderate their price concession reductions. Competitors without efficiency improvements reduce their market presence and transaction volumes. Consequently, the market structure would tend toward an oligopoly dominated by large intermediates exhibiting higher operational efficiency, while smaller intermediates would experience a reduction in both their physical presences and transaction volumes due to competitive pressures.

The next proposition considers the case where platform consolidation's effects on the equilibrium outcomes.

\begin{proposition}[Platform Consolidation: Enhanced Coverage and Lower Concessions] \label{prop:platform_consolidation}
  Suppose that a set of intermediates $P$ consolidates onto a common platform, so that for each $i \in P$, the platform size $|P(i)|$ increases. Then, under Assumption \ref{claim:dominance} and the submodularity condition of $\pi_{km}$, in equilibrium:

  \begin{enumerate}[noitemsep,nolistsep]
    \item $\psi(c_{im}, |P(i)|)$ increases in $|P(i)|$ and hence $R_{im}$ and $\sigma_{im}$ increase.
    \item The enhanced aggregated coverage $\Sigma_m(P)$ makes the platform more attractive to both local and global searchers, thereby increasing equilibrium transactions $Q_{im}$.
    \item The total profit $\pi_{im}$ increases as a result of the optimization.
  \end{enumerate}
\end{proposition}

This proposition considers the case where several intermediates decide to join forces by consolidating onto a common platform. When they do so, their individual listings are pooled together and presented under a unified brand or system. This consolidation dramatically increases the overall visibility and coverage of the platform in the market. 

From the buyers' perspective, a platform with a broader set of listings is much more attractive. As a consequence, the consolidated platform draws in more buyers, leading to higher overall transaction volumes for its member intermediates. Lastly, the intermediate can also adjust its price concessions to maximize its total profit. If the offline network and price concessions are strategic complements, the intermediate can set higher price concessions to attract more transactions. Conversely, if the two strategic variables are strategic substitutes, the intermediate might reduce its price concessions. In either case, the intermediate's total profit increases as a result of the optimization.

The next proposition considers the coexistence of big and small intermediates in the market.

\begin{proposition}[Coexistence of Big and Small intermediates] \label{prop:coexistence}
  Assume that intermediates differ in their intrinsic efficiency parameters $\alpha_i$ and cost structures $C_{im}(n_{im})$. Then there exists a Nash equilibrium in which:

  \begin{enumerate}[noitemsep,nolistsep]
    \item Under the Assumption \ref{claim:dominance}, efficient intermediate (with higher $\alpha_i$ or lower $C_{im}(\cdot)$ optimally adopt larger offline networks ($\mathbf{n}_i$) and capture a larger share of listings and earn higher profits.
    \item The large intermediates can not monopolize the market even if they have larger network effect.
  \end{enumerate} 
  
  Consequently, the market accommodates a heterogeneous equilibrium where both big and small intermediates coexist, each serving different segments of the buyer-seller market.
\end{proposition}

This proposition helps explain why in many real-estate or intermediate markets, we observe a few dominant players with extensive networks and brand power, while smaller shops also manage to survive by focusing on certain geographic or demographic segments. Moreover, even with digital platforms, the physical presence plays a key role in capturing listings. The more a intermediate invests in offices, the stronger its presence—and thus the more likely it is to gain sellers' business.

Lastly, we can consider the last proposition which considers the intermediates' strategic behavior in terms of the local and global searchers in the market.

\begin{proposition}[Local-Global Segmentation and Heterogeneous intermediates] \label{prop:searchers}
  Consider a segment $m$ in which there is an exogenous increase in the mass of searchers:

  \begin{itemize}[noitemsep,nolistsep]
    \item Increase in Global Searchers $\lambda_m^G$
    \item Increase in Local Searchers $\lambda_m^L$
  \end{itemize}

  Then, in each case, the equilibrium outcome satisfies:

  \begin{enumerate}[noitemsep,nolistsep]
    \item All intermediates weakly increase their branch coverage $n_{im}$ in the affected segment $m$.
    \item The large intermediate (with higher $\alpha_i$) increases its branch coverage by a weakly larger amount than any small intermediate.
    \item Moreover, for the global-searcher increase $(\lambda_m^G)$, the incremental expansion by the large intermediate is weakly larger than for a comparable increase in local searchers $(\lambda_m^L)$.
  \end{enumerate}
  
\end{proposition}

An increase in local searchers $\lambda_m^L$ intensifies demand exclusively within segment $m$, inducing each intermediate to expand its branch network in that segment to capture the additional local business. In contrast, an increase in global searchers $\lambda_m^G$ raises demand not only in $m$ but also in adjacent segments, since these searchers explore a broader geographic area. Consequently, a intermediate maintaining a strong network in $m$ and its neighboring segments can secure a disproportionate share of the newly available transactions, further elevating the returns to opening additional branches in $m$. Moreover, a large intermediate with higher intrinsic efficiency ($\alpha_i$) can leverage this expansion more effectively than smaller competitors, thereby increasing its market share. This reinforcing mechanism is especially pronounced when the growth in demand arises from global rather than local searchers.

\subsection{Welfare Analysis}

We examine the welfare effects of largest intermediates' strategies involving online market consolidation and offline expansion. Buyers continue to pay the net price $\mu_{m} - c_{im}$ and obtain a gross benefit of $v_{m}$ per transaction. Thus the per-transaction buyer surplus is:

\begin{equation*}
   BS_{im} = v_m - (\mu_m - c_{im}) 
\end{equation*}

and total surplus in segment $m$ becomes:

\begin{equation*}
    W_{m}^B = \sum_{i \in \mathcal{I}} Q_{im} (v_m - (\mu_m - c_{im})) 
\end{equation*}

With the commission fee, sellers receive only a fraction $(1 - \beta)$ of the net sale price. Therefore, the per-transaction seller surplus is modified to:

\begin{equation*}
    SS_{im} = (1 - \beta)(\mu_m - c_{im}) - r_m
\end{equation*}

where $r_m$ is the reserved utility, and the total surplus in segment $m$ is given by:

\begin{equation*}
    W_m^S = \sum_{i \in \mathcal{I}} Q_{im} [(1 - \beta)(\mu_m - c_{im}) - r_m] 
\end{equation*}

In terms of the intermediate, the fee revenue amounts to:

\begin{equation*}
    R_m^{fee} = \sum_{i \in \mathcal{I}} Q_{im} \beta (\mu_m - c_{im})
\end{equation*}

in segment $m$, and the total welfare is the sum of buyer and seller surplus plus of sellers' revenue minus the operation cost of the intermediate:

\begin{equation*}
    W^{total} = \sum_{m = 1}^M [W_m^B + W_m^S + R_m^{fee}] - \sum_{m = 1}^M C_{im}(n_{im})
\end{equation*}

Substituting the expressions above yields:

\begin{equation*}
    W^{total} = \sum_{m=1}^M \sum_{i \in \mathcal{I}} Q_{im}(v_{im} - r_{m}) - \sum_{m=1}^M C_{im}(n_{im})
\end{equation*}

\begin{assumption}[Large Firm Dominance] \label{claim:large_dominance}
    Consider an exogenous parameter $\theta$, independent of $\mathbf{n}_i$ and $\mathbf{c}_i$, which directly affects intermediate $i$'s effective attractiveness $R_{im}$. Suppose $\theta$ exogenous increases. In equilibrium, intermediates optimally adjust their price concessions from $c_{km}$ to $c_{km} + \Delta c_{km}$. Define the equilibrium total transaction volumes before the increase as $Q_{km}$. Then intermediate $i$, the intermediate with highest intrinsic efficiency parameter $\alpha_i$, exhibits large-firm dominance if:

    \begin{equation*}
        Q_{im} \left| \Delta c_{im} \right| > \sum_{k \neq i} Q_{km} \Delta c_{km} \ \forall m \in \mathcal{M}
    \end{equation*}
\end{assumption}

This assumption implies that when the most efficient intermediate (intermediate 1) improves its efficiency, it reduces its price concessions dramatically when scaled by its high transaction volume. In effect, intermediate 1's efficiency gains generate a disproportionate competitive edge that reinforces its dominant position. Our empirical evidence from the real estate market supports this mechanism in moderately concentrated settings, as Lianjia's entry led to substantial reductions in price concessions and a marked increase in its market dominance. However, our findings also indicate that in highly competitive markets or in markets with extreme concentration, the impact of efficiency gains on price concessions and market share is more nuanced, suggesting that the assumption is most robust in moderately concentrated markets. Then considering this assumption and we can take a look at the last proposition regarding the market welfare.

\begin{proposition}[Impact of intermediate Efficiency on Total Welfare] \label{prop:welfare}
    Assume that $v_m > r_m$ for all $m$ and suppose that intermediate $i$ with highest intrinsic efficiency $\alpha_i > \alpha_j \ \forall j \neq i$, experiences an exogenous increase in efficiency $\alpha_i$.  Under the condition of submodularity of the profit function $\pi_{im}$ with respect to $(n_{im}, c_{im})$ and Assumption \ref{claim:dominance}, the equilibrium satisfies: 

    \begin{enumerate}[noitemsep,nolistsep]
        \item If the dominant condition (Assumption \ref{claim:large_dominance}) is satisfied, then the total surplus of sellers $W^S$ increase, but the buyers' surplus $W^B$ is indeterminate.
        \item In the absence of the Assumption \ref{claim:large_dominance}, the change in $W^S$ is ambiguous, but the buyers' surplus $W^B$ increase.
        \item With Assumption \ref{claim:large_dominance}, the total welfare in the market $W^{total}$ increase. Without the Assumption \ref{claim:large_dominance}, the total welfare is indeterminate.
    \end{enumerate}
\end{proposition}

We can also consider the simultaneous efficiency increases across multiple intermediates including the largest the intermediate.

\begin{theorem}[Simultaneous Efficiency Increases and Welfare Implications] \label{theorem:simul_welfare}
  Assume that $v_m > r_m$ for all segments $m \in \mathcal{M}$. Consider intermediates partitioned into two sets:

  \begin{itemize}[noitemsep, nolistsep]
  \item Efficiency-increasing intermediates $\mathcal{E} \subseteq \mathcal{I}$ (including the intermediate that processes with highest intrinsic efficiency) simultaneously increase intrinsic efficiency parameters from $\{\alpha_i\}_{i \in \mathcal{E}}$ to $\{\alpha_i + \Delta \alpha_i\}_{i \in \mathcal{E}}$ where each $\Delta \alpha_i > 0$.
  \item Remaining competitors $\mathcal{C} = \mathcal{I} \setminus \mathcal{E}$ have unchanged efficiencies.
  \end{itemize}

  Under the submodularity condition, Assumption \ref{claim:dominance} and Assumption \ref{claim:large_dominance}, the equilibrium satisfies the following properties:

  \begin{enumerate}[nolistsep, noitemsep]
  \item Seller surplus weakly increases relative to the initial equilibrium: $W^{S, \mathcal{E}} - W^{S, init} \geq 0$;
  \item Total market welfare strictly increases relative to the initial equilibrium: $W^{total, \mathcal{E}} - W^{total, init} \geq 0$.
  \end{enumerate}
\end{theorem}

When multiple intermediates simultaneously improve their operational efficiencies, they expand their market presence and reduce their price concessions significantly, which is beneficial for sellers. Due to intensified competition, inefficient competitors shrink their market share and presence, thereby reallocating market transactions toward more efficient intermediates. This reallocation increases overall market effectiveness, driving up total transaction volumes and reducing aggregate intermediate costs. Thus, sellers benefit from higher net prices or sustained transaction volumes, resulting in a clear improvement in seller surplus. Simultaneously, total market welfare strictly improves due to increased efficiency, lower costs, and more optimal allocation of resources toward intermediates capable of converting listings to successful transactions more effectively.

\begin{proposition}[Platform Consolidation Welfare Implications] \label{prop:platform_welfare}
  Consider a set of intermediates $P \subseteq \mathcal{I}$ consolidating onto a single online platform (including the intermediate that processes with highest intrinsic efficiency), thus increasing the platform size $|P(i)|$ for each intermediate $i \in P$. Under the Assumption \ref{claim:dominance} and the submodularity condition of intermediate profits $\pi_{im}$, the resulting equilibrium satisfies the following welfare implications:

  \begin{enumerate}[noitemsep, nolistsep]
    \item If the dominant condition (Assumption \ref{claim:large_dominance}) is satisfied, then the total surplus of sellers $W^S$ increase, but the buyers' surplus $W^B$ is indeterminate.
    \item In the absence of the Assumption \ref{claim:large_dominance}, the change in $W^S$ is ambiguous, but the buyers' surplus $W^B$ increase.
    \item With Assumption \ref{claim:large_dominance}, the total welfare in the market $W^{total}$ weakly increase. Without the Assumption \ref{claim:large_dominance}, the total welfare is indeterminate.
  \end{enumerate}
\end{proposition}

The platform consolidation effect makes the intermediates' combined market visibility and attractiveness significantly increase, enabling these intermediates to capture more transactions and strategically adjust their price concessions. Under the assumption of large-firm dominance, the most efficient intermediate significantly reduces its price concessions, raising the net prices sellers receive, thereby enhancing seller surplus. However, this reduction in concessions could negatively impact buyers, making the buyer surplus ambiguous. Conversely, without large-firm dominance, smaller intermediates competitively lower their concessions to attract buyers, thus benefiting buyers, though the net impact on sellers becomes unclear. Ultimately, total welfare tends to increase clearly when dominant intermediates lead price adjustments, but without dominant intermediate leadership, the overall welfare implications remain uncertain.
% \ref{prop:platform_consolidation}

\subsection{Model Summary and Implications} \label{subsec:model_summary}

The theoretical framework developed above provides a structured approach for analyzing intermediate competition and market dynamics within segmented real estate markets, specifically reflecting the features of China's local real estate intermediate sector. Central to this analysis is the intermediate's dual strategy of offline branch expansion and price concession setting, shaped by intrinsic efficiency and platform affiliation. The model highlights how intermediates leverage physical presence not only to secure local market share but also to exploit cross-segment spillovers, thereby increasing their effective attractiveness and transaction volumes.

An increase in operational efficiency by dominant intermediates, characterized by a high intrinsic efficiency parameter, enables these firms to expand their physical presence strategically. This expansion amplifies market coverage both directly within local segments and indirectly via spillover effects in adjacent segments. The resulting higher market share reinforces dominant firms' positions, such as Lianjia, enhancing their effective attractiveness to buyers and sellers and establishing a reinforcing cycle of market dominance. The analysis extends to scenarios where multiple intermediates simultaneously improve their operational efficiencies. When a subset of intermediates enhances its intrinsic efficiency, these firms expand their offline presence and achieve higher transaction volumes, while competitors without efficiency gains adjust by reducing their branch coverage.

Platform consolidation further magnifies these effects by aggregating individual intermediates' listings and strengthening the overall visibility and attractiveness of the consolidated brand. Consequently, member intermediates optimize their price concessions to maximize transaction volumes and total profits. Under conditions of submodularity between branches and price concessions, dominant intermediates reduces concessions to earn more profits, further adjusting market equilibrium.

Finally, our welfare analysis reveals nuanced outcomes: when the dominant intermediate substantially enhances efficiency (large firm dominance condition), the market concentration gets more intensified, and seller surplus increases significantly due to higher transaction volumes, although buyer surplus changes are ambiguous due to potentially offsetting price and volume effects. Without large firm dominance, buyer surplus clearly benefits from reduced prices, but seller surplus changes are ambiguous. Crucially, total market welfare improves when efficiency gains are significant enough to drive pronounced transaction increases, indicating enhanced overall market efficiency. Conversely, smaller efficiency improvements yield indeterminate welfare outcomes due to intensified market competition.

Building on these insights, our empirical analysis will explore several key questions:

Offline Expansion and Listings Capture: Does an exogenous improvement in a intermediate's offline productivity lead to an expansion of its branch network, increased market coverage, and a greater share of listings? We examine the correlation between physical presence and transaction volumes, as well as the corresponding adjustments in price concessions.

Platform Consolidation Effects: How does the consolidation of intermediates onto common platforms affect their collective visibility, listings capture, and pricing strategies? We assess whether platforms with larger aggregated coverage indeed draw more buyer attention and how this influences individual intermediate behavior.

The following empirical sections will deploy detailed data from the Chinese real estate intermediate market to test these hypotheses. By linking our theoretical predictions with observed market behavior, we aim to provide a comprehensive picture of how technological innovation and network effects are reshaping the industry, enhancing overall market efficiency while presenting new challenges for regulatory oversight.

\section{Data and Descriptive Evidence} \label{sec:data}

\subsection{Background} \label{subsec:background}

In recent years, China's secondary housing market has expanded rapidly, with second-hand housing transactions rising to 37.1\% in 2023. Unlike the primary market, the secondary market hinges on seller-buyer interactions, making transaction costs a central concern. Real estate brokers help lower these costs by offering critical market information, negotiation assistance, and legal support. Although China's legal and regulatory framework has become more standardized, offline-dominated transactions still prevail, largely because of persistent information asymmetry. Indeed, the inherent heterogeneity and ``thin'' nature of real estate goods \citep{glaeser_real_2017, HAN2015813} underscore the importance of brokers' local expertise and market knowledge.

China's real estate brokerage market operates under a bilateral agency model, meaning that both buyers and sellers typically engage separate agents who negotiate on their behalf. This structure contrasts with the United States, where the Multiple Listing Service (MLS) provides centralized, transparent data, allowing for greater transaction efficiency \citep{hendel_relative_2009}. In China, the absence of a unified listing system and reliance on offline transactions further reinforces the importance of intermediaries, making the market less efficient compared to the unilateral agency model prevalent in other countries.

Historically, China's real estate brokerage market was characterized by intense offline competition, but online platforms have reshaped these dynamics. The brokerage's online platform consolidation---where large firms absorb smaller ones into a unified platform---now redefines how properties are bought and sold nationwide. Lianjia spearheaded this shift through its Beke platform, enabling rapid scaling by integrating offline stores and sharing resources. Initially, Beke concentrated on listings from Lianjia's own stores and attracting buyers. In 2014, Lianjia launched the Agent Cooperation Network (ACN) to foster resource sharing and diversify revenue streams,\footnote{The ACN model divides the transaction process into specialized tasks handled by individual agents or stores. These tasks include seller-side activities such as sourcing sellers, maintaining property listings, commission negotiations, and buyer-side activities like client acquisition, property-client matching, transaction facilitation, and financial services assistance.} which enhanced operational efficiency. In 2018, Lianjia expanded its ACN strategy by incorporating smaller brokerages and launching the Deyou franchise brand, further amplifying network effects and market power.

\begin{figure}[htb]
  \centering
  \includegraphics[width=0.6\textwidth]{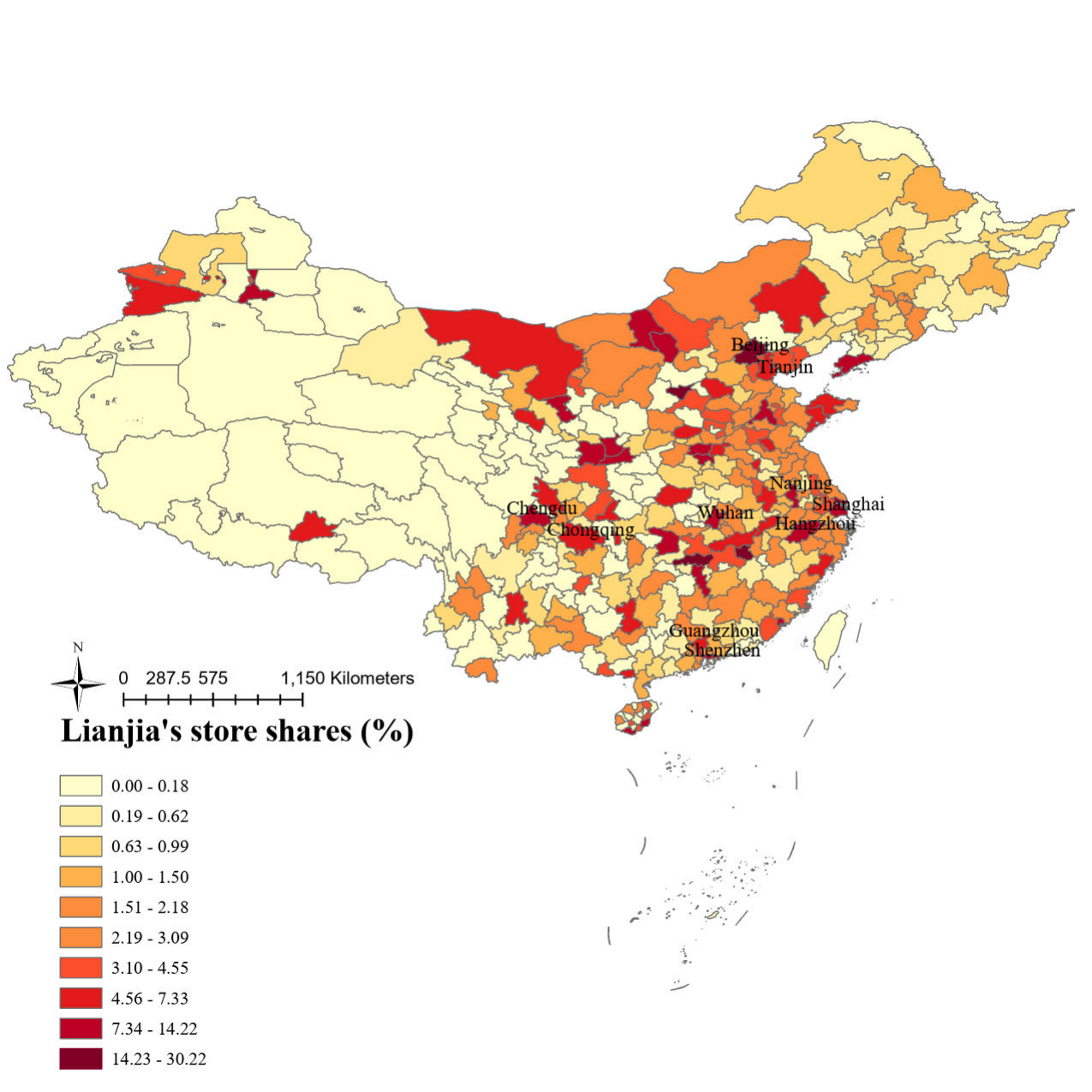}
  \caption{Distribution of Lianjia's Stores' Percentage across China}
  \label{fig:precise_proportion_contraction}
  Note: This plot displays the percentage of Lianjia's offline stores in each city relative to the total number of brokerages in that city, based on data from AutoNavi Map. Gray areas indicate invalid information, while white areas denote cities without any Lianjia stores. Additionally, Sansha city is featured in the bottom right corner of the graph. The base map is sourced from the AutoNavi Map.
\end{figure}

By the platform consolidation, Lianjia transforms internal competition within the system into overall system competitiveness, thereby enhancing the brand's reputation and market influence. More importantly, this strategic allows Lianjia to cooperate with other previously competitors by consolidating resources. According to Beke's IPO prospectus, Lianjia held approximately a 20.8\% share of China's second-hand housing market in 2020, yet its nationwide store presence was below 5\%. In major cities such as Beijing and Shanghai, Lianjia's share exceeded 45\% and 30\%, respectively, while offline store shares in these cities were only 25\% (Beijing) and 10\% (Shanghai). This discrepancy highlights the powerful market penetration achieved by Lianjia's integrated platform and ACN model. Figure \ref{fig:precise_proportion_contraction} illustrates the uneven spatial distribution of Lianjia's offline stores, showing a heavy concentration in first-tier cities and sparser coverage in other regions.\footnote{In the Appendix Figure \ref{fig:precise_proportion_contraction}, we also plotted the geospatial distribution of Lianjia's offline stores and the corresponding relationship between the housing price.} The data also indicates that Lianjia's offline store presence is majorly clustered in major urban centers and sparse distribution in other regions, reflecting strategic market positioning and underlying economic and demographic factors influencing store locations.

Under China's bilateral agency model, sellers strategically choose brokerages to balance transaction speed, price efficiency, and brokerage fees, reflecting their efforts to overcome informational asymmetry inherent in real estate markets. Although theoretically sellers would be indifferent among brokerages in perfectly competitive conditions, China's segmented market structure fosters monopolistic competition, leading sellers to prefer dominant brokerages that offer wider buyer networks and better transaction efficiency despite potentially higher fees. This preference aligns with the theoretical insights from \citep{bergemann_data_2024}, suggesting that platforms with significant market power and information advantages selectively attract high-quality sellers by promising a larger buyer pool and lower expected listing prices. Consequently, this drives a reinforcing cycle where larger brokerages, such as Lianjia, increasingly dominate the market, raising important efficiency and welfare considerations. While enhanced platform competitiveness improves market matching efficiency and consumer outcomes, it simultaneously raises concerns over potential monopolistic power and diminished alternatives for sellers, highlighting the nuanced trade-offs inherent in the evolving platform-based competition landscape.

To better understand these dynamics, this study utilizes comprehensive and granular micro-level transaction data from China's leading real estate brokerage platform, Lianjia. Our detailed dataset, collected across ten major Chinese cities from 2016 to 2022, allows us to empirically evaluate the nuanced effects of Lianjia's offline store expansion and its strategic shift towards online-mediated platform consolidation.

\subsection{Data Collection and Processing} \label{subsec:data_collection}

This study focuses on the housing markets in ten major cities in China, namely Beijing, Shanghai, Chongqing, Tianjin, Shenzhen, Guangzhou, Chengdu, Hangzhou, Wuhan and Nanjing. These cities are not only pivotal to China's economic development, but also serve as exemplars of the broader trends and characteristics inherent in the China's real estate dynamics.% Spanning from 2016 to 2022, the research period encapsulates a pivotal era in China's real estate sector. During the first phase of the study, from 2016 to 2019, the housing markets in these cities experienced a remarkable boom. This period was characterized by significant growth in property prices, supported by robust economic expansion and increased demand in these urban centers. However, the final phase of our study, from 2020 to 2022, paints a contrasting picture. During this period, China's overall economic growth rate has been slower significantly, which has also reflected in a slowdown in the real estate markets of these major cities. In addition, the Chinese government has implemented strict rules in Covid-19 protection and an unprecedented suite of new policies to prevent house prices from falling, so the real estate agents in these major cities are significantly affected.
The second hand housing transaction data was collected form \href{https://www.ke.com/city/}{beke.com} for ten cities ranging from 2016 to 2022.\footnote{Due to government policy, Lianjia was unable to disclose transaction prices in Beijing, Shenzhen, and Wuhan for the years 2021 and 2022, as well as in Chengdu for 2021. Consequently, we have excluded this period of data for these cities from our analysis.} Initially, we filtered out transaction records exhibiting unusually high prices, identifying them as outliers that could skew the analysis. We then removed records with missing values to maintain the integrity of our dataset. We also removed any records that were listed duplicated. We finally have a data with length 1,778,647 second-hand houses.\footnote{Due to government restrictions, four of these cities did not list the transaction price for each transaction during the study period.} After cleaning the data, we constructed two research samples: one at the individual transaction level and the other at the neighborhood level. The individual transaction sample contains detailed information on each transaction, including the transaction price, transaction date, price concessions, and the transaction period. The neighborhood-level sample aggregates transaction data at the community level, encompassing variables such as the average transaction price and the average number of house tours. Additionally, we calculated the annual number of house transactions to capture Lianjia's transaction activity within each neighborhood. We also calculated the revenue of each neighborhood for each year by the formula: $\text{annual transaction number} \times \text{average transaction price} \times \text{brokerage fee}$ where the brokerage fee is set at 2.7\% as the majority of houses in our research sample are charged at this uniform rate. This data enables us to carry out a comprehensive analysis of the impact of offline stores on housing transactions across both individual and neighborhood dimensions.

To gather additional characteristic information, Point of Interest (POI) data was extracted from the AutoNavi map using a web-scraping Python program.\footnote{AutoNavi, a leading mapping application in China with a user base exceeding 700 million, is renowned for its detailed and accurate POI data as well as precise public transportation information. These features underscore AutoNavi's leadership in the digital mapping sector, highlighting its ability to provide unparalleled navigation accuracy and comprehensive urban mobility solutions. Our extracted AutoNavi dataset includes over one million POIs per city annually.} The extracted POIs were then classified into various categories, as detailed in Table \ref{tab:statistical_district} and Table \ref{tab:statistical_individual} and it primarily represents living facilities, entertainment venues, restaurants, hospitals, and other public amenities. This classification is crucial for understanding the urban infrastructure and amenities available in the vicinity of the analyzed properties.

Each type of POI, excluding brokerages offline stores information, was matched to our data within a 500-meter radius, a distance typically covered by walking and consistent with urban planning standards for accessible urban design.\footnote{We chose a 500-meter radius because the existing literature does not specifically document whether these types of points of interest (POIs) adhere strictly to the 5-minute walk policy. Furthermore, several studies have utilized a 500-meter radius to examine the influence of various types of POIs, including: \citep{LI2019165, doi:10.1287/mnsc.2019.3550}} This radius reflects the immediate urban environment influencing residential desirability and property values, as most of these POIs provide recreational services. Additionally, geo-informational data, including annual GDP data from \citet{zhao_forecasting_2017}, nighttime lights data from \citet{elvidge_annual_2021}, and air pollution data from \citet{doi:10.1021/acs.est.1c05309}, was integrated into our research. The centroid of the neighborhood-level data's polygon was generated, and values from the geo-informational data were extracted. By merging these data sources with our research panel, a comprehensive research sample was constructed.

\subsection{Statistical Summary} \label{subsec:Statistical_Summary}

To effectively capture the segmented market in China's real estate industry, it is crucial to determine the optimal radius of influence for brokerage stores. To effectively capture the segmented market in China's real estate industry, it is essential to determine the optimal radius of influence for brokerage stores. This radius defines the extent of a store's market reach, allowing for a precise understanding of localized demand. 

The key consideration is that most of the real estate brokerages in China adopt a pedestrian-shed strategy, commonly referred to as the 5-minute walking distance principle (410 meters). This strategy is designed to minimize customer travel costs and enhance accessibility, ensuring that potential buyers and sellers can conveniently reach a brokerage without relying on motorized transportation.\footnote{The effectiveness of the 5-minute walking radius is supported by both urban planning principles and empirical observations. The government's urban development guidelines advocate for essential living facilities to be located within a five-minute walking distance, reinforcing the idea that accessibility significantly influences consumer behavior \citep{GB50180-2018}. Furthermore, research in other domains has demonstrated that distances exceeding this threshold often lead to significant declines in engagement and service utilization \citep{AZMI2012406, liu2023chrono-urbanism}.} Additionally, the spatial distribution of brokerage stores also aligns with the commercial density of urban areas, where multiple streets and residential communities typically fall within a two-kilometer radius. However, within this broader commercial district, the density of brokerage branches plays a crucial role in maintaining service coverage. A sufficient number of stores within a given area ensures continuous customer flow and stable transaction volumes, reinforcing the necessity of localized service networks. The 5-minute walk strategy thus serves as a practical benchmark for assessing the extent of a store's influence and optimizing the placement of new locations to sustain revenue generation.

\begin{figure}
    \centering
    \includegraphics[width=0.7\textwidth]{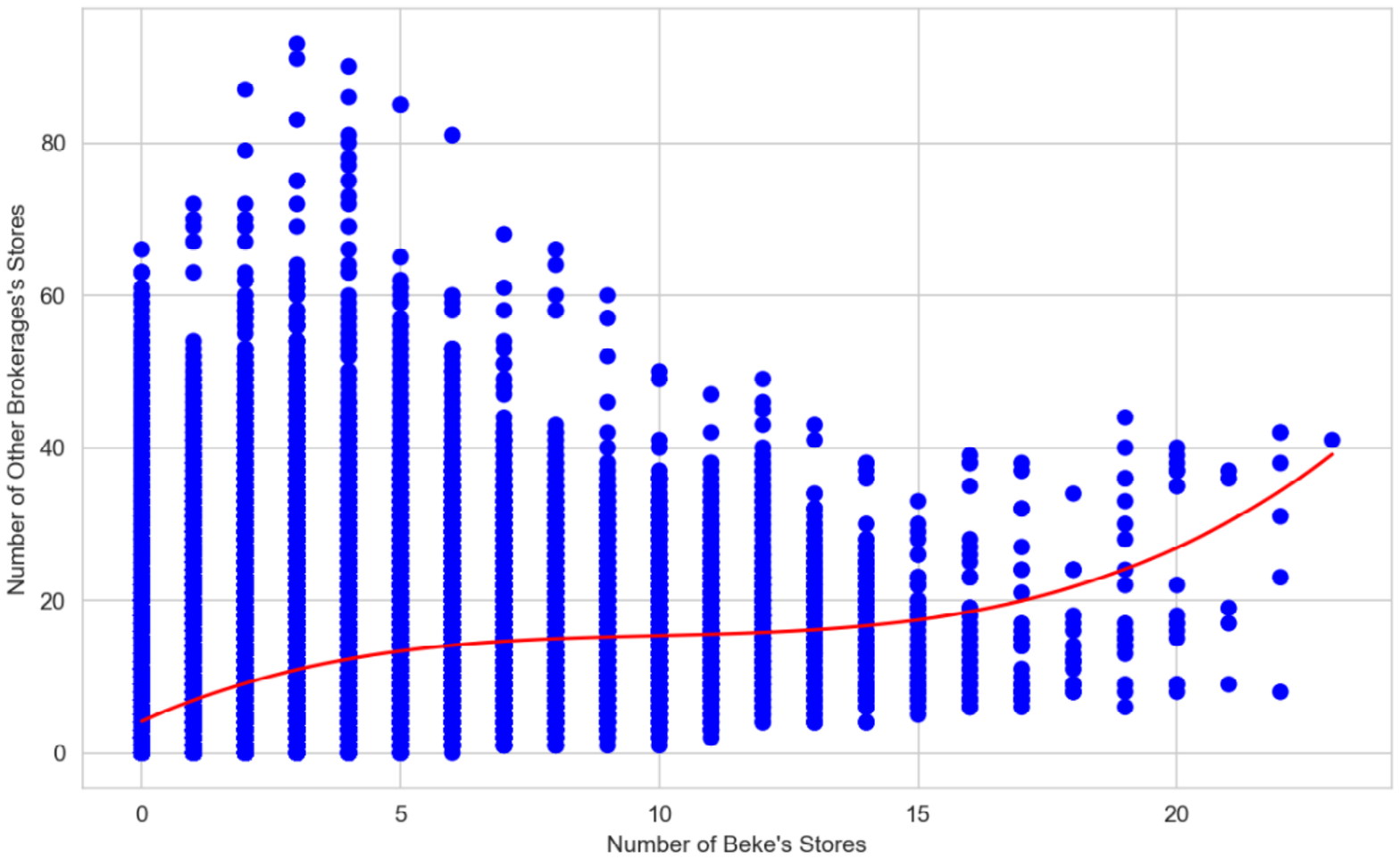}
    \caption{Tendency between Two Types of Brokerages}
    \label{fig:same_distribution}
    Note: the x-axis is the number of Lianjia's stores and the y-axis is the number of other brokerages' stores. The fit is a cubic polynomial fit.
\end{figure}

After constructing our optimal radius, we recalculate the number of Lianjia and other brokerages' stores within this radius. To check the robustness of the data, we divide our data to those with Lianjia and those without Lianjia and to check whether Lianjia's offline stores have influential effect on the transaction effect in the neighborhoods. We can see that for transaction numbers, income, number of house tours and price concession are all significantly different in those neighborhoods with or without Lianjia. In addition, we find that the other brokerages also have the same tendency that they typically open stores with the same strategy as Lianjia, which suggests that Lianjia does not have the market power to exclude competent companies from entering the market, and it also suggests that the market is not monopolized by Lianjia. We plot the relationship between the number of other brokerages's stores and the number of Lianjia's stores and the figure is shown in Figure \ref{fig:same_distribution}. The figure shows that the the number of other brokerages' stores is positively correlated and the general trend tends to be linear, which further suggests that the market is not monopolized by Lianjia.

From Table \ref{tab:statistical_district} we can see that in our metro areas, the neighborhoods with Lianjia within the influential radius tends to have higher number of sales, and the final transaction price is also \textyen 910,000 (approximately 35\%) higher than those neighborhoods without Lianjia. Moreover, the number of other stores within the influential distance is also significantly more than 7.3, which also aligns with our previous intuition that the market is not monopolized by Lianjia. This significance difference suggests that if we treat our sample as a cross sectional data and estimate the result with static model without considering the individual fixed effect, we may get a biased result. Besides, our model may suffer from endogeneity issue, since the number of Lianjia's stores may be endogenous to the transaction price and the number of stores. 

\begin{table}[htb!]
    \centering
    \begin{tiny}
    \caption{Statistical Summary for the Neighborhoods-Data with Lianjia and without Lianjia}
    \begin{tabular}{llllll}
\toprule
Name & Mean without lianjia & SD without lianjia & Mean with lianjia & SD with lianjia & Difference \\
\midrule
\multicolumn{6}{l}{\textbf{Panel 1: }Transaction property} \\
income & 44.44 & 73.85 & 77.68 & 117.3 & -33.24 (-76.638***) \\
lead\_times & 13.59 & 14.56 & 17.09 & 17.50 & -3.501 (-49.823***) \\
price\_concession & -0.0367 & 0.0318 & -0.0351 & 0.0294 & -0.00200 (-12.088***) \\
\multicolumn{6}{l}{\textbf{Panel 2: }Brokerage property} \\
density & 0 & 0 & 0.206 & 0.166 & -0.206 (-384.429***) \\
broker\_410 & 5.362 & 6.871 & 12.66 & 8.410 & -7.300 (-217.614***) \\
watching\_people & 17.61 & 29.11 & 20.29 & 29.40 & -2.682 (-21.214***) \\
end\_price & 260.5 & 254.9 & 351.9 & 292.1 & -91.48 (-76.662***) \\
non\_online\_effect & 0.195 & 0.396 & 0.233 & 0.423 & -0.0380 (-21.384***) \\
watched\_times & 1121 & 1827 & 1233 & 1969 & -112.4 (-13.636***) \\
nego\_times & 4.769 & 7.686 & 5.683 & 10.78 & -0.914 (-22.178***) \\
nego\_period & 150.5 & 187.5 & 166.6 & 239.1 & -16.06 (-17.075***) \\
\multicolumn{6}{l}{\textbf{Panel 3: }Hedonic information property} \\
jiadian & 1.489 & 3.566 & 1.942 & 4.365 & -0.453 (-26.004***) \\
kind & 8.665 & 6.229 & 11.76 & 6.081 & -3.097 (-116.630***) \\
hotel & 3.211 & 5.287 & 5.428 & 6.315 & -2.217 (-87.264***) \\
shop\_mall & 4.554 & 7.539 & 6.698 & 8.489 & -2.144 (-61.405***) \\
museum & 0.617 & 1.533 & 1.023 & 1.869 & -0.406 (-54.395***) \\
old & 0.894 & 1.628 & 1.339 & 1.950 & -0.446 (-56.870***) \\
ktv & 5.179 & 7.853 & 7.305 & 7.759 & -2.126 (-63.096***) \\
mid & 2.059 & 2.329 & 3.368 & 2.853 & -1.309 (-115.077***) \\
prim & 2.812 & 2.846 & 4.343 & 3.205 & -1.531 (-116.172***) \\
west\_food & 3.880 & 7.709 & 7.317 & 10.02 & -3.437 (-87.756***) \\
super & 3.155 & 3.374 & 4.499 & 3.686 & -1.344 (-87.586***) \\
sub & 0.683 & 0.945 & 1.111 & 1.099 & -0.429 (-96.038***) \\
park & 3.422 & 4.575 & 4.569 & 3.944 & -1.147 (-62.708***) \\
\multicolumn{6}{l}{\textbf{Panel 4: }House property} \\
area & 90.82 & 48.57 & 84.97 & 39.15 & 5.845 (31.050***) \\
bedroom & 2.338 & 0.816 & 2.202 & 0.730 & 0.136 (40.958***) \\
toilet & 1.306 & 0.579 & 1.246 & 0.455 & 0.0600 (26.915***) \\
house\_age & 18.08 & 11.68 & 20.73 & 11.77 & -2.651 (-52.316***) \\
floor\_level & 1.854 & 0.975 & 1.933 & 0.931 & -0.0790 (-19.324***) \\
green\_ratio & 0.309 & 0.237 & 0.300 & 0.107 & 0.00900 (12.288***) \\
total\_building & 26.88 & 56.66 & 20.51 & 50.54 & 6.371 (27.653***) \\
total\_floor\_number & 12.75 & 8.454 & 12.97 & 8.485 & -0.221 (-6.035***) \\
living\_room & 1.445 & 0.504 & 1.341 & 0.492 & 0.104 (48.345***) \\
elevator\_ratio & 0.453 & 0.413 & 0.408 & 0.279 & 0.0450 (30.475***) \\
kitchen & 0.983 & 0.150 & 0.983 & 0.125 & 0 (-0.459) \\
floor\_ratio & 4.942 & 329.1 & 2.702 & 9.385 & 2.240 (2.367**) \\
\multicolumn{6}{l}{\textbf{Panel 5: }Regional property} \\
total\_resident & 995.8 & 1079 & 901.8 & 956.5 & 93.97 (21.484***) \\
pm25 & 44.08 & 13.29 & 45.74 & 13.89 & -1.664 (-28.266***) \\
pop & 15506 & 16336 & 24634 & 18866 & -9100 (-118.769***) \\
light & 32.62 & 13.59 & 38.41 & 11.92 & -5.782 (-105.485***) \\
\bottomrule
\end{tabular}

    \label{tab:statistical_district}

    Note that the code book of the variables can be seen in the Appendix \ref{tab:codebook} and with or without Lianjia represent whether there are Lianjia's offline stores within the influential radius.
    \end{tiny}
\end{table}

\begin{table}[htb!]
  \centering
  \begin{tiny}
  \caption{Statistical Summary for the Individual-Data with Lianjia and without Lianjia}
  \begin{tabular}{llllll}
\toprule
Name & Mean without lianjia & SD without lianjia & Mean with lianjia & SD with lianjia & Difference \\
\midrule
\multicolumn{6}{l}{\textbf{Panel 1: }Transaction property} \\
income & 0.493 & 0.656 & 0.797 & 0.974 & -0.304 (-226.763***) \\
lead\_times & 16.83 & 26.13 & 20.47 & 31.23 & -3.648 (-80.341***) \\
price\_concession & -0.028 & 2.984 & -0.027 & 2.776 & -0.001 (-19.908***) \\
\multicolumn{6}{l}{\textbf{Panel 2: }Brokerage property} \\
density & 0 & 0 & 0.235 & 0.182 & -0.235 (-1.1e+03***) \\
broker\_410 & 4.620 & 6.135 & 11.39 & 7.999 & -6.765 (-596.183***) \\
watching\_people & 18.49 & 59.53 & 22.13 & 49.02 & -3.647 (-44.332***) \\
end\_price & 229.8 & 197.1 & 305.3 & 237.0 & -75.50 (-219.538***) \\
non\_online\_effect & 0.232 & 0.422 & 0.238 & 0.426 & -0.00600 (-9.241***) \\
watched\_times & 1128 & 2226 & 1235 & 2387 & -106.8 (-29.718***) \\
nego\_times & 6.275 & 15.16 & 6.636 & 17.65 & -0.361 (-13.948***) \\
nego\_period & 139.0 & 197.5 & 141.8 & 225.4 & -2.840 (-8.544***) \\
\multicolumn{6}{l}{\textbf{Panel 3: }Hedonic information property} \\
jiadian & 0.299 & 1.492 & 0.379 & 1.683 & -0.0800 (-32.152***) \\
kind & 2.253 & 2.136 & 3.347 & 2.421 & -1.094 (-305.752***) \\
hotel & 0.586 & 1.413 & 1.022 & 1.762 & -0.436 (-172.389***) \\
shop\_mall & 0.896 & 2.355 & 1.463 & 3.000 & -0.567 (-132.449***) \\
museum & 0.103 & 0.492 & 0.146 & 0.526 & -0.0430 (-53.977***) \\
old & 0.194 & 0.582 & 0.273 & 0.690 & -0.0790 (-78.187***) \\
ktv & 1.021 & 2.439 & 1.611 & 2.726 & -0.590 (-145.734***) \\
mid & 0.468 & 0.851 & 0.753 & 1.079 & -0.285 (-184.969***) \\
prim & 0.666 & 0.996 & 1.034 & 1.146 & -0.368 (-218.091***) \\
west\_food & 0.716 & 1.942 & 1.423 & 2.640 & -0.707 (-190.757***) \\
super & 2.515 & 2.959 & 3.748 & 3.356 & -1.233 (-248.593***) \\
sub & 0.143 & 0.365 & 0.247 & 0.461 & -0.104 (-158.165***) \\
park & 0.671 & 1.319 & 0.916 & 1.292 & -0.245 (-121.784***) \\
\multicolumn{6}{l}{\textbf{Panel 4: }House property} \\
area & 86.14 & 40.03 & 81.62 & 36.37 & 4.518 (77.381***) \\
bedroom & 2.285 & 0.873 & 2.145 & 0.854 & 0.140 (105.281***) \\
toilet & 1.261 & 0.554 & 1.222 & 0.481 & 0.0400 (50.222***) \\
house\_age & 15.47 & 10.88 & 18.36 & 11.45 & -2.888 (-166.420***) \\
floor\_level & 1.851 & 0.977 & 1.890 & 0.959 & -0.0390 (-25.910***) \\
green\_ratio & 0.325 & 0.168 & 0.319 & 0.0942 & 0.00600 (31.022***) \\
total\_building & 34.11 & 64.21 & 30.05 & 67.64 & 4.061 (39.620***) \\
total\_floor\_number & 15.81 & 9.744 & 15.47 & 9.637 & 0.348 (23.296***) \\
living\_room & 1.436 & 0.574 & 1.319 & 0.575 & 0.117 (131.401***) \\
elevator\_ratio & 0.430 & 0.376 & 0.390 & 0.250 & 0.0400 (84.381***) \\
kitchen & 0.988 & 0.144 & 0.988 & 0.136 & 0 (0.385) \\
floor\_ratio & 2.923 & 124.3 & 2.738 & 6.971 & 0.184 (1.555) \\
\multicolumn{6}{l}{\textbf{Panel 5: }Regional property} \\
total\_resident & 1899 & 1750 & 1903 & 1666 & -4.785 (-1.824*) \\
pm25 & 45.49 & 13.47 & 48.01 & 14.35 & -2.519 (-116.310***) \\
pop & 12369 & 14407 & 19596 & 17151 & -7200 (-289.518***) \\
light & 31.41 & 12.98 & 36.56 & 11.92 & -5.155 (-270.639***) \\
\bottomrule
\end{tabular}

  \label{tab:statistical_individual}

  Note that the code book of the variables can be seen in the Appendix \ref{tab:codebook} and with or without Lianjia represent whether there are Lianjia's offline stores within the influential radius.
  \end{tiny}  
\end{table}

In our analysis, we focus on two dependent variables: \emph{the natural logarithm of Lianjia's transaction number} and \emph{the price concession}. We decided to winsorize the sample for the natural logarithms of Lianjia's transaction number at the 1st and 99th percentiles to mitigate the influence of extreme values. The price concession is defined as $\frac{\text{listing price per $m^2$} - \text{transaction price per $m^2$}}{\text{listing price per $m^2$}} \times 100\%$.

\section{Estimation of Offline Expansion Effect and Online-Mediated Consolidation Effect} \label{sec:mechanism_design}

\subsection{Does Lianjia's Entry influence the segmented market?} \label{subsec:entry_effect}

We adopted the Difference-in-Difference (DID) estimation to estimate the offline's store's entry's influence on the market performance. From the market's performance, the Lianjia's offline's store's entry into the segmented market is an exogenous shock to the two-sided customers, where sellers are more likely to be attracted by the brokerage and buyers are also more likely to be attracted by the more listing information in the neighborhood. Although Lianjia's entry is not a randomized event, the DID estimation method remains consistent in estimating the entry effect by comparing variations in outcome variables before and after Lianjia's entry in the segmented market. We begin by a standard Twoway Fixed Effect (TWFE) model:

\begin{equation}
  Y_{it} = \beta_0 + \beta_1 \text{entry} + \beta_2 \text{time} + \beta_3 * \text{entry * year} + \bf{\alpha} \bf{X}_{it} + \eta_{t} \times \text{bs\_code}_i + \mu_i + \epsilon_{it}.   \label{eq:entry_effect_standard_form}
\end{equation}

where $Y_{it}$ is the two main dependent variables, including $\log(\text{number})$, and price concession. The variable  $\text{entry}$ serves as an indicator representing the year in which Lianjia enters a segmented market. The variable $\text{time}$ denotes the years since Lianjia's market entry. The coefficient $\beta_3$ corresponds to the estimator of the difference-in-differences (DID) treatment effect. $\bf{X}_{it}$ are a set of control variables, including brokerage\_control, hedonic\_control, transaction\_control and region\_control, while $\eta_{t} \times \text{bs\_code}_i$ is the time dummy variable interacting with the fixed effect of the business area, $\mu_i$ is the neighborhood fixed effect and $\epsilon_{it}$ is the random error term. The standard errors are clustered at the each business area level.  To eliminate the effect of the well constructed effect, we drop all the variables that have Lianjia's offline stores before the year 2016 to better estimate the effect of Lianjia's entry on the market.

A simple TWFE model often assumes that the treatment effect is constant over time and homogeneous across units. However, in many empirical settings—especially those with staggered treatment adoption or effects that evolve over time---this assumption can lead to biased or misleading estimates. By contrast, a dynamic DID estimator explicitly models the time path of treatment effects, allowing us to capture both immediate impacts and how those impacts may intensify, diminish, or otherwise change in subsequent periods.

Accordingly, rather than relying on a single treatment indicator, we construct a sequence of indicator variables surrounding Lianjia's entry, delineated as $pre_2$, $pre_1$ (before entry), $entry$, $post_1$, $post_2$, and $post_3$ (successive post-entry intervals), which encapsulate the respective temporal epochs relative to the Lianjia's entry. These dummy variables serve as key independent variables with $pre_1$ as the control group, described in Equation \eqref{eq:entry_effect}:

\begin{equation}
  Y_{it} = \beta_0 +  \beta_1 pre_2 + \beta_2 entry + \beta_3 post_1 + \beta_4 post_2 + \beta_5 post_3 + \bf{\alpha} \bf{X}_{it} + \eta_{t} \times \text{bs\_code}_i + \mu_i + \epsilon_{it}. \label{eq:entry_effect}
\end{equation}

where our key independent variables are described above. The standard errors are clustered at the each business area level. The results are reported in Table \ref{tab:entry_effect}.

\begin{table}[htb!]
  \begin{center}
    \begin{scriptsize}
    \caption{Entry Effect}
    \label{tab:entry_effect}
    {
\def\sym#1{\ifmmode^{#1}\else\(^{#1}\)\fi}
\begin{tabular}{l*{4}{c}}
\toprule
            &\multicolumn{1}{c}{(1)}&\multicolumn{1}{c}{(2)}&\multicolumn{1}{c}{(3)}&\multicolumn{1}{c}{(4)}\\
            &\multicolumn{1}{c}{log(number)}&\multicolumn{1}{c}{log(number)}&\multicolumn{1}{c}{price\_concession}&\multicolumn{1}{c}{price\_concession}\\
\midrule
entry\_flag  &       0.096\sym{***}&                     &      -0.009\sym{***}&                     \\
            &     (0.011)         &                     &     (0.000)         &                     \\
\addlinespace
pre2        &                     &      -0.007         &                     &       0.001         \\
            &                     &     (0.014)         &                     &     (0.000)         \\
\addlinespace
entry       &                     &       0.095\sym{***}&                     &      -0.011\sym{***}\\
            &                     &     (0.012)         &                     &     (0.000)         \\
\addlinespace
post1       &                     &       0.052\sym{***}&                     &      -0.009\sym{***}\\
            &                     &     (0.012)         &                     &     (0.000)         \\
\addlinespace
post2       &                     &       0.013         &                     &      -0.008\sym{***}\\
            &                     &     (0.013)         &                     &     (0.000)         \\
\addlinespace
post3       &                     &       0.015         &                     &      -0.006\sym{***}\\
            &                     &     (0.015)         &                     &     (0.000)         \\
\addlinespace
Brokerage Control &  \checkmark         &  \checkmark         &  \checkmark         &  \checkmark         \\
\addlinespace
Hedonic Control &  \checkmark         &  \checkmark         &  \checkmark         &  \checkmark         \\
\addlinespace
Transaction Control &  \checkmark         &  \checkmark         &  \checkmark         &  \checkmark         \\
\addlinespace
Regional Control &  \checkmark         &  \checkmark         &  \checkmark         &  \checkmark         \\
\midrule
\(N\)       &      103966         &      103966         &      845904         &      845904         \\
R-squared   &       0.816         &       0.816         &       0.285         &       0.286         \\
\bottomrule
\multicolumn{5}{l}{\footnotesize Standard errors in parentheses}\\
\multicolumn{5}{l}{\footnotesize \sym{*} \(p<0.1\), \sym{**} \(p<0.05\), \sym{***} \(p<0.01\)}\\
\end{tabular}
}

    Note: we omit all the control variables in the regression model due to limited space, and detailed descriptions can be seen from Table \ref{tab:statistical_district} and Table \ref{tab:statistical_individual}. Standard Errors are clustered at the business area level.
    \end{scriptsize}
  \end{center}
\end{table}

Table \ref{tab:entry_effect} reports the estimated impact of Lianjia's offline store entry on outcome measures. All regressions include Brokerage, Hedonic, Transaction, and Regional controls. For transaction number, Column (1) shows that the presence of an offline store is associated with a statistically significant increase of 9.6\% in the $\log(\text{number})$. In terms of the dynamic estimator, we find that pre-entry period (pre2) is statistically insignificant, suggesting the absence of anticipation effects prior to entry. In the entry period, the coefficient is 0.095 and statistically significant at the 1\% level, implying an immediate increase of roughly 9.5\% in transaction activity. This effect then attenuates in the first post-entry period (post1) to about 5.2\% and further diminishes in the subsequent periods. This pattern illustrates the dynamic competitive equilibrium: as additional brokerages enhance their efficiency, increased competitive pressures prompt a redistribution of market share, thereby moderating the marginal gains from these efficiency improvements over time. This outcome is consistent with the predictions of Theorem \ref{theorem:simul_efficiency}.

Regarding price concessions, Column (3) indicates that Lianjia's entry to the market yields a negative and significant effect, with a coefficient of -0.009. When dynamics are explicitly modeled in Column (4), the pre-entry period again shows no significant effect. In contrast, during the entry period, there is a robust decline in price concessions of approximately 1.1\%. This significant effect persists into the subsequent post-entry periods, although its magnitude gradually attenuates. Such a pattern suggests that while Lianjia's strategic adjustment in concession levels remains robust in the wake of its expansion, competitive improvements in offline efficiency among rival brokerages may moderate the effect over time, without reversing the overall downward trend in price concessions.

These empirical findings align with the predictions of our theoretical framework. Specifically, the expansion of Lianjia's offline network enhances matching efficiency and strengthens its bargaining power, thereby reducing the need for high price concessions to secure listings. The immediate and significant decline in price concessions following entry is consistent with the model's implication that increased offline presence facilitates higher transaction volumes and improved market positioning. Over time, as competitors adjust their strategies in response to these dynamics, the initial impact attenuates, reflecting the endogenous equilibrium adjustments anticipated by the model.

To verify the robustness of our results, we calculated the Herfindahl-Hirschman Index (HHI) for each segmented market, defined as $HHI = \sum_{i=1}^N (s_i)^2$ where $s_i$ is the market share of firm $i$ expressed as a percentage. Higher HHI values indicate greater market concentration. To further validate the entry effect, we classified the sample into three groups based on HHI values: low HHI ($0 \leq HHI \leq 1,000$), moderate HHI ($1,000 \leq HHI \leq 2,500$) and high HHI ($2,500 \leq HHI \leq 10,000$). This classification allows us to assess the impact of Lianjia's entry across markets with varying levels of competition and concentration, ensuring that our findings are not driven by specific market conditions. 

Table \ref{tab:entry_effect_robustness} presents the estimated effects of Lianjia's offline store entry on key outcome variables, with results stratified by market concentration. In terms of the transaction number effect, the entry coefficient is statistically significant across all market concentration groups. In both the lower and higher concentration markets, the entry period is associated with an approximate 8.1\% increase in transaction numbers, while the moderate concentration group exhibits a slightly smaller, yet significant, increase of 6.5\%. Moreover, in the lower concentration market, the positive effect persists into the first post-entry period (5.2\%), though it attenuates in later periods. In contrast, the dynamic effects in the moderate and high concentration markets are not statistically significant after the entry period, suggesting that the initial income boost gradually dissipates as market participants adjust.

Turning to price concessions, the estimates reveal that Lianjia's offline expansion exerts a statistically significant negative effect on price concessions during the entry period. In particular, the entry period corresponds to an approximate 1.1\% decline in price concessions (columns (4)-(6)). The diminishing effect magnitude is observed consistently across markets with varying levels of competition, suggesting that the effect is global. Overall, these results imply that Lianjia's offline expansion delivers an immediate and robust increase in income across diverse market structures, while the price concession effects are more homogeneous but they are gradually overwhelmed by competitors' gains in efficiencies.

To further check the robustness of our results, we conducted another three tests. In the Appendix, we conducted the same estimation but without additional control variables and the results are reported in Table \ref{tab:robustness_check_platform_1} and Table \ref{tab:robustness_check_platform_2}. The estimated results are shown consistent without additional control variables. In the Appendix we also conducted the robustness check by classifying the market with low and high nighttime light areas and the results are reported in Appendix Table \ref{tab:robustness_nighttime_light_entry}.

We also conducted a placebo test, as illustrated in Appendix Figure \ref{fig:placebo_concession_entry}. For this test, we employed a neighborhood sample with randomly generated treatment effects. Additionally, to assess the impact of heteorogeneity across years, we generated interactions between year and the dummy random treatment effect to determine the significance of these effects. The results indicate that none of the treatment effects are statistically significant, suggesting that our estimates are not influenced by other confounding factors.
% This delayed impact may be attributed to the offline stores gradually improving market transparency and information dissemination, which eventually leads to better price concessions. For the higher HHI group, the entry of Lianjia's offline stores consistently has a significant effect on price concessions. This reinforces the notion that in less competitive markets, offline stores play a crucial role in facilitating better matches between buyers and sellers and improving overall market efficiency. In the Appendix we also conducted the robustness check by classifying the group by matured market and less matured market.

\begin{table}[H]
  \begin{center}
    \begin{scriptsize}
      \caption{Robustness Check of Entry Effect}
      \label{tab:entry_effect_robustness}
      {
\def\sym#1{\ifmmode^{#1}\else\(^{#1}\)\fi}
\begin{tabular}{l*{6}{c}}
\toprule
            &\multicolumn{1}{c}{(1)}&\multicolumn{1}{c}{(2)}&\multicolumn{1}{c}{(3)}&\multicolumn{1}{c}{(4)}&\multicolumn{1}{c}{(5)}&\multicolumn{1}{c}{(6)}\\
            &\multicolumn{1}{c}{log(number)}&\multicolumn{1}{c}{log(number)}&\multicolumn{1}{c}{log(number)}&\multicolumn{1}{c}{price concession}&\multicolumn{1}{c}{price concession}&\multicolumn{1}{c}{price concession}\\
\midrule
pre2        &      -0.032         &      -0.032         &       0.005         &       0.000         &      -0.002         &      -0.001         \\
            &     (0.029)         &     (0.033)         &     (0.029)         &     (0.001)         &     (0.001)         &     (0.001)         \\
\addlinespace
entry       &       0.081\sym{***}&       0.065\sym{***}&       0.081\sym{***}&      -0.011\sym{***}&      -0.011\sym{***}&      -0.011\sym{***}\\
            &     (0.029)         &     (0.023)         &     (0.024)         &     (0.001)         &     (0.001)         &     (0.001)         \\
\addlinespace
post1       &       0.052\sym{*}  &       0.014         &       0.028         &      -0.010\sym{***}&      -0.008\sym{***}&      -0.009\sym{***}\\
            &     (0.029)         &     (0.024)         &     (0.026)         &     (0.001)         &     (0.001)         &     (0.001)         \\
\addlinespace
post2       &       0.007         &      -0.005         &       0.030         &      -0.008\sym{***}&      -0.006\sym{***}&      -0.008\sym{***}\\
            &     (0.031)         &     (0.025)         &     (0.029)         &     (0.001)         &     (0.001)         &     (0.001)         \\
\addlinespace
post3       &       0.012         &       0.003         &       0.011         &      -0.005\sym{***}&      -0.004\sym{***}&      -0.008\sym{***}\\
            &     (0.034)         &     (0.026)         &     (0.034)         &     (0.001)         &     (0.001)         &     (0.001)         \\
\addlinespace
Brokerage Control &  \checkmark         &  \checkmark         &  \checkmark         &  \checkmark         &  \checkmark         &  \checkmark         \\
\addlinespace
Hedonic Control &  \checkmark         &  \checkmark         &  \checkmark         &  \checkmark         &  \checkmark         &  \checkmark         \\
\addlinespace
Transaction Control &  \checkmark         &  \checkmark         &  \checkmark         &  \checkmark         &  \checkmark         &  \checkmark         \\
\addlinespace
Regional Control &  \checkmark         &  \checkmark         &  \checkmark         &  \checkmark         &  \checkmark         &  \checkmark         \\
\midrule
\(N\)       &       31741         &       23336         &       34022         &      284794         &      236617         &      320111         \\
R-squared   &       0.849         &       0.880         &       0.851         &       0.313         &       0.310         &       0.311         \\
\bottomrule
\multicolumn{7}{l}{\footnotesize Standard errors in parentheses}\\
\multicolumn{7}{l}{\footnotesize \sym{*} \(p<0.1\), \sym{**} \(p<0.05\), \sym{***} \(p<0.01\)}\\
\end{tabular}
}

    Note: we omit all the control variables in the regression model, and detailed descriptions can be seen from Table \ref{tab:statistical_district} and Table \ref{tab:statistical_individual}. Standard Errors are clustered at the business area level.
    \end{scriptsize}
  \end{center}
\end{table}

\subsection{Estimate Lianjia's platform consolidation effect} \label{subsec:acn_strategy}

To empirically estimate the effect of Lianjia's platform strategy, we consider an exogenous shock that occurred during our study period: Lianjia's implementation of a downstream consolidation strategy. This strategy involves the integration of offline stores with online platforms, leveraging the advantages of Lianjia's ACN strategy. The ACN strategy subdivides the entire process of buying and selling a house into distinct parts, with each part managed by a specific agent or store. By sharing transaction dividends among multiple stores, Lianjia fosters cooperation with other market competitors and integrates their resources to enhance service quality for customers. This approach is designed to improve efficiency and customer satisfaction by combining online and offline resources, thus providing a comprehensive and streamlined service experience. 

This strategy is a significant change in Lianjia's business model, and it is expected to have a significant impact on the market. In addition to this, Lianjia also opened up the form of franchises and gradually started platform integration. To empirically measure the effect of platform consolidation on offline store operations, we first counted the number of all non-Lianjia stores on Lianjia's Beke platform within a radius of 410 meters. This allowed us to generate a dummy variable, $\text{Treatment}_{it}$, which equals one if the ratio $\frac{\text{Lianjia}}{\text{Beke}} < 0.8$ in this area and the year is 2018 or later.  This ratio is selected because if Lianjia accounts for more than 80\% of the Beke's offline stores, the strategy's effect is negligible in this segmented market. 

We first consider the standard TWFE model:

\begin{equation}
    Y_{it} = \beta_0 + \beta_1 \text{Treatment}_{it} + \beta_2 \text{time}_{it} + \beta_3 \text{Treatment}_{it} * \text{time}_{it} + \bf{\alpha} \bf{X}_{it} + \eta_{t} \times \text{bs\_code}_i + \mu_i + \epsilon_{it}. \label{eq:platform_consolidation_twfe}
\end{equation}

Similar to the estimation of the entry effect, we also consider the dynamic DID estimation model. We generate a set of dummy variable $\text{post\_j\_treatment}_{it}$, where $j \in \{1, 2, 3\}$ and $\text{pre\_treatment}_{it}$ and include them in our regression model. We then consider the following regression model:

\begin{equation}
  \begin{aligned}
    Y_{it} & = \beta_0 + \beta_1 \text{prev\_treatment}_{it} + \beta_2 \text{treatment}_{it} + \sum_{j=3}^5 \beta_{i} \text{post\_j\_treatment}_{it} + \\
           & \bf{\alpha} \bf{X}_{it} + \eta_{t} \times \text{bs\_code}_i + \mu_i + \epsilon_{it}. \label{eq:acn_assessment}
  \end{aligned}
\end{equation}

where our key independent variables are described above. Other settings are consistent with previous Table \ref{tab:entry_effect} and the result is reported in Table \ref{tab:platform_consolidation}.

\begin{table}[H]
  \begin{center}
    \begin{scriptsize}
      \caption{Platform Consolidation Effect}
      \label{tab:platform_consolidation}
      {
\def\sym#1{\ifmmode^{#1}\else\(^{#1}\)\fi}
\begin{tabular}{l*{4}{c}}
\toprule
            &\multicolumn{1}{c}{(1)}&\multicolumn{1}{c}{(2)}&\multicolumn{1}{c}{(3)}&\multicolumn{1}{c}{(4)}\\
            &\multicolumn{1}{c}{log(number)}&\multicolumn{1}{c}{log(number)}&\multicolumn{1}{c}{price concession}&\multicolumn{1}{c}{price concession}\\
\midrule
treatment\_flag&       0.021\sym{**} &                     &      -0.009\sym{***}&                     \\
            &     (0.009)         &                     &     (0.000)         &                     \\
\addlinespace
pre2\_treatment&                     &      -0.012         &                     &      -0.000         \\
            &                     &     (0.009)         &                     &     (0.000)         \\
\addlinespace
treatment   &                     &      -0.001         &                     &      -0.008\sym{***}\\
            &                     &     (0.008)         &                     &     (0.000)         \\
\addlinespace
post1\_treatment&                     &       0.058\sym{***}&                     &      -0.009\sym{***}\\
            &                     &     (0.010)         &                     &     (0.000)         \\
\addlinespace
post2\_treatment&                     &       0.059\sym{***}&                     &      -0.010\sym{***}\\
            &                     &     (0.012)         &                     &     (0.000)         \\
\addlinespace
post3\_treatment&                     &       0.027\sym{*}  &                     &      -0.010\sym{***}\\
            &                     &     (0.015)         &                     &     (0.001)         \\
\addlinespace
Brokerage Control &  \checkmark         &  \checkmark         &  \checkmark         &  \checkmark         \\
\addlinespace
Hedonic Control &  \checkmark         &  \checkmark         &  \checkmark         &  \checkmark         \\
\addlinespace
Transaction Control &  \checkmark         &  \checkmark         &  \checkmark         &  \checkmark         \\
\addlinespace
Regional Control &  \checkmark         &  \checkmark         &  \checkmark         &  \checkmark         \\
\midrule
\(N\)       &      133420         &      133420         &     1246299         &     1246299         \\
R-squared   &       0.853         &       0.853         &       0.290         &       0.290         \\
\bottomrule
\multicolumn{5}{l}{\footnotesize Standard errors in parentheses}\\
\multicolumn{5}{l}{\footnotesize \sym{*} \(p<0.1\), \sym{**} \(p<0.05\), \sym{***} \(p<0.01\)}\\
\end{tabular}
}

    Note: we omit all the control variables in the regression model, and detailed descriptions can be seen from Table \ref{tab:statistical_district} and Table \ref{tab:statistical_individual}. Standard Errors are clustered at the business area level.
    \end{scriptsize}
  \end{center}
\end{table}

Table \ref{tab:platform_consolidation} provides robust evidence on the effects of Lianjia's platform consolidation strategy. In our dynamic DID framework, the immediate (treatment) period shows no statistically significant change in either transaction volume or price concessions. However, in subsequent post-treatment periods, significant positive effects emerge.

For transaction volume, the dynamic specification reveals that, after consolidation, there is a statistically significant increase of approximately 5.8\% in the first post-treatment period and 5.9\% in the second period. Although the effect declines to 2.7\% in the third period, it remains marginally significant, indicating a sustained improvement in market activity following consolidation. In terms of price concessions, the results indicate that consolidation is associated with a statistically significant reduction. Specifically, Column (3) shows that the treatment flag is associated with a 0.9\% decrease in price concessions, while the dynamic specification in Column (4) reveals an immediate decline of approximately 0.8\% in the treatment period. This reduction persists in subsequent periods, with estimated decreases of about 0.9\% in the first period and 1.0\% later periods. Its persistence suggests that consolidation exerts a robust and lasting influence on reducing price concessions. These empirical results align with our theoretical predictions that platform consolidation enhances market coverage and bargaining power, enabling brokerages to lower their concession levels to capture improved margins.

These results align with our theoretical predictions that platform consolidation enhances market performance by improving information aggregation and increasing transaction efficiency. The delayed onset of significant effects suggests that, although consolidation does not produce an immediate boost, it yields a more pronounced and persistent impact on transaction effectiveness over time compared to traditional offline store entry strategies. Moreover, the observed reduction in price concessions suggests that heightened seller competition may prompt adjustments in listing prices, reflecting increased bargaining pressure in a more transparent market environment. Notably, in contrast to the transient effects observed with offline expansion---where the initial impact on price concessions attenuates over time---the effects of platform consolidation are more persistent. This sustained reduction underscores the strategic advantage of consolidation in strengthening a brokerage's market position, while simultaneously imposing potential trade-offs for sellers.

We further examine how market concentration influences the effectiveness of the platform consolidation strategy by partitioning the sample into groups based on their HHI—low, moderate, and high—and estimating the model for each group. Table \ref{tab:heter_platform_did} presents the regression estimates for two key outcome measures: the logarithm of the number of transactions (columns 1-3) and price concessions (columns 4-6).

In the low HHI markets, the consolidation strategy is associated with robust increases in transaction volume. For example, the post-treatment coefficients are positive and statistically significant—ranging from approximately 4.7\% to 7.5\% increases in log-transactions---suggesting that the consolidation strategy successfully drives customer inflows and market activity. In contrast, the corresponding estimates for the high HHI group do not reach statistical significance, implying that in more concentrated markets the strategy does not yield appreciable benefits for Lianjia's transaction performance. With respect to price concessions, the findings indicates a more homogeneous effect across specifications. Specifically, the results consistently demonstrate that platform consolidation is associated with a statistically significant reduction in price concessions across all post-treatment periods. This contrasts with earlier mixed findings and highlights a distinct platform effect that robustly improves the efficiency of the buyer-seller matching process.

Overall, these updated results suggest that Lianjia's platform consolidation strategy is most effective in less concentrated markets, where it significantly boosts transaction numbers. Platform consolidation yields a consistent, statistically significant reduction in price concessions. This finding implies that while the strategy effectively expands the customer base, its impact on negotiation outcomes is distinct from that of enhanced offline efficiency by competing brokerages, which tends to attenuate the effect when adopted more broadly.

\begin{table}
  \begin{center}
    \begin{scriptsize}
      \caption{Robustness Check of Online Consolidation Effect}
      \label{tab:heter_platform_did}
      {
\def\sym#1{\ifmmode^{#1}\else\(^{#1}\)\fi}
\begin{tabular}{l*{6}{c}}
\toprule
            &\multicolumn{1}{c}{(1)}&\multicolumn{1}{c}{(2)}&\multicolumn{1}{c}{(3)}&\multicolumn{1}{c}{(4)}&\multicolumn{1}{c}{(5)}&\multicolumn{1}{c}{(6)}\\
            &\multicolumn{1}{c}{log(number)}&\multicolumn{1}{c}{log(number)}&\multicolumn{1}{c}{log(number)}&\multicolumn{1}{c}{price concession}&\multicolumn{1}{c}{price concession}&\multicolumn{1}{c}{price concession}\\
\midrule
pre2\_treatment&       0.003         &      -0.018         &      -0.030         &      -0.000         &      -0.000         &       0.000         \\
            &     (0.015)         &     (0.013)         &     (0.025)         &     (0.000)         &     (0.000)         &     (0.001)         \\
\addlinespace
treatment   &      -0.022         &      -0.002         &       0.006         &      -0.010\sym{***}&      -0.008\sym{***}&      -0.008\sym{***}\\
            &     (0.015)         &     (0.014)         &     (0.029)         &     (0.001)         &     (0.000)         &     (0.001)         \\
\addlinespace
post1\_treatment&       0.047\sym{**} &       0.069\sym{***}&       0.037         &      -0.010\sym{***}&      -0.009\sym{***}&      -0.008\sym{***}\\
            &     (0.019)         &     (0.017)         &     (0.034)         &     (0.001)         &     (0.001)         &     (0.001)         \\
\addlinespace
post2\_treatment&       0.067\sym{***}&       0.075\sym{***}&       0.018         &      -0.010\sym{***}&      -0.009\sym{***}&      -0.010\sym{***}\\
            &     (0.020)         &     (0.020)         &     (0.040)         &     (0.001)         &     (0.001)         &     (0.001)         \\
\addlinespace
post3\_treatment&       0.002         &       0.068\sym{***}&       0.034         &      -0.009\sym{***}&      -0.010\sym{***}&      -0.008\sym{***}\\
            &     (0.024)         &     (0.024)         &     (0.070)         &     (0.001)         &     (0.001)         &     (0.001)         \\
\addlinespace
Brokerage Control &  \checkmark         &  \checkmark         &  \checkmark         &  \checkmark         &  \checkmark         &  \checkmark         \\
\addlinespace
Hedonic Control &  \checkmark         &  \checkmark         &  \checkmark         &  \checkmark         &  \checkmark         &  \checkmark         \\
\addlinespace
Transaction Control &  \checkmark         &  \checkmark         &  \checkmark         &  \checkmark         &  \checkmark         &  \checkmark         \\
\addlinespace
Regional Control &  \checkmark         &  \checkmark         &  \checkmark         &  \checkmark         &  \checkmark         &  \checkmark         \\
\midrule
\(N\)       &       41849         &       53806         &       24273         &      327800         &      576496         &      338586         \\
R-squared   &       0.851         &       0.871         &       0.895         &       0.315         &       0.300         &       0.313         \\
\bottomrule
\multicolumn{7}{l}{\footnotesize Standard errors in parentheses}\\
\multicolumn{7}{l}{\footnotesize \sym{*} \(p<0.1\), \sym{**} \(p<0.05\), \sym{***} \(p<0.01\)}\\
\end{tabular}
}

    Note: we omit all the control variables in the regression model, and detailed descriptions can be seen from Table \ref{tab:statistical_district} and Table \ref{tab:statistical_individual}. Standard Errors are clustered at the business area level.
    \end{scriptsize}
  \end{center}
\end{table}

Appendix Figure \ref{fig:treatment_consolidation} illustrates the distribution of income across the treatment and control groups. Prior to the intervention, the two groups exhibit parallel trends, indicating no significant differences. The treatment effect becomes significant only after the first period, aligning with our empirical findings. 

Same as the Section \ref{subsec:entry_effect}, we performed the estimation without incorporating additional control variables, as documented in Appendix Table \ref{tab:robustness_check_platform_1} and Appendix Table \ref{tab:robustness_check_platform_2}. The findings remain consistent with our initial estimates. In the Appendix Table \ref{tab:robustness_nighttime_light_consolidation}, we also categorize the sample into areas with low and high nighttime light. The results from this stratification further corroborate the robustness of our original conclusions.

We also conducted a placebo test, as illustrated in Appendix Figure \ref{fig:placebo_concession_plat} like our previous estimation of the offline store expansion. The test result also shows that there is no other confounding factors affecting our analysis.

\section{Discussion and Conclusion} \label{sec:conclusion}

Our study highlights the transformative effects of strategic expansion and platform consolidation in intermediary markets,  supported by a comprehensive theoretical framework and validated using micro-level transaction data from China's real estate sector. By analyzing the dual strategies of offline network growth and online platform integration, we identify distinct but complementary mechanisms through which intermediaries enhance their market position and influence overall market outcomes. The case of Lianjia, a leading real estate brokerage, serves as a key example to illustrate how these strategies operate in practice, offering broader insights into the dynamics of digital and physical expansion across platform-based industries.

Empirically, Lianjia's entry into segmented markets generates an immediate and substantial increase in transaction volumes (approximately 9-10\%) and reduces price concessions by about 1\%, reflecting improved matching efficiency and enhanced bargaining power driven by localized network effects. However, these positive effects attenuate over subsequent periods as market participants adjust to the new equilibrium, particularly when competing brokerages also realize efficiency gains and expand their networks. In contrast, our analysis of platform consolidation—exemplified by ACN reveals a delayed yet sustained improvement in market performance. Specifically, consolidation is associated with a 5-6\% increase in transaction volumes, especially in less concentrated markets, along with a persistent reduction in price concessions. 

Our theoretical model accurately predicts these empirical patterns, demonstrating that offline expansions enhance brokerage attractiveness by leveraging local and cross-segment spillovers, while platform consolidation strengthens brokerage visibility and overall market coverage. Importantly, the model highlights the conditions under which offline presence and price concessions act as strategic substitutes, inducing brokerages to offer lower price concessions following physical expansion and digital consolidation. Welfare analysis further underscores that these strategies enhance overall market efficiency and sellers welfare, although increased market concentration raises critical questions regarding long-term competition dynamics.

Overall, our findings provide nuanced insights into the complementary roles of offline presence and online platform integration, illustrating how dominant brokerages strategically leverage both physical and digital assets to reinforce market power, enhance transaction effectiveness, and sustain competitive advantages.

These findings carry several important implications. Firstly, it indicates that with the advent of new technologies, the online consolidation effect is becoming increasingly important for real estate companies. This shift underscores the need for firms to enhance their digital strategies to remain competitive. Secondly, our findings have broader significance for other industries where online consolidation can potentially replace traditional offline clustering. For example, in the retail sector, online marketplaces such as Alibaba in China have significantly outpaced physical stores by consolidating various vendors into a single, accessible platform. Thirdly, this study demonstrates that, under the platform's consolidation effect, offline stores can enhance revenue and improve transaction efficiency. These findings can be extended to other similar fields, as discussed by \citep{10.1257/jep.28.2.3}. Lastly, our paper use empirical data to shows the broad implication of the research \citep{bergemann_data_2024}, where the platform can benefits from increased bargaining power vis-\'a-vis sellers by exploiting its information advantage. % \citep{gilbukh_goldsmith-pinkham_2019}

Future research could explore several potential directions. First, investigating the long-term impacts of online and offline integration on market competition and consumer behavior in various real estate markets globally could provide comparative insights. Second, examining the role of emerging technologies, such as artificial intelligence and blockchain, in further enhancing the efficiency and transparency of real estate transactions would be valuable. Third, assessing the socio-economic implications of platform monopolization, particularly in terms of access to affordable housing and regional economic disparities, could offer important policy implications. Lastly, exploring the adaptability of the online-offline integration model in other sectors, such as healthcare and education, could extend the applicability of these findings and contribute to a broader understanding of digital transformation across industries.

% main structure:
% 1. conclusions
% 2. why the exercise is important
% 3. policy implications
% 4. further research directions

%%%%%%%%%%%%%%%%%%%%%%%%%%%%%%%%%%%%%%%%%%%%%%%%%
\clearpage
\begin{singlespace}
%\bibliographystyle{plainnat}
%\bibliographystyle{chicago}
% \bibliographystyle{aer}
% \bibliography{our-cites.bib}

\end{singlespace}
%%%%%%%%%%%%%%%%%%%%%%%%%%%%%%%%%%%%%%%%%%%%%%%%%

%%%%%%%%%%%%%%%%%%%%%%%%%%%%%%%%%%%%%%%%%%%%%%%%%%%%%%%%%%%%%%%%%%%%%%%%%%%%%%%%

% APPENDIX

%%%%%%%%%%%%%%%%%%%%%%%%%%%%%%%%%%%%%%%%%%%%%%%%%
%%%%% These commands start the appendix and change the Table & Figure numbering
\newpage
\appendix
\setcounter{table}{0}
\renewcommand{\tablename}{Appendix Table}
\renewcommand{\figurename}{Appendix Figure}
\renewcommand{\thetable}{\thesection. \arabic{table}}
\setcounter{figure}{0}
\renewcommand{\thefigure}{\thesection. \arabic{figure}}
%%%%%%%%%%%%%%%%%%%%%%%%%%%%%%%%%%%%%%%%%%%%%%%%%

\section{Proofs}

\begin{proof}[Proof of Lemma \ref{lemma:monotonicity_coverage}]
  Let the vector of price concessions $\mathbf{c}_i$ be fixed and fixed platform membership $|P(i)|$. Consider two branch configurations for brokerage $i$ in segment $m$ and its adjacent segments. Let:

  \begin{equation*}
    \mathbf{n}_i = (n_{im}, (n_{im^\prime})_{m^\prime \in \mathcal{N}(m)})
  \end{equation*}

  and suppose that we consider two alternative configurations $\mathbf{n}_i^\prime = (n_{im}^\prime, (n_{im^\prime}^\prime)_{m^\prime \in \mathcal{N}(m)})$, satisfying $n_{im}^\prime \geq n_{im}$ and $n_{im^\prime}^\prime \geq n_{im^\prime}$ for all $m^\prime \in \mathcal{N}(m)$.

  Since the function $h: \mathbb{Z}_+ \mapsto \mathbb{R}_+$ is non-decreasing, we have for each relevant segment:

  \begin{equation*}
    h(n_{im}^\prime) \geq h(n_{im}) \text{ and } h(n_{im^\prime}^\prime) \geq h(n_{im^\prime}) \text{ for all } m^\prime \in \mathcal{N}(m)
  \end{equation*}

  Thus, by definition of the effective store presence function,

  \begin{equation*}
    \begin{aligned}
      f(n_{im^\prime}, n_{i, \mathcal{N}(m)}^\prime) & = h(n_{im}^\prime) + \gamma \sum_{m^\prime \in \mathcal{N}(m)} \omega_{m, m^\prime} h(n_{im^\prime}^\prime) \\
      & \geq h(n_{im}) + \gamma \sum_{m^\prime \in \mathcal{N}(m)} \omega_{m, m^\prime} h(n_{im^\prime}) \\
      & = f(n_{im}, n_{i, \mathcal{N}(m)})
    \end{aligned}
  \end{equation*}

  Multiplying both sides by the positive constants $\alpha_i > 0$ and $\psi(c_{im}, |P(i)|) > 0$, we obtain:

  \begin{equation*}
    R_{im}^\prime = \alpha_i f(n_{im}^\prime, n_{i, \mathcal{N}(m)}^\prime) \psi(c_{im}, |P(i)|) \geq \alpha_i f(n_{im}, n_{i, \mathcal{N}(m)}) \psi(c_{im}, |P(i)|) = R_{im} 
  \end{equation*}

  Thus, the per-segment attractiveness $R_{im}$ is weakly increasing in the branch configuration.
  
  Next, consider the captured share of listings:

  \begin{equation*}
    \sigma_{im} = \dfrac{R_{im}}{\sum_{k \in \mathcal{I}} R_{km}} \cdot L_m
  \end{equation*}

  Assume that the branch configuration for brokerage $i$ changes as above, while the branch configurations and concession levels of all other brokerages remain fixed. We denote $x = R_{im}$ and $T = \sum_{k \in \mathcal{I} \setminus \{i\}} R_{km}$, so that $\sigma_{im} = \frac{x}{x + T} \cdot L_m$. Since the function is weakly increasing in $x$ for any fixed $T \geq 0$, we have: $\frac{x^\prime}{x^\prime + T} \geq \frac{x}{x + T}$. Therefore, the share of listings $\sigma_{im}$ is weakly increasing in $R_{im}$, and consequently, in the branch configuration $\mathbf{n}_i$.

  Now, fix the branch configuration $\mathbf{n}_i$ and consider two concession levels in segment $m$: an initial concession $c_{im}$ and an increased concession $c_{im}^\prime$ with:

  \begin{equation*}
    c_{im}^\prime \geq c_{im}
  \end{equation*}

  Because the function $\psi: \mathbb{R}_+ \times \mathbb{N} \mapsto \mathbb{R}_+$ is continuously differentiable with respect to its first argument, and $\frac{\partial \psi}{\partial c_{im}}(c_{im}, |P(i)|) > 0$ then we have: $\psi(c_{im}^\prime, |P(i)|) \geq \psi(c_{im}, |P(i)|)$. Since $f(n_{im}, n_{i, \mathcal{N}(m)})$ and $\sigma_{im}$ are unchanged, we have:

  \begin{equation*}
    R_{im}^\prime = \alpha_i f(n_{im}, n_{i, \mathcal{N}(m)}) \psi(c_{im}^\prime, |P(i)|) \geq \alpha_i f(n_{im}, n_{i, \mathcal{N}(m)}) \psi(c_{im}, |P(i)|) = R_{im}
  \end{equation*}
  
  Using the same argument regarding the function $g(x) = \frac{x}{x + T}$, an increase in $R_{im}$ implies the that the captured share $\sigma_{im}$ is weakly increasing.\footnote{$g^\prime(x) = \frac{T}{(x + T)^2} > 0$ for all $x \geq 0$ and $T \geq 0$, which implies that $g(x)$ is increasing in $x$.} 

  The last part of the lemma follows from the fact that $\psi(\cdot)$ is increasing in $|P(i)|$.
\end{proof}

\begin{proof}[Proof of lemma \ref{lemma:existence_best_response}]

  We denote the offline branch configuration: $S_i^n = \prod_{m=1}^M \{0, 1, \ldots, \overline{n}_{im}\}$ and denote price concession offers: $S_i^c = \prod_{m=1}^M [0, \overline{c}_{im}]$.

  For each segment $m$, the set $\{0, 1, \ldots, \overline{n}_{im}\}$ is finite. In any discrete topology, every finite set is compact. Since a finite Cartesian product of compact sets is compact, it follows that $S_i^n$ is compact.

  Each interval $\left[0, \overline{c}_{im}\right]$ is closed and bounded subset of $\mathbb{R}$ and is compact by the Heine-Borel theorem. Since the finite Cartesian product of compact sets in $\mathbb{R}$ is also compact, $S_i^c$ is compact.

  Consequently, the overall strategy space is compact.

  The profit function for brokerage $i$ is given by:

  \begin{equation*}
    \pi_{i}(s_i, s_{-i}) = \sum_{m=1}^M p_{im} Q_{im} - \left[ \sum_{m=1}^M F_i n_{im} + H(\sum_{m=1}^M n_{im}) \right]
  \end{equation*}

  It is easy to see that the per-transaction profit is linear in $c_{im}$ and the $\psi(c_{im}, |P(i)|)$ in the definition is continuously differentiable in its first argument. Moreover, we have that all other functions entering the definition of $Q_{im}$ and the cost functions are continuous in the concession variables. Hence, for any fixed branch configuration and fixed strategies $s_{-i}$ of the other brokerages, the mapping

  \begin{equation*}
    \mathbf{c}_i \mapsto \pi_i ((\mathbf{n}_i, \mathbf{c}_i), s_{-i})
  \end{equation*}

  is continuous.
  
  Since the overall strategy space $S_i$ is compact and the profit function $\pi_i(s_i, s_{-i})$ is continuous in $\mathbf{c}_i$, the Weierstrass extreme value theorem guarantees that the profit function attains its maximum on $S_i$.That is, for any fixed strategies $s_{-i}$ of the other brokerages, there exists at least one best response

  \begin{equation*}
    s_i^\star = (\mathbf{n}_i^\star, \mathbf{c}_i^\star) \in S_i \text{ such that } \pi_i(s_i^\star, s_{-i}) \geq \pi_{i}(s_i, s_{-i}) \text{ for all } s_i \in S_i
  \end{equation*}

  Define the best response correspondence:

  \begin{equation*}
      BR_i(s_{-i}) = \arg\max_{s_i \in S_i} \pi_i(s_i, s_{-i})
  \end{equation*}

  Since each $BR_i(s_{-i})$ is non-empty, compact-valued, and upper-hemicontinuous (by the Maximum Theorem), the product best response correspondence:

  \begin{equation*}
      BR(s) = \prod_{i \in \mathcal{I}} BR_i(s_{-i})
  \end{equation*}

  is also non-empty, compact-valued, and upper-hemicontinuous. By Kakutani's Fixed Point Theorem, there exists at least one Nash Equilibrium $\mathbf{s}^\star \in \prod_{i \in \mathcal{I}} S_i$ such that $\mathbf{s}^\star \in BR(\mathbf{s}^\star)$. This completes the proof of equilibrium existence.
\end{proof}

\begin{proof}[Proof of Lemma \ref{lemma:exogenous_c_alpha}]
    We first prove that $\frac{\partial c_{im}}{\partial \alpha_i} < 0$ under the submodularity condition of $\pi_{im}$. Note that in equilibrium, the first order condition of $\pi_{im}$ yields:

    \begin{equation*}
        \frac{\partial \pi_{im}}{\partial c_{im}} = 0 \implies \frac{\partial Q_{im}}{\partial c_{im}} p_{im} + \frac{\partial p_{im}}{\partial c_{im}} Q_{im} = 0
    \end{equation*}

    where $p_{im} = \beta \cdot (\mu_m - c_{im})$ and note that $\frac{\partial p_{im}}{\partial c_{im}} = -\beta$ and this implies:

\begin{equation*}
    (\mu_m - c_{im}) \frac{\partial Q_{im}}{\partial c_{im}} = Q_{im}
\end{equation*}

Then define the implicit function:

\begin{equation*}
    F(c_{im}, n_{im}; \alpha_i) \equiv (\mu_m - c_{im}) \frac{\partial Q_{im}(c_{im}, n_{im}; \alpha_i)}{\partial c_{im}} - Q_{im}(c_{im}, n_{im}; \alpha_i)
\end{equation*}

Then applying the implicit function theorem, we have:

\begin{equation*}
    \frac{\partial c_{im}}{\partial \alpha_i} = -\frac{\frac{\partial F_{im}}{\partial \alpha_i}}{\frac{\partial F_{im}}{\partial c_{im}}}
\end{equation*}

Differentiating with respect to $c_{im}$ we have:

\begin{equation*}
    \frac{\partial F_{im}}{\partial c_{im}} = -\frac{Q_{im}}{\partial c_{im}} + (\mu_m - c_{im}) \frac{\partial^2 Q_{im}}{\partial c_{im}^2}
\end{equation*}

and note that $\frac{Q_{im}}{\partial c_{im}} > 0$ and $\frac{\partial^2 Q_{im}}{\partial c_{im}^2} < 0$ then we have: $\frac{\partial F_{im}}{\partial c_{im}} < 0$;

Note that $F_{im} = \frac{1}{\beta} \cdot \frac{\partial \pi_{im}}{\partial c_{im}}$ and then under the condition of submodularity and the proof of Proposition \ref{prop:offline_expansion} part (i) we have:

\begin{equation*}
  \begin{aligned}
    & \frac{\partial \pi_{im}}{\partial c_{im}}(n_{im } + \Delta n_{im}, c_{im}, \alpha_i) - \frac{\partial \pi_{im}}{\partial c_{im}}(n_{im}, c_{im}, \alpha_i) < 0 \ \forall \Delta n_{im} > 0 \text{ and } n_{im}^\star(\alpha_i) \text{ non-decreasing in } \alpha_i \\
    & \implies \frac{\partial \pi_{im}}{\partial c_{im}}(n_{im}^\star(\alpha_i^\prime), c_{im}, \alpha_i^\prime) - \frac{\partial \pi_{im}}{\partial c_{im}}(n_{im}^\star(\alpha_i), c_{im}, \alpha_i) < 0 \ \forall \alpha_i^\prime > \alpha_i \\
    & \implies \frac{\partial \pi_{im}}{\partial c_{im}} \text{ is decreasing in $\alpha_i$ } \implies \frac{\partial F_{im}}{\partial \alpha_i} < 0
  \end{aligned}
\end{equation*}

Therefore, $\frac{\partial c_{im}}{\partial \alpha_i} < 0$. Next, we argue that $\frac{\partial c_{km}}{\partial \alpha_i} > 0$ for $k \neq i$. Similarly to the proof above, we can define the implicit function $G_{km}(c_{km}, n_{km}, c_{im}, n_{im}; \alpha_i) \equiv \frac{\partial \pi_{km}}{\partial c_{km}} = 0$ and then based on the implicit function theorem, we have:

\begin{equation*}
    \frac{\partial c_{km}}{\partial \alpha_i} = -\frac{\frac{\partial G_{km}}{\partial \alpha_i}}{\frac{\partial G_{km}}{\partial c_{km}}}
\end{equation*}

Since $\frac{\partial G_{km}}{\partial c_{km}} < 0$ and under the condition of submodularity we have:

\begin{equation*}
    \frac{\partial G_{km}}{\partial \alpha_i} = \frac{\partial^2 \pi_{km}}{\partial c_{km}\partial \alpha_i} = \frac{\partial G_{km}}{\partial \alpha_i}
\end{equation*}

Note that $\frac{\partial \pi_{km}}{\partial c_{km}}(n_{km} + \Delta n_{km}) - \frac{\partial \pi_{km}}{\partial c_{km}}(n_{km}) > 0$ for all $\Delta n_{km} > 0$ and $n_{km}^\star(\alpha_i)$ is non-increasing in $\alpha_i$. Then for $\alpha_i^\prime > \alpha_i$ we have: $\frac{\partial \pi_{km}}{\partial c_{km}}(n_{km}^\star(\alpha_i^\prime), c_{km}, \alpha_i) - \frac{\partial \pi_{km}}{\partial c_{km}}(n_{km}^\star(\alpha_i), c_{km}, \alpha_i) > 0$ and this implies $\frac{\partial \pi_{km}}{\partial c_{km}}$ is non-decreasing in $\alpha_i \implies \frac{\partial G_{km}}{\partial \alpha_i} > 0$

and this results $\frac{\partial c_{km}}{\partial \alpha_i} > 0 \ \forall k \neq i$.
\end{proof}

\begin{proof}[Proof of Proposition \ref{prop:offline_expansion}]
  Consider an increase from $\alpha_i^0$ to $\alpha_i^1$ and because the profit function is increasing in $\alpha_i$ we have for fixed $n_{im}$:
  
  \begin{equation*}
    \pi_{im}(n_{im}, \alpha_i^1) > \pi_{im}(n_{im}, \alpha_i^0)
  \end{equation*}

  Because the marginal increment of transactions increases as $\alpha_i$ increases, we have the profit function exhibits increasing differences in $(n_{im}, \alpha_i)$, i.e., for $n_{im}^\prime \geq n_{im}$ and $\alpha_i^1 > \alpha_i^0$ we have:

  \begin{equation*}
    \pi_{im}(n_{im}, \alpha_i^1) - \pi_{im}(n_{im}, \alpha_i^0) \leq \pi_{im}(n_{im}^\prime, \alpha_i^1) - \pi_{im}(n_{im}^\prime, \alpha_i^0)
  \end{equation*}

  Standard monotone comparative statics \citep{10.2307/2951479} applied to the best response problem imply that the optimal solution $n_{im}^\star(\alpha_i)$ is non-decreasing in $\alpha_i$:

  \begin{equation*}
    \alpha_i^1 > \alpha_i^0 \implies n_{im}^\star(\alpha_i^1) \geq n_{im}^\star(\alpha_i^0)
  \end{equation*}

  This establishes (i).

  (ii)

  $\frac{\partial R_{im}}{\partial \alpha_i} > 0$ directs follow from the the definition and model setups. Differentiating $\sigma_{im}$ with respect to $\alpha_i$ yields:

    \begin{equation*}
        \begin{aligned}
        \frac{\partial \sigma_{im}}{\partial \alpha_i} & = \frac{\frac{\partial R_{im}}{\partial \alpha_i} \sum_{k \in \mathcal{I}} R_{km} - R_{im} \frac{\partial (\sum_{k \in \mathcal{I}} R_{km})}{\partial \alpha_i}}{(\sum_{k \in \mathcal{I}} R_{km})^2} \cdot L_m \\
            & = \underbrace{\frac{\frac{\partial R_{im}}{\partial \alpha_i} (\sum_{k \neq i} R_{km})}{(\sum_{k \in \mathcal{I}} R_{km})^2} \cdot L_m}_{\text{Direct Effect}} - \underbrace{\frac{R_{im} \frac{\partial (\sum_{k \neq i} R_{km})}{\partial \alpha_i}}{(\sum_{k \in \mathcal{I}} R_{km})^2} \cdot L_m}_{\text{Indirect Effect}}
        \end{aligned}
    \end{equation*}

    Under the Assumption \ref{claim:dominance}, we have the direct effect is positive and the indirect effect is negative, and therefore, the overall impact is positive. This establishes (ii). 
    
    part (iii) follows from Lemma \ref{lemma:exogenous_c_alpha}.
\end{proof}

Before we prove for Theorem \ref{theorem:simul_efficiency}, we first prove the following corollary of the Proposition \ref{prop:offline_expansion}:

\begin{corollary}[Simultaneous Efficiency Increase] \label{coro:simultaneous_expansion}
  Suppose brokerage differ in intrinsic efficiency parameters, and there exist brokerage $i$ and $j$ such that $\alpha_i > \alpha_j$. If there is an exogenous efficiency increase, i.e., $\alpha_i \to \alpha_i + \Delta \alpha$ and $\alpha_j \to \alpha_j + \Delta \alpha$, then under the submodularity condition and Assumption \ref{claim:dominance}, we have:

  \begin{itemize}[noitemsep,nolistsep]
    \item Brokerages $i$ and $j$ weakly increases their branch networks $\mathbf{n}_i$ and $\mathbf{n}_j$ and their transaction volumes $Q_{im}$ and $Q_{jm}$.
    \item Brokerages $k \neq i, j$ weakly decrease their branch networks $\mathbf{n}_k$ and their transaction volumes $Q_{km}$. 
    \item Brokerages $i$ and $j$ reduce their price concessions, but the magnitudes of their reduction is diminished compared to the scenario where only a single brokerage efficiency increases.
  \end{itemize}
\end{corollary}

\begin{proof}[Proof of Corollary \ref{coro:simultaneous_expansion}]
  (i)

  We have brokerage $i, j, k \in \mathcal{I}$ with intrinsic efficiency parameters $\alpha_i, \alpha_j, \alpha_k$ and $\alpha_i > \alpha_j$. Consider the initial equilibrium $(\mathbf{n}, \mathbf{c}) = \{n_{im}^\star, c_{im}^\star\}_{i \in \mathcal{I}}$ with efficiencies $\alpha_i, \alpha_j, \{\alpha_k\}_{k \neq i, j}$ and new equilibrium $(\hat{\mathbf{n}}, \hat{\mathbf{c}}) = \{\hat{n}_{im}, \hat{c}_{im}\}_{i \in \mathcal{I}}$.

  From the proof of the Proposition \ref{prop:offline_expansion}, we know that the brokerage profit function $\pi_{im}(n_{im}, \alpha_i)$ exhibits increasing differences in $(n_{im}, \alpha_i)$: for $n_{im}^\prime \geq n_{im}, \alpha_i^1 > \alpha_i^0$:

  \begin{equation*}
    \pi_{im}(n_{im}^\prime, \alpha_i^1) - \pi_{im}(n_{im}, \alpha_i^0) > \pi_{im}(n_{im}^\prime, \alpha_i^0) - \pi_{im}(n_{im}, \alpha_i^0)
  \end{equation*}

  Applying the Topkis's Monotonicity Theorem, we have:

  \begin{equation*}
    \alpha_i + \Delta \alpha > \alpha_i \implies \hat{n}_{im}, n_{im}^\star, \alpha_j + \Delta \alpha \implies \hat{n}_{jm} \geq n_{jm}^\star
  \end{equation*}

  By the same analogy, we have: $\hat{\sigma}_{im} > \sigma_{im}^\star$ and $\hat{\sigma}_{jm} > \sigma_{jm}^\star$. Then we have: $\hat{Q}_{im} > Q_{im}^\star$ and $\hat{Q}_{jm} > Q_{jm}^\star$.

  (ii)

  From part (i), we observe that $R_{im}(\alpha_i + \Delta \alpha) > R_{im}(\alpha_i)$ and $R_{jm}(\alpha_j + \Delta \alpha) > R_{jm}(\alpha_j)$, which implies that for brokerage $k$, we have:

  \begin{equation*}
    \sigma_{km}(n_{km}, \alpha_i + \Delta \alpha, \alpha_j + \Delta \alpha) < \sigma_{km}(n_{km}, \alpha_i, \alpha_j)
  \end{equation*}

  Because the brokerage $k$'s profit function strictly increases in its captured market share $\sigma_{km}$, the reduction in market share due to increased competitor attractiveness implies a strict reduction in brokerage $k$'s profit at every possible level $n_{km}$:

  \begin{equation*}
    \pi_{km}(n_{km}, \alpha_i + \Delta \alpha, \alpha_j + \Delta \alpha) < \pi_{km}(n_{km}, \alpha_i, \alpha_j)
  \end{equation*}

  Hence, after the simultaneous efficiency increase of $i, j$, each possible configuration $n_{km}$ for brokerage $k$ is strictly less profitable than it was before. Initially, $n_{km}^\star$ is the best response for the profit function under the initial condition. We prove that after the efficiency increase, every potential state $n_{km}^\prime \geq n_{km}^\star$ becomes even less profitable relatively to $n_{km}^\star$. Consider $n_{km}^\prime \geq n_{km}^\star$:

  The initial condition satisfies:

  \begin{equation*}
    \pi_{km}(n_{km}^\star, \alpha_i, \alpha_j) \geq \pi_{km}(n_{km}^\prime, \alpha_i, \alpha_j)
  \end{equation*}

  After the efficiency increases, this inequality strictly strengthens:

  \begin{equation*}
    \pi_{km}(n_{km}^\star, \alpha_i + \Delta \alpha, \alpha_j + \Delta \alpha) - \pi_{km}(n_{km}^\prime, \alpha_i + \Delta \alpha, \alpha_j + \Delta \alpha) > \pi_{km}(n_{km}^\star, \alpha_i, \alpha_j) - \pi_{km}(n_{km}^\prime, \alpha_i, \alpha_j) \geq 0
  \end{equation*}

  Thus, any state greater than the original equilibrium $n_{km}^\star$ becomes strictly suboptimal after the simultaneous efficiency increase. Therefore, brokerage $k$'s new equilibrium choice $\hat{n}_{km}$ must weakly decrease or stay the same:

  \begin{equation*}
    \hat{n}_{km} \leq n_{km}^\star, \quad k \neq i, j
  \end{equation*}

  (iii)

  Consider the single increase scenario (only $\alpha_i$ increases):

  \begin{equation*}
    \frac{\partial \pi_{im}}{\partial \alpha_i}(\tilde{c}_{im}, \alpha_i + \Delta \alpha, \alpha_j) = 0
  \end{equation*}

  and correspondingly the simultaneous increase scenario:

  \begin{equation*}
    \frac{\partial \pi_{im}}{\partial c_{im}}(\hat{c}_{im}, \alpha_i + \Delta \alpha, \alpha_j + \Delta \alpha) = 0
  \end{equation*}

  Since we already established the complementarity condition in price concessions across competitors:

  \begin{equation*}
    \frac{\partial^2 \pi_{im}}{\partial c_{im} \partial \alpha_j} > 0
  \end{equation*}

  which implies that:

  \begin{equation*}
    \hat{c}_{im} \geq \tilde{c}_{im}
  \end{equation*}
  
  By the same analogy, we can show that if there is a single increase in $\alpha_j$, then we would have:

  \begin{equation*}
    \hat{c}_{jm} \geq \tilde{c}_{jm}
  \end{equation*}

  Thus, brokerage $i, j$'s concessions decrease, but less than under the single-efficiency-increase scenario, due to internal moderation of competition.
\end{proof}

\begin{proof}[Proof of Theorem \ref{theorem:simul_efficiency}]
  This Theorem easily follows from the Proposition \ref{prop:offline_expansion} and Corollary \ref{coro:simultaneous_expansion}.

  The base case is trivial and follows from the Proposition \ref{prop:offline_expansion} and Corollary \ref{coro:simultaneous_expansion}. Assume the statement holds for every subset $S \subseteq \mathcal{E}$ with $|S| = t$ $(1 \leq t < |\mathcal{E}|)$. Choose $S \subseteq \mathcal{E}$ with $|S| = t$ and add one more firm $j \in \mathcal{E} \setminus |S|$. Fix $i \in S$. Relative to the equilibrium $s(S)$ we now simultaneously raise the efficiencies of $i$ and $j$ by the same marginal increments. Corollary \ref{coro:simultaneous_expansion} implies that $j$ weakly expands branches, strictly raises transactions and lowers concessions. Every outsider $k \notin S \cup \{j\}$ weakly shrinks branches and volumes. Combining these new inequalities with the inductive hypothesis for members of $S$ delivers the required relations for the enlarged set $S \cup \{j\}$.
\end{proof}

\begin{proof}[Proof of Proposition \ref{prop:platform_consolidation}]
Let brokerage $i$ be a member of a platform $P$. Thus, if the platform consolidates so that the platform size increases from, say, $|P(i)|$ to $|P(i)^\prime|$ with $|P(i)^\prime| > |P(i)|$, we have:

\begin{equation*}
  \psi(c_{im}, |P(i)^\prime|) > \psi(c_{im}, |P(i)|)
\end{equation*}

which by the definition of $R_{im}$ and Lemma \ref{lemma:monotonicity_coverage} implies: $R_{im}^\prime > R_{im}$ when holding $\mathbf{n}_i, \mathbf{c}_i$ fixed. Based on the proof of the Lemma \ref{lemma:monotonicity_coverage}, the captured share $\sigma_{im}$ is also increasing in $R_{im}$, so that $\sigma_{im}^\prime > \sigma_{im}$. 

Moreover, in equilibrium, brokerages adjust their strategies $(n_{km}, c_{km})$ in response to brokerage $i$'s increased attractiveness. The increase in brokerage $i$'s initial attractiveness $R_{im}$ induces competitor brokerage $k \neq i$ to optimally respond by potentially reducing offline presence $(\Delta n_{km}) \leq 0)$ and increasing concessions $(\Delta c_{km} \geq 0)$, while brokerage $i$ optimally adjusts by possibly increasing offline presence $(\Delta n_{im} \geq 0)$ and decreasing concessions $(\Delta c_{im} \leq 0)$ , as implied by the strategic incentives. Applying the Assumption \ref{claim:dominance}, we see that $R_{im}^{\star \prime} > R_{im}^\star$. Then applying the same logic, we have $\sigma_{im}^\star$ is also increases, which establishes (i).

The aggregated coverage of the platform in segment $m$ is:

\begin{equation*}
  \Sigma_m(P) = \sum_{i \in P} \sigma_{im}
\end{equation*}

Since for each $i \in P$ we have shown that $\sigma_{im}$ increases when $|P(i)|$ increases, it follows that the aggregated coverage over segment $m$ becomes:

\begin{equation*}
  \Sigma_m(P)^\prime = \sum_{i \in P} \sigma_{im}^\prime > \sum_{i \in P} \sigma_{im} = \Sigma_m(P)
\end{equation*}

We denote $X = \Sigma_m(P(i))$ and $X^\prime = \Sigma_m(P^\prime)$ after consolidation, with $X^\prime > X$. Similarly, let the total aggregated coverage across all platforms in segment $m$ be

\begin{equation*}
  T = \sum_{Q \in \mathcal{P}} \Sigma_m(Q)
\end{equation*}

We assume that consolidation affects only platform $P$, so that the coverage of other platforms remain fixed; thus $T^\prime = X^\prime + K > T$ where $K$ is a constant term. Then the local searcher's allocation to platform $P$ is given by:

\begin{equation*}
  \nu_{im}^L = \lambda_m^L \frac{X}{T} \text{ and } \tilde{\nu}_{im}^L = \lambda_m^L \frac{X^\prime}{T^\prime}
\end{equation*}

We can show that:

\begin{equation*}
  \begin{aligned}
    \tilde{\nu}_{im}^L - \nu_{im}^L & = \lambda_m^L \left( \frac{X^\prime}{X^\prime + K} - \frac{X}{X + K} \right) \\
    & = \lambda_m^L \left( \frac{X^\prime(X + K) - X(X^\prime + K)}{(X^\prime + K)(X + K)} \right) \\
    & = \lambda_m^L \left( \frac{X^\prime K - X K}{(X^\prime + K)(X + K)} \right) \\
    & = \lambda_m^L \left( \frac{K(X^\prime - X)}{(X^\prime + K)(X + K)} \right) > 0
  \end{aligned}
\end{equation*}

Similarly, let the global searchers' allocation be

\begin{equation*}
  \nu_{im}^G = \lambda_m^G \frac{X + Y}{Z}
\end{equation*}

where $Y$ denotes the platform's aggregated coverage in neighboring segments and $Z$ is the total coverage across all platforms.  The same algebra shows that $\tilde{\nu}_{im}^{G} > \nu_{im}^G$. Thus, the overall matching fraction allocated to platform $P$ increases. Finally, the transactions allocated to brokerage $i$ are

\begin{equation*}
  Q_{im} = [\nu_{im}^L + \nu_{im}^G] \cdot \tau_{im}(\nu_{im}^L + \nu_{im}^G) \cdot \frac{\sigma_{im}}{X}
\end{equation*}

If consolidation increases both $\sigma_{im}$ and $X$, the ratio remains at least weakly higher, so that $Q_{im}^\prime > Q_{im}$. This establishes (ii).

The part (iii) follows from the fact that the profit function $\pi_{im}$ satisfies the submodularity condition in $n_{im}$ and $c_{im}$, as shown in the proof of Proposition \ref{prop:offline_expansion}.

\end{proof}

\begin{proof}[Proof of Proposition \ref{prop:coexistence}]
  Without loss of generality, suppose brokerage $i$ has an intrinsic efficiency parameter $\alpha_i > \alpha_j$ for some brokerage $j$. 
  In terms of the effective attractiveness in segment $m$ with other competitors' strategies fixed, i.e., $R_{im}$, we have:

  \begin{equation*}
    R_{im}(n^\prime_{im}, \mathbf{s}_{-i}, \alpha_i) - R_{im}(n_{im}, \mathbf{s}_{-i}, \alpha_i) \geq R_{im}(n_{im}^\prime, \mathbf{s}_{-i}, \alpha_j) - R_{im}(n_{im}, \mathbf{s}_{-i}, \alpha_j)
  \end{equation*}
  
  Then, by applying Topkis's Monotonicity Theorem, the maximal best response is nondecreasing in $\alpha_i$, i.e., $R_{im}$ is nondecreasing in $\alpha_i$. Furthermore, applying the same argument to the captured share $\sigma_{im}$, we have that $\sigma_{im}$ is nondecreasing in $\alpha_i$.

  Now consider $\mathbf{s}_{-i}$ also varies with $\alpha_i$. That is, we compare two equilibria $\mathbf{s}^\star(\alpha_i)$ and $\mathbf{s}^\star(\alpha_j)$ for $\alpha_i > \alpha_j$. Then given the Assumption \ref{claim:dominance}, we can see that that with $n_{im}^\star(\alpha_i) > n_{im}^\star(\alpha_j)$ and $c_{im}^\star(\alpha_i) < c_{im}^\star(\alpha_j)$, we have the effect of $n_{im}^\star$ dominates the effect of $c_{im}^\star$ in the profit function $R_{im}^\star$. Therefore, $R_{im}$ and $\sigma_{im}$ remains larger.
  
  Moreover, the equilibrium transactions $Q_{im}$ are increasing in $\sigma_{im}$, so that $Q_{im}$ is nondecreasing in $\alpha_i$. Lastly, since the cost structure does not depend on $\alpha_i$, we have the profit function $\pi_{im}$ is nondecreasing in $\alpha_i$. The equilibrium profit increases directly follows from the Submodularity condition.
    
  Now consider the case where brokerage $i$ has $C_{im}(n) < C_{jm}(n)$ for all integers $n \geq 0$ and for each segment $m$. Denote the change of $\pi_{im}$ with $n_{im}^\prime > n_{im}$ as:

  \begin{equation*}
    \Delta \pi_{im}(\alpha_i) = [\underbrace{\beta (\mu_i - c_{im}) Q_{im}(n_{im}^\prime, \alpha_i)}_{\text{new revenue}} - C_{im}(n_{im}^\prime)] - [\beta(\mu_i - c_{im}) Q_{im}(n_{im}, \alpha_i) - C_{im}(n_{im})]
  \end{equation*}

  And also consider respectively for $\Delta \pi_{jm}(n_{jm})$ with $n_{jm}^\prime > n_{jm}$. Consider the case where $n_{im} = n_{jm}$ and $n_{im}^\prime - n_{jm}^\prime$. By assumption, for the same number of branches $n$,

  \begin{equation*}
    C_{im}(n_{im}^\prime) - C_{im}(n_{im}) < C_{jm}(n_{jm}^\prime) - C_{jm}(n_{jm})
  \end{equation*}
  
  Hence, brokerage $i$ faces a strictly lower incremental cost and pushes $n_{im}$ to be weakly higher than the best response of brokerage $j$. Then all other results follow from the same logic as above. This completes the part (i).

  (ii): As $i$ contemplates covering every segment, the total cost $\sum_{m=1}^M C_{im}(n_{im})$ escalates quickly. Yet the added profit from conquering those smaller segments may not keep pace. Let $\Delta \pi_{im}$  denote brokerage $i$'s net gain from covering an \emph{additional} segment which it chooses to stay out, and because costs eventually become prohibitive, $\Delta \pi_{im} < 0$ for small or peripheral segments. Thus it is optimal for $i$ to stay out or maintain minimal presence in those segments. Hence, large brokerage $i$ strictly prefers the partial strategy: covering profitable segments, leaving some segments un- or under-covered by itself. However in terms of small brokerages, they can open stores in segments that are not covered by large brokerages and they can also set up little amount of stores in segements covered by large brokerages but setting up high price concessions $c_{km}$ to attract customers.  Thus, a smaller player can sustain a positive profit in those segments, even if big brokerage $i$ could have overshadowed them in principle. This completes the proof.
\end{proof}

\begin{proof}[Proof of Proposition \ref{prop:searchers}]

  Consider brokerage $i$, initial equilibrium strategies $(n_{im}^\star, c_{im}^\star)$, and all other brokerages strategies $\mathbf{s}_{-i}$ remain fixed initially. From the matching structure and Lemma \ref{lemma:monotonicity_coverage}, an increase in searcher volume from $\lambda_m$ to $\lambda + \Delta \lambda > 0$ strictly increases the potential transaction volume at every existing branch configuration. Thus, for $\Delta n_{im} \geq 0$:

  \begin{equation*}
    Q_{im}(n_{im}^\star + \Delta n_{im}; \lambda + \Delta \lambda) - Q_{im}(n_{im}^\star \lambda + \Delta \lambda) \geq Q_{im}(n_{im}; \lambda) \geq Q_{im}(n_{im}^\star + \Delta n_{im}; \lambda) - Q_{im}(n_{im}^\star; \lambda)
  \end{equation*}
  
  Since the cost function remains unchanged by $\lambda$, the increment profits satisfies:

  \begin{equation*}
    \begin{aligned}
    & \pi_{im}(n_{im}^\star + \Delta n_{im}, c_{im}, \mathbf{s}_{-i}, \lambda + \Delta \lambda) - \pi_{im}(n_{im}^\star, c_{im}, \mathbf{s}_{-i}, \lambda + \Delta \lambda) \\
    & \geq \pi_{im}(n_{im}^\star + \Delta n_{im}, c_{im}, \mathbf{s}_{-i}, \lambda) - \pi_{im}(n_{im}^\star, c_{im}, \mathbf{s}_{-i}, \lambda)
    \end{aligned}
  \end{equation*}

  Thus, by Topkis' monotone comparative statics theorem, the best-response branch count $n_{im}^\star$ is weakly increasing in searcher volume $\lambda$.

  We now incorporate rival brokerages' responses. Consider brokerage $k \neq i$.  By identical logic, each rival brokerage also faces strictly increasing incentives to increase branches as searchers increase. Thus, rival brokerage branch configurations $n_{km}^\star$ are weakly increasing in $\lambda$. Due to the monotonicity of coverage (Lemma \ref{lemma:monotonicity_coverage}), when rivals increase their branch coverage, brokerage $i$ faces intensified competition. However, given the dominance condition, the strategic complementarity from higher branch coverage of rivals means brokerage $i$ benefits even more strongly from increasing its own branches to retain or expand market share: Formally, brokerage $i$'s incremental profit for adding branches increases further after rivals' branch expansions:

  \begin{equation*}
    \pi_{im}(n_{im}^\star + \Delta n_{im}, c_{im}; \mathbf{s}_{-i} + \Delta \mathbf{s}_{-i}, \lambda + \Delta \lambda) - \pi_{im}(n_{im}^\star, c_{im}; \mathbf{s}_{-i} + \Delta \mathbf{s}_{-i}, \lambda + \Delta \lambda) \geq 0
  \end{equation*}

  Hence, equilibrium responses reinforce each brokerage's incentive to expand. Therefore, equilibrium branch coverage $n_{im}^\star$ weakly increases for all brokerages, establishing (i).

Now consider two brokerages $i$ and $j$ with $\alpha_i > \alpha_j$. Starting from initial equilibrium $n_{im}^\star(\lambda)$ and $n_{jm}^\star(\lambda)$, examine incremental responses to increased searchers.

Due to higher efficiency $\alpha$, brokerage $i$'s marginal profit gain from expanding branches is strictly greater than the brokerage $j$'s, as efficiency enhances the attractiveness $R_{im}$ with $\Delta n_{im} = \Delta n_{jm}$:

\begin{equation*}
  \begin{aligned}
  \pi_{im}(n_{im}^\star + \Delta n_{im}, \alpha_i; \lambda + \Delta \lambda) - \pi_{im}(n_{im}^\star, \alpha_i; \lambda + \Delta \lambda) \\
  \geq \pi_{jm}(n_{jm}^\star + \Delta n_{jm}, \alpha_j; \lambda + \Delta \lambda) - \pi_{jm}(n_{jm}^\star, \alpha_j; \lambda + \Delta \lambda)
  \end{aligned}
\end{equation*}

Thus, by Topkis's monotone comparative statics, we have: $n_{im}^\star(\lambda + \Delta \lambda) - n_{im}^\star(\lambda) \geq n_{jm}^\star(\lambda + \Delta \lambda) - n_{jm}^\star(\lambda)$, establishing (ii).

Consider separately global $\lambda_m^G$ and local $\lambda_m^L$ searcher shocks, each increased by equal increments. Global searchers search across multiple segments $(\{m\} \cup \mathcal{N}(m))$, thus amplifying marginal returns of branch expansions beyond the affected segment $m$, due to spillovers captured by function $f$. Formally, incremental profit for global searcher increase satisfies:

\begin{equation*}
  \begin{aligned}
  & \pi_{im}(n_{im}^\star + \Delta n_{im}, \alpha_i; \lambda_m^G + \Delta \lambda_m^G) - \pi_{im}(n_{im}^\star, \alpha_i; \lambda_m^G + \Delta \lambda_m^G) \\
  & \geq \pi_{im}(n_{im}^\star + \Delta n_{im}, \alpha_i; \lambda_m^L + \Delta \lambda_m^L) - \pi_{im}(n_{im}^\star, \alpha_i; \lambda_m^G + \Delta \lambda_m^G)
  \end{aligned}
\end{equation*}

Thus, equilibrium branch expansion is greater under global than under comparable local shocks, establishing (iii).
\end{proof}

\begin{proof}[Proof of Proposition \ref{prop:welfare}]
  
Now, we consider the total surplus of the sellers:

\begin{equation*}
\begin{aligned}
    \frac{\partial W^S_m}{\partial \alpha_i} & = \underbrace{\frac{\partial Q_{im}}{\partial \alpha_i} [(1 - \beta)(\mu_m - c_{im}) - r_m]}_{\text{Brokerage $i$ volume effect} (*)} + \underbrace{Q_{im}(1 - \beta) \frac{\partial (\mu_m - c_{im})}{\partial \alpha_i}}_{\text{Brokerage $i$'s price effect} (**)} \\
    & + \underbrace{\sum_{k \neq i} \frac{\partial Q_{km}}{\partial \alpha_i} [(1 - \beta)(\mu_m - c_{km}) - r_m]}_{\text{Competitor displacement effect}(***)} + \underbrace{\sum_{k \neq i} Q_{km}(1 - \beta) \frac{\partial (\mu_m - c_{km})}{\partial \alpha_i}}_{\text{Competitor's indirect price effect}(****)}
\end{aligned}
\end{equation*}

where $\frac{\partial Q_{km}}{\partial \alpha_i} \leq 0$. Let the total segment transaction explicitly defined by:

\begin{equation*}
    T_m = \sum_{k \in \mathcal{I}} Q_{km} = \sum_{k \in \mathcal{I}} (\nu_m^L + \nu_m^G) \tau_{km}(\nu_m^L + \nu_m^G) \frac{\sigma_{km}}{\Sigma_m(P(k))}
\end{equation*}

and the transaction probability $\tau_{im}(\nu_{im}^L + \nu_{im}^G)$ has increasing returns to scale, i.e., $\frac{\partial \tau_{km}(x)}{\partial x} > 0, \frac{\partial^2 \tau_{km}(x)}{\partial x^2} > 0$. Define the function $\ell(\nu_{im}^L + \nu_{im}^G) = (\nu_{im}^L + \nu_{im}^G) \cdot \tau_{im}(\nu_{im}^L + \nu_{im}^G)$, and it is easy to see that $\ell(x)$ is also has: $\frac{\partial \ell}{\partial x} > 0, \frac{\partial^2 \ell}{\partial x^2} > 0$. Therefore, consider the differentiation with respect to $\alpha_i$

\begin{equation*}
\begin{aligned}
    \frac{\partial T_m}{\partial \alpha_i} & = \underbrace{\sum_{k \in \mathcal{I}} \left[ \frac{\partial (\nu_{km}^L + \nu_{km}^G)}{\partial \alpha_i} \tau_{km}(\nu_{km}^L + \nu_{km}^G) + (\nu_{km}^L + \nu_{km}^G) \frac{\partial \tau_{km}(\nu_{km}^L + \nu_{km}^G)}{\partial \alpha_i}\right] \frac{\sigma_{km}}{\Sigma_m(P(k))}}_{\text{redistribution effect}} \\
    & + \underbrace{\sum_{k \in \mathcal{I}}(\nu_{km}^L + \nu_{km}^G) \tau_{km}(\nu_{km}^L + \nu_{km}^G) \left(\frac{\partial}{\partial \alpha_i} (\frac{\sigma_{km}}{\Sigma_m(P(k))})\right)}_{\text{market share adjustment effect}}
\end{aligned}
\end{equation*}

Since $\sum_{k \in \mathcal{I}} x_{km} = \lambda_m^L + \lambda_m^G = \text{constant}$ and $x\tau(x)$ is a convex function, then by Jensen's inequality, we have a more concentrated allocation of searchers would yield a more-than-proportional increase in $\ell(x)$, and therefore, the redistribution effect is $> 0$. In terms of the market share adjustment effect, define $M_{km} \equiv \frac{\sigma_{km}}{\Sigma_m(P(k))}$ with $\sum_{k \in \mathcal{I}} M_{km} = 1$. Note that brokerage $i$'s effective attractiveness clearly increases with respect to $\alpha_i$, i.e., $\frac{\partial M_{im}}{\partial \alpha_i} > 0$, and competitor's market shares explicitly fall clearly to preserve total market share sum explicitly equal to 1: $\frac{\partial M_{km}}{\partial \alpha_i} < 0, \ \forall k \neq i$. Then we have $\frac{\partial \sum_{k \in \mathcal{I}} M_{km}}{\partial \alpha_i} = \sum_{k \in \mathcal{I}}  \frac{\partial M_{km}}{\partial \alpha_i}= 0$. Since $\alpha_i > \alpha_k \ \forall k \neq i$ then we have $\ell(\nu_{im}^L + \nu_{im}^G) > \ell(\nu_{km}^L + \nu_{km}^G), \ \forall k \neq i$. Therefore, we have:

\begin{equation*}
    \ell(\nu_{im}^L + \nu_{im}^G) \frac{\partial M_{im}}{\partial \alpha_i} + \sum_{k \neq i}  \ell(\nu_{km}^L + \nu_{km}^G) \frac{\partial M_{km}}{\partial \alpha_i} > \sum_{k \in \mathcal{I}} \ell(\nu_{km}^L + \nu_{km}^G)  \frac{\partial M_{km}}{\partial \alpha_i} = 0
\end{equation*}

Therefore, we have the market share adjustment effect $> 0$. Therefore, $\frac{\partial T_{m}}{\partial \alpha_i} = \sum_{k \in \mathcal{I}} \frac{\partial Q_{km}}{\partial \alpha_i} > 0$

Thus explicitly: $\frac{\partial Q_{im}}{\partial \alpha_i} > -\sum_{k \neq i} \frac{\partial Q_{km}}{\partial \alpha_i}$.

Since for any $j$, we already have $\frac{\partial c_{jm}}{\partial \alpha_j} < 0$ and given brokerage $i$ has highest intrinsic efficiency, we have $c_{im} < c_{km}$ for all $k \neq i$. \footnote{To rigorously prove this, we can define the function $F_k(c_{km}, n_{km}; \alpha_k, \alpha_{-k}) = \frac{\partial \pi_{km}}{\partial c_{km}} = 0$ and then the profit function has increasing differences in $(\alpha_k, -c_{km})$; equivalently, the marginal benefit of lowering $c_{km}$ is larger when the firm’s efficiency $\alpha_k$ is higher. For each brokerage $k$, the first-order condition $F_k$ is continuously differentiable and the equilibrium $c_{km}^\star$ is unique. Then by Topkis's Monotonicity Theorem, the unique equilibrium price concession $c_{km}^\star$ is a decreasing function of the firm's own efficiency parameter $\alpha_k$. In particular, for $\alpha_i > \alpha_k$ we have $c_{im}^\star < c_{km}^\star$ for all $k \neq i$.} Therefore, we have

\begin{equation*}
    \text{(*) $+$ (***)} > \left[(1 - \beta)(\mu_m - c_{im} - r_m \right] \left(\frac{\partial Q_{im}}{\partial \alpha_i} + \sum_{k \neq i}\frac{\partial Q_{km}}{\partial \alpha_i} \right) > 0.
\end{equation*}

Now given the Assumption \ref{claim:large_dominance}, we have (**) + (****) $>$ 0. Therefore, we have $\frac{\partial W_m^S}{\partial \alpha_i} > 0$, which further implies that $\frac{\partial W^S}{\partial \alpha_i} > 0$.

Now we can consider the effect on buyer's surplus:

   \begin{equation*}
       W_m^B = \sum_{k \in \mathcal{I}} Q_{km} (v_m - (\mu_m - c_{km}))
   \end{equation*}

   Differentiating with respect to $\alpha_i$ yields:

   \begin{equation*}
       \frac{\partial W_m^B}{\partial \alpha_i} = \sum_{k \in \mathcal{I}} \frac{\partial Q_{km}}{\partial \alpha_i} (v_m - (\mu_m - c_{km})) + \sum_{k \in \mathcal{I}} Q_{km}\frac{\partial c_{km}}{\partial \alpha_i}
   \end{equation*}

Under the Assumption \ref{claim:large_dominance}, we have the second effect is negative, and this results in the effect is ambiguous.

    (ii):

    Consider without the condition of Assumption \ref{claim:large_dominance}, and in this case, we have the sign of the (**) + (****) to be undetermined, and therefore, we have $\frac{\partial W_m^S}{\partial \alpha_i}$ is undetermined, which further implies that $\frac{\partial W^S}{\partial \alpha_i}$ is undetermined. In terms of the buyer's surplus, we see that the second effect is positive, and this results in the $\frac{\partial W_m^B}{\partial \alpha_i} > 0 \implies \frac{\partial W^B}{\partial \alpha_i} > 0$.

(iii):

   The total welfare is given by:

   \begin{equation*}
    W_m^{total} = \sum_{k \in \mathcal{I}} Q_{km}(v_m - r_m) - \sum_{k \in \mathcal{I}} C_{km}(n_{km})
   \end{equation*}

  We focus on the effect of exogenously increasing of efficiency of brokerage $i$ from $\alpha_i$ to $\alpha_i + \Delta \alpha_i > \alpha_i$. We show that $\Delta W_m^{total} > 0$ and hence $\Delta W^{total} > 0$. Allow brokerages—including $i$ itself—to \emph{update} their discrete branch configurations, from $n_{km}$ to $n_{km}^\prime$ and for brokerage $i$ we have $\Delta n_{im} = n_{im}^\prime - n_{im} > 0$ and for brokerage $k \neq i$ we have $\Delta n_{km} = n_{km}^\prime - n_{km} < 0$. Similarly, denote $c_{km}^\prime$ as the new response for the updated profile choice of brokerage $k$, and then by the first order condition of the real estate brokerage's profit we have:

  \begin{equation*}
    \begin{aligned}
    \Delta W_m^{total} & = W_m^{total}(\alpha_i + \Delta \alpha_i) - W_m^{total}(\alpha_i) \\
    & = \sum_{k \in \mathcal{I}} \left[ Q_{km}(n_{km}^\prime, c_{km}^{\prime}, \alpha_i + \Delta \alpha_i) - Q_{km}(n_{km}, c_{km}, \alpha)\right] (v_m - r_m) - \sum_{k \in \mathcal{I}} \left[ C_{km}(n_{km}^\prime) - C_{km}(n_{km}) \right] \\
    & = \underbrace{\sum_{k \in \mathcal{I}} \left[ Q_{km}(n_{km}^\prime, c_{km}^{\prime}, \alpha_i + \Delta \alpha_i) - Q_{km}(n_{km}, c_{km}, \alpha + \Delta \alpha)\right] (v_m - r_m) - \sum_{k \in \mathcal{I}} \left[ C_{km}(n_{km}^\prime) - C_{km}(n_{km}) \right]}_{\text{branch-and-cost adjustment effect}} \\
    & + \underbrace{\sum_{k \in \mathcal{I}} \left[ Q_{km}(n_{km}, c_{km}, \alpha_i + \Delta \alpha_i) - Q_{km}(n_{km}, c_{km}, \alpha)\right] (v_m - r_m)}_{\text{direct efficiency effect}}
    \end{aligned}
  \end{equation*}

For the direct efficiency effect, we can see that at fixed branches, the increased efficiency of brokerage $i$ enhances transactions for itself and redistributes market shares from less efficient firms. Such reallocation necessarily raises total welfare, as efficiency implies a higher net surplus per transaction. 

For the branch and cost adjustment effect, we have for brokerage $i$, $\Delta (n_{im}) > 0$ and it increases its transaction $Q_{im}$ thereby raising welfare. However, it also incurs increased branch costs $C_{im}(n_{im}^\prime) > C_{im}(n_{im})$. Other brokerage $k \neq i$, we have $\Delta (n_{im}) < 0$ but it also reduces the costs.

  Brokerage $k$'s profit optimization condition we have:

  \begin{equation*}
    \begin{aligned}
    & \pi_{im}(n_{im}^\prime, c_{im}^{\prime}, \alpha + \Delta \alpha) - \pi_{im}(n_{im}, c_{im}, \alpha + \Delta \alpha) \geq 0 \\
    & \implies \beta (\mu_m - c_{im}^{\prime})Q_{im}(n_{im}^\prime, c_{im}^{\prime}, \alpha + \Delta \alpha) - \beta(\mu_m - c_{im}) Q_{im}(n_{im}, c_{im}, \alpha + \Delta \alpha) \geq C_{im}(n_{im}^\prime) - C_{im}(n_{im})
    \end{aligned}
  \end{equation*}

  and respectively for $k \neq i$ we have:

  \begin{equation*}
    \begin{aligned}
    & \pi_{km}(n_{km}^\prime, c_{km}^{\prime}, \alpha + \Delta \alpha) - \pi_{km}(n_{km}, c_{km}, \alpha + \Delta \alpha) \geq 0 \\
    & \implies \beta (\mu_m - c_{km}^{\prime})Q_{km}(n_{km}^\prime, c_{km}^{\prime}, \alpha + \Delta \alpha) - \beta(\mu_m - c_{km}) Q_{km}(n_{km}, c_{km}, \alpha + \Delta \alpha) \geq C_{km}(n_{km}^\prime) - C_{km}(n_{km})
    \end{aligned}
  \end{equation*}
  
  Then we have:

  \begin{equation*}
    \begin{aligned}
    - \sum_{k \in \mathcal{I}} \left[C_{km}(n_{km}^\prime) - C_{km}(n_{km})\right] \geq - \sum_{k \in \mathcal{I}}  & \left\{ \beta (\mu_m - c_{km}^{\prime})Q_{km}(n_{km}^\prime, c_{km}^{\prime}, \alpha + \Delta \alpha) \right. - \\
    & \left. \beta(\mu_m - c_{km}) Q_{km}(n_{km}, c_{km}, \alpha + \Delta \alpha)\right\} 
    \end{aligned}
  \end{equation*}

  Denote $p_{km} = \beta(\mu_m - c_{km})$ and $p_{km}^{\prime} = \beta(\mu_m - c_{km}^{\prime})$. Then the overall branch-and-cost adjustment effect is: 

  \begin{equation*}
    \begin{aligned}
      & \sum_{k \in \mathcal{I}}\left\{ \left[(v_m - r_m) - p_{km}^{\prime}\right] Q_{km}(n_{km}^\prime, c_{km}^{\prime}, \alpha + \Delta \alpha) - \left[(v_m - r_m) - p_{km}\right] Q_{km}(n_{km}, c_{km}, \alpha + \Delta \alpha) \right\} \\
      & = \sum_{k \in \mathcal{I}} \left\{ \left[p_{km}^{\prime} - p_{km} \right] Q_{km}(n_{km}^\prime, c_{km}^{\prime}, \alpha + \Delta \alpha) \right\} \\
      & + \sum_{k \in \mathcal{I}} \left\{ \left[(v_m - r_m) - p_{km}\right] \left[ Q_{km}(n_{km}^\prime, c_{km}^{\prime}, \alpha + \Delta \alpha) - Q_{km}(n_{km}, c_{km}, \alpha + \Delta \alpha) \right] \right\}
    \end{aligned}
  \end{equation*}

  The first effect is positive because of the Assumption \ref{claim:large_dominance}, where $c_{im}^\prime < c_{im}$ and therefore, the changes of $p_{im}^\prime - p_{im} > 0$ and this effect is larger than the rest of price change effet of brokerage $k \neq i$. The second effect is positive is from the firm's best response profile $\mathbf{s}^\prime = (\mathbf{n}^\prime, \mathbf{s}^\prime)$ and the Assumption \ref{claim:dominance}. Therefore, the overall welfare effect is positive.

  In the absence of the Assumption \ref{claim:large_dominance}, then the first effect is negative but the second effect remains the same, and as a result, we have the overall effect is ambiguous.

\end{proof}

\begin{proof}[Proof of Theorem \ref{theorem:simul_welfare}]
  The proof is established based on the Proposition \ref{prop:welfare} and Theorem \ref{theorem:simul_efficiency}.
\end{proof}

\begin{proof}[Proof of Proposition \ref{prop:platform_welfare}]
  (i):

  Consider a set of brokerages $P \subseteq \mathcal{I}$ consolidating onto a single online platform. Denote:

  \begin{equation*}
    |P(i)| \to |P(i)|^\prime \text{ with } |P(i)|^\prime > |P(i)|, \ \forall i \in P
  \end{equation*}

  We can check that the change before and after the consolidation of seller surplus as:

  \begin{equation*}
    \begin{aligned}
    \Delta W_{m}^S = W_{m}^{S, post} - W_{m}^{S, init} & = \sum_{k \in \mathcal{I}} \underbrace{(Q_{km}^{post} - Q_{km}^{init})\left[ (1 - \beta)(\mu_m - c_{km}^{post}) - r_m\right]}_{\text{Transaction Volume Effect}} \\
    & + \underbrace{(1 - \beta) \sum_{k \in \mathcal{I}} Q_{km}^{init} (c_{km}^{init} - c_{km}^{post})}_{\text{Price Concession Adjustment Effect}}
    \end{aligned}
  \end{equation*}

  The Transaction Volume Effect is positive can be shown in the proof of Proposition \ref{prop:welfare}. The second effect is positive when the Assumption \ref{claim:large_dominance} holds.  

  Part (ii) and (iii) can be shown similarly to the proof of Proposition \ref{prop:welfare}.
\end{proof}

\section{Codebook} \label{sec:codebook}

\begin{table}[H]
  \begin{center}
    \begin{scriptsize}
    \caption{Codebook}
    \label{tab:codebook}
    \begin{tabular}{lll}
\toprule
Name & Label & Dimension \\
\midrule
income & The income lianjia in this given district/housing. & $10^5 \times$ \textyen \\
lead\_times & The time it takes before a deal is made. & counts \\
price\_concession & price changes (ending price - starting price) / starting price & \% \\
density & percentage of lianjia to all brokerages & \% / 100 \\
broker\_410 & number of other brokerages within 410 meters, which is the cutoff of RD & counts \\
watching\_people & The number of people watching this listing. & counts \\
end\_price & The final agreed price. & \textyen \\
non\_online\_effect & without online platformization influence & bool indicator \\
watched\_times & The number of times a listing is watched. & counts \\
nego\_times & The number of times a negotiation was held. & counts \\
nego\_period & The period over which negotiations took place. & days \\
jiadian & Referring to electronic shops. & counts \\
kind & Referring to proximity to kindergartens & counts \\
hotel & Referring to proximity to hotels & counts \\
shop\_mall & Referring to shopping mall. & counts \\
museum & Distance to the nearest museum. & counts \\
old & Referring to old care systems. & counts \\
ktv & Referring to KTV and some entertainment venues. & counts \\
mid & Referring to middle schools. & counts \\
prim & Referring to primary schools. & counts \\
west\_food & Referring to the availability of western food nearby. & counts \\
super & Referring to proximity to supermarkets (measured by number within given distance & counts \\
sub & Referring to proximity to subway stations. & counts \\
park & Referring to parks. & counts \\
area & The area of a property. & $m^2$ \\
bedroom & The number of bedrooms in a property. & counts \\
toilet & The number of toilets in a property. & counts \\
house\_age & The age of the house. & years \\
floor\_level & The level on which a particular room or apartment is, within a building. & categories \\
green\_ratio & The ratio of the green space to the total plot area. & \% / 100 \\
total\_building & The total number of buildings in an area. & counts \\
total\_floor\_number & The number of floors in a building. & counts \\
living\_room & The number of living rooms in a property. & counts \\
elevator\_ratio & The ratio of elevators to the total number of floors. & \% / 100 \\
kitchen & The number of kitchens in a property. & counts \\
floor\_ratio & The ratio of the floor area to the total plot area. & fraction \\
total\_resident & The total number of residents in an area. & counts \\
pm25 & Air quality measure. & mass/volume \\
pop & Population density. & people/$km^2$ \\
light & Night time lights. & lux \\
\bottomrule
\end{tabular}

    \end{scriptsize}
  \end{center}
\end{table}

\newpage

\section{Additional Robustness Check of Main Results} \label{sec:additional_robustness}

\subsection{Additional Robustness Check for the Offline Expansion Effect} \label{subsec:expansion_effect}

In this section, we report the robustness check of DID model. Table \ref{tab:robustness_check_additional_1}, Table \ref{tab:robustness_check_additional_2} display the estimation outcomes without the inclusion of additional control variables. By comparing these results with those obtained when controls are included, we observe that the estimates remain consistent across different model specifications. This consistency indicates the robustness of our findings, reinforcing the validity of our conclusions.

\begin{table}[H]
  \begin{center}
    \begin{scriptsize}
      \caption{Robustness Check of Lianjia's Offline Expansion Effect}
      \label{tab:robustness_check_additional_1}
      {
\def\sym#1{\ifmmode^{#1}\else\(^{#1}\)\fi}
\begin{tabular}{l*{4}{c}}
\toprule
            &\multicolumn{1}{c}{(1)}&\multicolumn{1}{c}{(2)}&\multicolumn{1}{c}{(3)}&\multicolumn{1}{c}{(4)}\\
            &\multicolumn{1}{c}{log(number)}&\multicolumn{1}{c}{log(number)}&\multicolumn{1}{c}{log(number)}&\multicolumn{1}{c}{log(number)}\\
\midrule
pre2        &      -0.010         &      -0.008         &      -0.007         &      -0.007         \\
            &     (0.015)         &     (0.014)         &     (0.014)         &     (0.014)         \\
\addlinespace
entry       &       0.102\sym{***}&       0.096\sym{***}&       0.095\sym{***}&       0.096\sym{***}\\
            &     (0.012)         &     (0.012)         &     (0.012)         &     (0.012)         \\
\addlinespace
post1       &       0.060\sym{***}&       0.054\sym{***}&       0.053\sym{***}&       0.052\sym{***}\\
            &     (0.013)         &     (0.012)         &     (0.012)         &     (0.012)         \\
\addlinespace
post2       &       0.019         &       0.014         &       0.014         &       0.013         \\
            &     (0.013)         &     (0.013)         &     (0.013)         &     (0.013)         \\
\addlinespace
post3       &       0.018         &       0.015         &       0.015         &       0.014         \\
            &     (0.015)         &     (0.015)         &     (0.015)         &     (0.015)         \\
\addlinespace
Brokerage Control &                     &  \checkmark         &  \checkmark         &  \checkmark         \\
\addlinespace
Hedonic Control &                     &                     &  \checkmark         &  \checkmark         \\
\addlinespace
Transaction Control &                     &                     &                     &  \checkmark         \\
\midrule
\(N\)       &      103966         &      103966         &      103966         &      103966         \\
R-squared   &       0.806         &       0.816         &       0.816         &       0.816         \\
\bottomrule
\multicolumn{5}{l}{\footnotesize Standard errors in parentheses}\\
\multicolumn{5}{l}{\footnotesize \sym{*} \(p<0.1\), \sym{**} \(p<0.05\), \sym{***} \(p<0.01\)}\\
\end{tabular}
}

    Note: we omit all the control variables in the regression model, and detailed descriptions can be seen from Table \ref{tab:statistical_district} and Table \ref{tab:statistical_individual}.
    \end{scriptsize}
  \end{center}
\end{table}

\begin{table}[H]
  \begin{center}
    \begin{scriptsize}
      \caption{Robustness Check of Lianjia's Offline Expansion Effect (Continued)}
      \label{tab:robustness_check_additional_2}
      {
\def\sym#1{\ifmmode^{#1}\else\(^{#1}\)\fi}
\begin{tabular}{l*{4}{c}}
\toprule
            &\multicolumn{1}{c}{(1)}&\multicolumn{1}{c}{(2)}&\multicolumn{1}{c}{(3)}&\multicolumn{1}{c}{(4)}\\
            &\multicolumn{1}{c}{price concession}&\multicolumn{1}{c}{price concession}&\multicolumn{1}{c}{price concession}&\multicolumn{1}{c}{price concession}\\
\midrule
pre2        &       0.000         &       0.000         &       0.001         &       0.001         \\
            &     (0.000)         &     (0.000)         &     (0.000)         &     (0.000)         \\
\addlinespace
entry       &      -0.011\sym{***}&      -0.011\sym{***}&      -0.011\sym{***}&      -0.011\sym{***}\\
            &     (0.000)         &     (0.000)         &     (0.000)         &     (0.000)         \\
\addlinespace
post1       &      -0.009\sym{***}&      -0.009\sym{***}&      -0.009\sym{***}&      -0.009\sym{***}\\
            &     (0.000)         &     (0.000)         &     (0.000)         &     (0.000)         \\
\addlinespace
post2       &      -0.008\sym{***}&      -0.008\sym{***}&      -0.008\sym{***}&      -0.008\sym{***}\\
            &     (0.000)         &     (0.000)         &     (0.000)         &     (0.000)         \\
\addlinespace
post3       &      -0.006\sym{***}&      -0.006\sym{***}&      -0.006\sym{***}&      -0.006\sym{***}\\
            &     (0.000)         &     (0.000)         &     (0.000)         &     (0.000)         \\
\addlinespace
Brokerage Control &                     &  \checkmark         &  \checkmark         &  \checkmark         \\
\addlinespace
Hedonic Control &                     &                     &  \checkmark         &  \checkmark         \\
\addlinespace
Transaction Control &                     &                     &                     &  \checkmark         \\
\midrule
\(N\)       &      845904         &      845904         &      845904         &      845904         \\
R-squared   &       0.266         &       0.282         &       0.282         &       0.285         \\
\bottomrule
\multicolumn{5}{l}{\footnotesize Standard errors in parentheses}\\
\multicolumn{5}{l}{\footnotesize \sym{*} \(p<0.1\), \sym{**} \(p<0.05\), \sym{***} \(p<0.01\)}\\
\end{tabular}
}

    Note: we omit all the control variables in the regression model, and detailed descriptions can be seen from Table \ref{tab:statistical_district} and Table \ref{tab:statistical_individual}.
    \end{scriptsize}
  \end{center}
\end{table}

Table \ref{tab:robustness_check_platform_1} and \ref{tab:robustness_check_platform_2} shows the results or our estimation with or without additional control variables, and the results are consistent.

\begin{table}[H]
  \begin{center}
    \begin{scriptsize}
      \caption{Robustness Check of Lianjia's Platform-Mediated Consolidation Effect}
      \label{tab:robustness_check_platform_1}
      {
\def\sym#1{\ifmmode^{#1}\else\(^{#1}\)\fi}
\begin{tabular}{l*{4}{c}}
\toprule
            &\multicolumn{1}{c}{(1)}&\multicolumn{1}{c}{(2)}&\multicolumn{1}{c}{(3)}&\multicolumn{1}{c}{(4)}\\
            &\multicolumn{1}{c}{log(number)}&\multicolumn{1}{c}{log(number)}&\multicolumn{1}{c}{log(number)}&\multicolumn{1}{c}{log(number)}\\
\midrule
pre2\_treatment&      -0.015\sym{*}  &      -0.012         &      -0.012         &      -0.012         \\
            &     (0.009)         &     (0.009)         &     (0.009)         &     (0.009)         \\
\addlinespace
treatment   &      -0.003         &      -0.000         &      -0.001         &      -0.000         \\
            &     (0.009)         &     (0.008)         &     (0.008)         &     (0.008)         \\
\addlinespace
post1\_treatment&       0.062\sym{***}&       0.058\sym{***}&       0.058\sym{***}&       0.058\sym{***}\\
            &     (0.011)         &     (0.010)         &     (0.010)         &     (0.010)         \\
\addlinespace
post2\_treatment&       0.064\sym{***}&       0.059\sym{***}&       0.059\sym{***}&       0.059\sym{***}\\
            &     (0.012)         &     (0.012)         &     (0.012)         &     (0.012)         \\
\addlinespace
post3\_treatment&       0.030\sym{**} &       0.028\sym{*}  &       0.027\sym{*}  &       0.027\sym{*}  \\
            &     (0.015)         &     (0.015)         &     (0.015)         &     (0.015)         \\
\addlinespace
Brokerage Control &                     &  \checkmark         &  \checkmark         &  \checkmark         \\
\addlinespace
Hedonic Control &                     &                     &  \checkmark         &  \checkmark         \\
\addlinespace
Transaction Control &                     &                     &                     &  \checkmark         \\
\midrule
\(N\)       &      133420         &      133420         &      133420         &      133420         \\
R-squared   &       0.847         &       0.853         &       0.853         &       0.853         \\
\bottomrule
\multicolumn{5}{l}{\footnotesize Standard errors in parentheses}\\
\multicolumn{5}{l}{\footnotesize \sym{*} \(p<0.1\), \sym{**} \(p<0.05\), \sym{***} \(p<0.01\)}\\
\end{tabular}
}

    Note: we omit all the control variables in the regression model, and detailed descriptions can be seen from Table \ref{tab:statistical_district} and Table \ref{tab:statistical_individual}.
    \end{scriptsize}
  \end{center}
\end{table}

\begin{table}
  \begin{center}
    \begin{scriptsize}
      \caption{Robustness Check of Lianjia's Platform-Mediated Consolidation Effect (Continued)}
      \label{tab:robustness_check_platform_2}
      {
\def\sym#1{\ifmmode^{#1}\else\(^{#1}\)\fi}
\begin{tabular}{l*{4}{c}}
\toprule
            &\multicolumn{1}{c}{(1)}&\multicolumn{1}{c}{(2)}&\multicolumn{1}{c}{(3)}&\multicolumn{1}{c}{(4)}\\
            &\multicolumn{1}{c}{price concession}&\multicolumn{1}{c}{price concession}&\multicolumn{1}{c}{price concession}&\multicolumn{1}{c}{price concession}\\
\midrule
pre2\_treatment&      -0.000         &      -0.000         &      -0.000         &      -0.000         \\
            &     (0.000)         &     (0.000)         &     (0.000)         &     (0.000)         \\
\addlinespace
treatment   &      -0.008\sym{***}&      -0.008\sym{***}&      -0.009\sym{***}&      -0.009\sym{***}\\
            &     (0.000)         &     (0.000)         &     (0.000)         &     (0.000)         \\
\addlinespace
post1\_treatment&      -0.009\sym{***}&      -0.009\sym{***}&      -0.009\sym{***}&      -0.009\sym{***}\\
            &     (0.000)         &     (0.000)         &     (0.000)         &     (0.000)         \\
\addlinespace
post2\_treatment&      -0.010\sym{***}&      -0.010\sym{***}&      -0.010\sym{***}&      -0.010\sym{***}\\
            &     (0.000)         &     (0.000)         &     (0.000)         &     (0.000)         \\
\addlinespace
post3\_treatment&      -0.010\sym{***}&      -0.010\sym{***}&      -0.010\sym{***}&      -0.010\sym{***}\\
            &     (0.001)         &     (0.001)         &     (0.001)         &     (0.001)         \\
\addlinespace
Brokerage Control &                     &  \checkmark         &  \checkmark         &  \checkmark         \\
\addlinespace
Hedonic Control &                     &                     &  \checkmark         &  \checkmark         \\
\addlinespace
Transaction Control &                     &                     &                     &  \checkmark         \\
\midrule
\(N\)       &     1246299         &     1246299         &     1246299         &     1246299         \\
R-squared   &       0.272         &       0.286         &       0.286         &       0.289         \\
\bottomrule
\multicolumn{5}{l}{\footnotesize Standard errors in parentheses}\\
\multicolumn{5}{l}{\footnotesize \sym{*} \(p<0.1\), \sym{**} \(p<0.05\), \sym{***} \(p<0.01\)}\\
\end{tabular}
}

    Note: we omit all the control variables in the regression model, and detailed descriptions can be seen from Table \ref{tab:statistical_district} and Table \ref{tab:statistical_individual}.
    \end{scriptsize}
  \end{center}
\end{table}

\clearpage

\begin{figure}[H]
    \centering
    \includegraphics[width=0.7\textwidth]{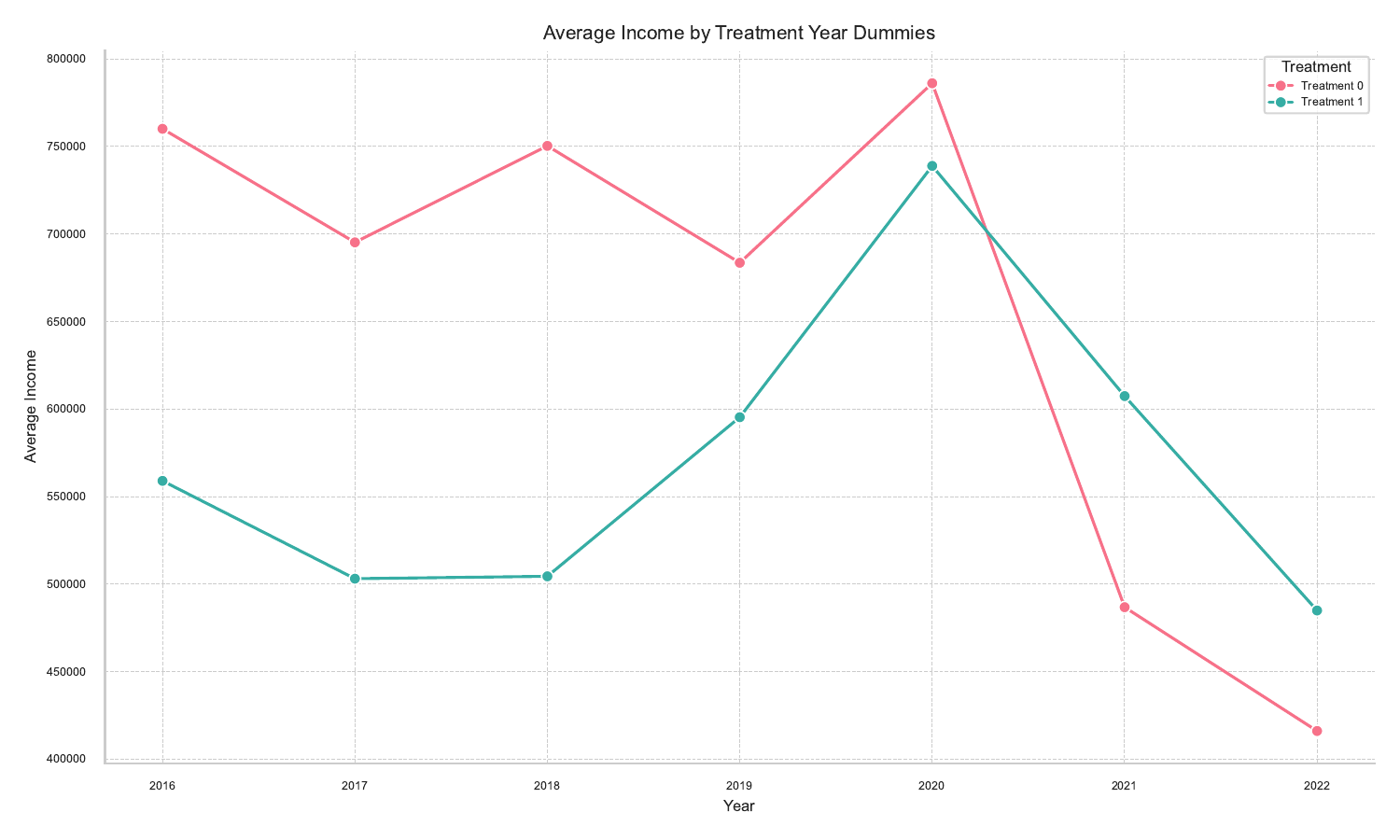}
    \caption{Treatment Effect of Platform Consolidation}
    \label{fig:treatment_consolidation}
    Note: the x-axis is year and the y-axis is the average income of Lianjia in each year. The graph uses the neighborhood-level data.
\end{figure}

Figure \ref{fig:treatment_consolidation} shows the statistical summary of the Lianjia's summary with the treatment group and the control grouls. We can see that the general trend is similar to the estimation results.

\newpage

\subsection{Robustness Check by Dividing the Samples to Low and High Nighttime Lights} \label{subsec:nighttime_light_robustness_check}

Table \ref{tab:robustness_nighttime_light_entry} and Table \ref{tab:robustness_nighttime_light_consolidation} show the additional results of the DID model by classifying the samples to low and high nighttime light areas. The results show that there is heterogeneity effect across different areas, but the general estimation results are in line with our main model.

\begin{table}[H]
  \begin{center}
    \begin{scriptsize}
      \caption{Robustness Check for Offline Expansion Effect Using Nighttime Lights}
      \label{tab:robustness_nighttime_light_entry}
      {
\def\sym#1{\ifmmode^{#1}\else\(^{#1}\)\fi}
\begin{tabular}{l*{4}{c}}
\toprule
            &\multicolumn{1}{c}{(1)}&\multicolumn{1}{c}{(2)}&\multicolumn{1}{c}{(3)}&\multicolumn{1}{c}{(4)}\\
            &\multicolumn{1}{c}{log(number)}&\multicolumn{1}{c}{log(number)}&\multicolumn{1}{c}{price concession}&\multicolumn{1}{c}{price concession}\\
\midrule
pre2        &      -0.017         &       0.034         &       0.001         &      -0.001         \\
            &     (0.017)         &     (0.030)         &     (0.000)         &     (0.001)         \\
\addlinespace
entry       &       0.096\sym{***}&       0.099\sym{***}&      -0.012\sym{***}&      -0.008\sym{***}\\
            &     (0.015)         &     (0.022)         &     (0.001)         &     (0.001)         \\
\addlinespace
post1       &       0.043\sym{***}&       0.058\sym{**} &      -0.010\sym{***}&      -0.006\sym{***}\\
            &     (0.016)         &     (0.023)         &     (0.001)         &     (0.001)         \\
\addlinespace
post2       &      -0.003         &       0.035         &      -0.009\sym{***}&      -0.005\sym{***}\\
            &     (0.017)         &     (0.024)         &     (0.000)         &     (0.001)         \\
\addlinespace
post3       &      -0.010         &       0.051\sym{**} &      -0.007\sym{***}&      -0.004\sym{***}\\
            &     (0.020)         &     (0.026)         &     (0.000)         &     (0.001)         \\
\addlinespace
Brokerage Control &  \checkmark         &  \checkmark         &  \checkmark         &  \checkmark         \\
\addlinespace
Hedonic Control &  \checkmark         &  \checkmark         &  \checkmark         &  \checkmark         \\
\addlinespace
Transaction Control &  \checkmark         &  \checkmark         &  \checkmark         &  \checkmark         \\
\addlinespace
Regional Control &  \checkmark         &  \checkmark         &  \checkmark         &  \checkmark         \\
\midrule
\(N\)       &       69400         &       28889         &      616702         &      227215         \\
R-squared   &       0.831         &       0.828         &       0.295         &       0.317         \\
\bottomrule
\multicolumn{5}{l}{\footnotesize Standard errors in parentheses}\\
\multicolumn{5}{l}{\footnotesize \sym{*} \(p<0.1\), \sym{**} \(p<0.05\), \sym{***} \(p<0.01\)}\\
\end{tabular}
}

    Note: we omit all the control variables in the regression model, and detailed descriptions can be seen from Table \ref{tab:statistical_district} and Table \ref{tab:statistical_individual}.
    \end{scriptsize}
  \end{center}
\end{table}

\begin{table}[H]
  \begin{center}
    \begin{scriptsize}
      \caption{Robustness Check for Platform Consolidation Effect using Nighttime Lights}
      \label{tab:robustness_nighttime_light_consolidation}
      {
\def\sym#1{\ifmmode^{#1}\else\(^{#1}\)\fi}
\begin{tabular}{l*{4}{c}}
\toprule
            &\multicolumn{1}{c}{(1)}&\multicolumn{1}{c}{(2)}&\multicolumn{1}{c}{(3)}&\multicolumn{1}{c}{(4)}\\
            &\multicolumn{1}{c}{log(number)}&\multicolumn{1}{c}{log(number)}&\multicolumn{1}{c}{price concession}&\multicolumn{1}{c}{price concession}\\
\midrule
pre2\_treatment&      -0.008         &      -0.004         &       0.000         &      -0.000         \\
            &     (0.012)         &     (0.012)         &     (0.000)         &     (0.000)         \\
\addlinespace
treatment   &       0.014         &      -0.015         &      -0.008\sym{***}&      -0.009\sym{***}\\
            &     (0.013)         &     (0.013)         &     (0.001)         &     (0.000)         \\
\addlinespace
post1\_treatment&       0.068\sym{***}&       0.040\sym{**} &      -0.009\sym{***}&      -0.009\sym{***}\\
            &     (0.016)         &     (0.016)         &     (0.001)         &     (0.001)         \\
\addlinespace
post2\_treatment&       0.052\sym{***}&       0.054\sym{***}&      -0.010\sym{***}&      -0.009\sym{***}\\
            &     (0.019)         &     (0.019)         &     (0.001)         &     (0.001)         \\
\addlinespace
post3\_treatment&       0.017         &       0.015         &      -0.010\sym{***}&      -0.009\sym{***}\\
            &     (0.024)         &     (0.025)         &     (0.001)         &     (0.001)         \\
\addlinespace
Brokerage Control &  \checkmark         &  \checkmark         &  \checkmark         &  \checkmark         \\
\addlinespace
Hedonic Control &  \checkmark         &  \checkmark         &  \checkmark         &  \checkmark         \\
\addlinespace
Transaction Control &  \checkmark         &  \checkmark         &  \checkmark         &  \checkmark         \\
\addlinespace
Regional Control &  \checkmark         &  \checkmark         &  \checkmark         &  \checkmark         \\
\midrule
\(N\)       &       69000         &       56419         &      756404         &      487353         \\
R-squared   &       0.871         &       0.849         &       0.295         &       0.318         \\
\bottomrule
\multicolumn{5}{l}{\footnotesize Standard errors in parentheses}\\
\multicolumn{5}{l}{\footnotesize \sym{*} \(p<0.1\), \sym{**} \(p<0.05\), \sym{***} \(p<0.01\)}\\
\end{tabular}
}

    Note: we omit all the control variables in the regression model, and detailed descriptions can be seen from Table \ref{tab:statistical_district} and Table \ref{tab:statistical_individual}.
    \end{scriptsize}
  \end{center}
\end{table}

\subsection{Placebo Test} \label{subsec:placebo_test}

\begin{figure}[H]
  \centering
  \subfloat[Placebo Test to transaction number]{\includegraphics[width=0.45\textwidth]{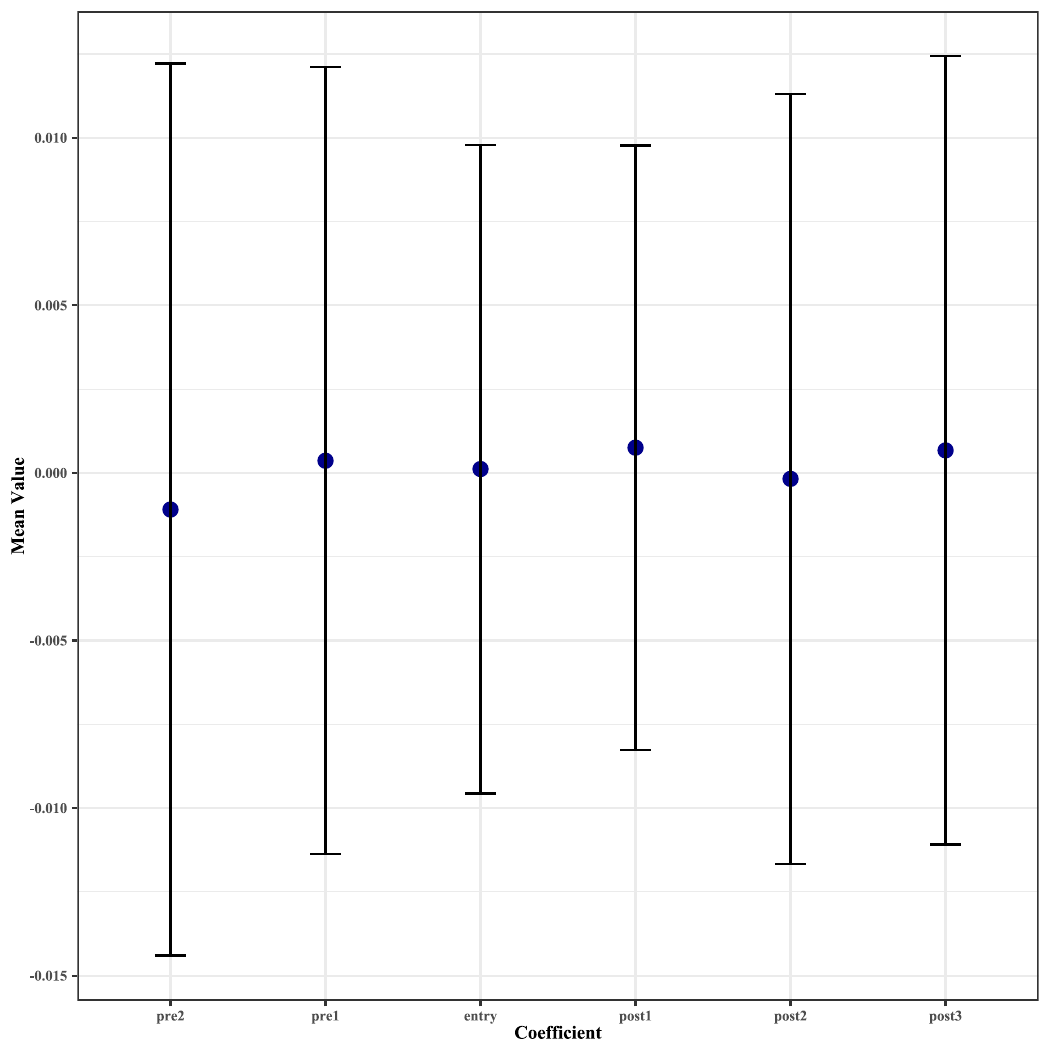}\label{fig:placebo_num_entry}}
  \subfloat[Placebo Test to number of house tours]{\includegraphics[width=0.45\textwidth]{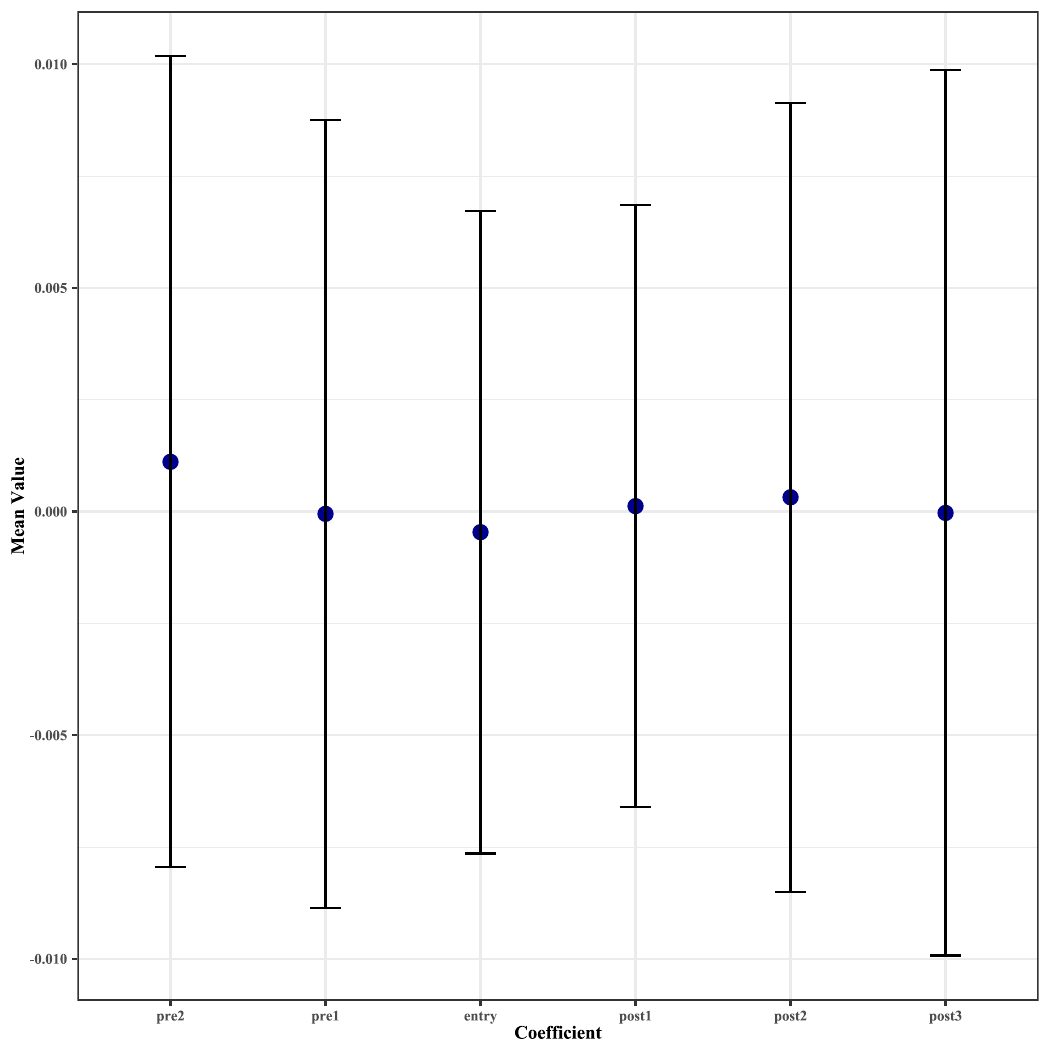}\label{fig:placebo_lead_entry}}
  \hfill % Adds horizontal space between figures
  \subfloat[Placebo Test to transaction period]{\includegraphics[width=0.45\textwidth]{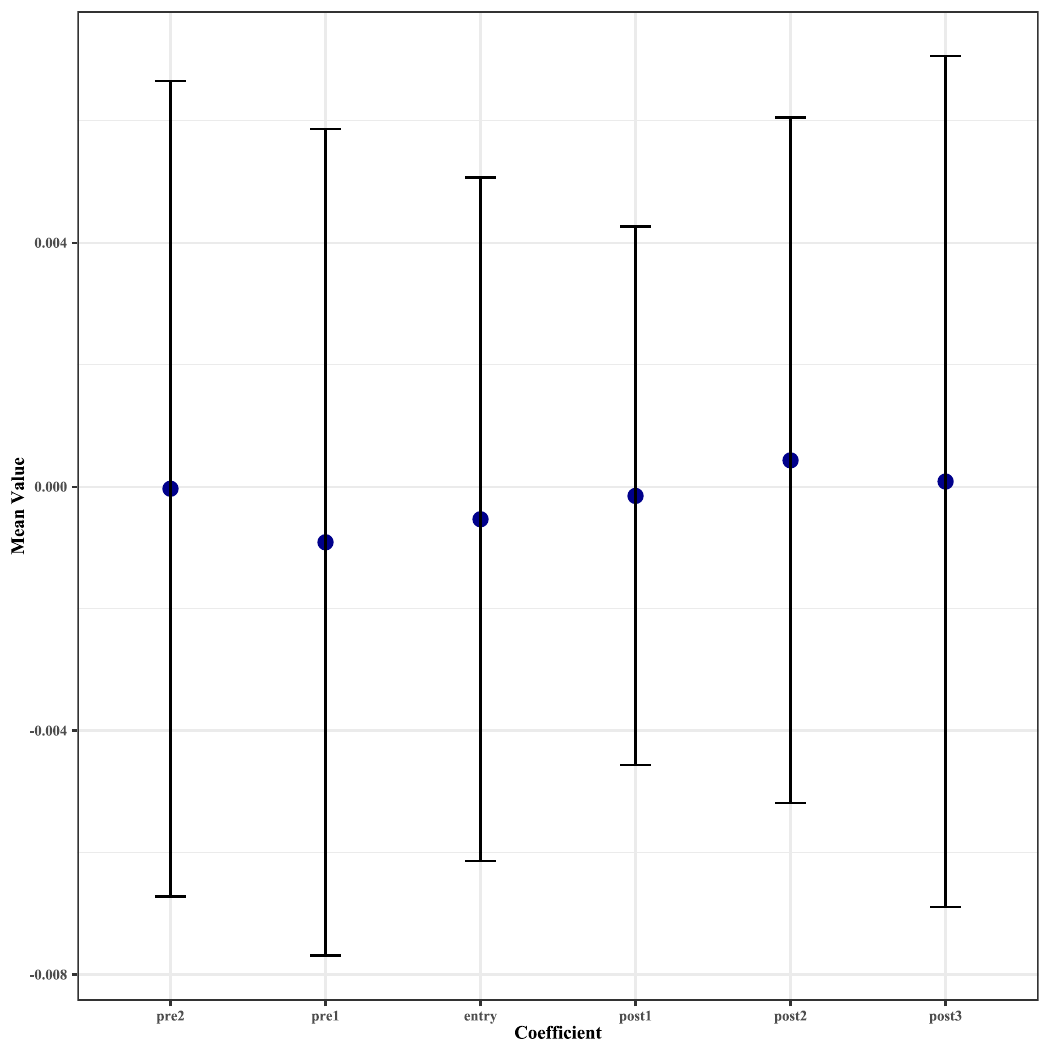}\label{fig:placebo_period_entry}}
  \subfloat[Placebo Test to price concession]{\includegraphics[width=0.45\textwidth]{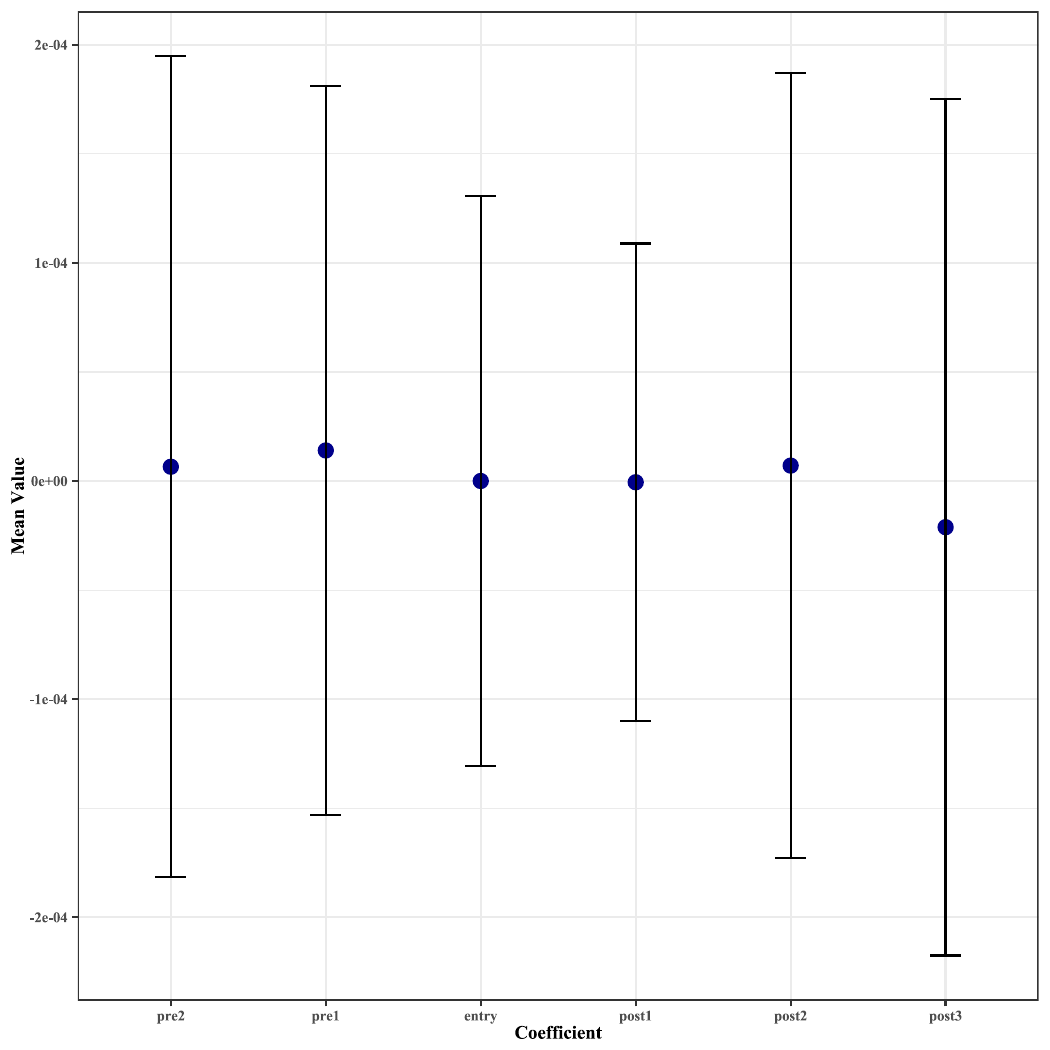}\label{fig:placebo_concession_entry}}
  \caption{Placebo Test to Entry Effect}
  \label{fig:placebo_test_entry}

  Note: The x-axis is the coefficient of the 

  the model is described in Equation \eqref{eq:entry_effect}
  \end{figure}

\begin{figure}[H]
  \centering
  \subfloat[Placebo Test to transaction number]{\includegraphics[width=0.45\textwidth]{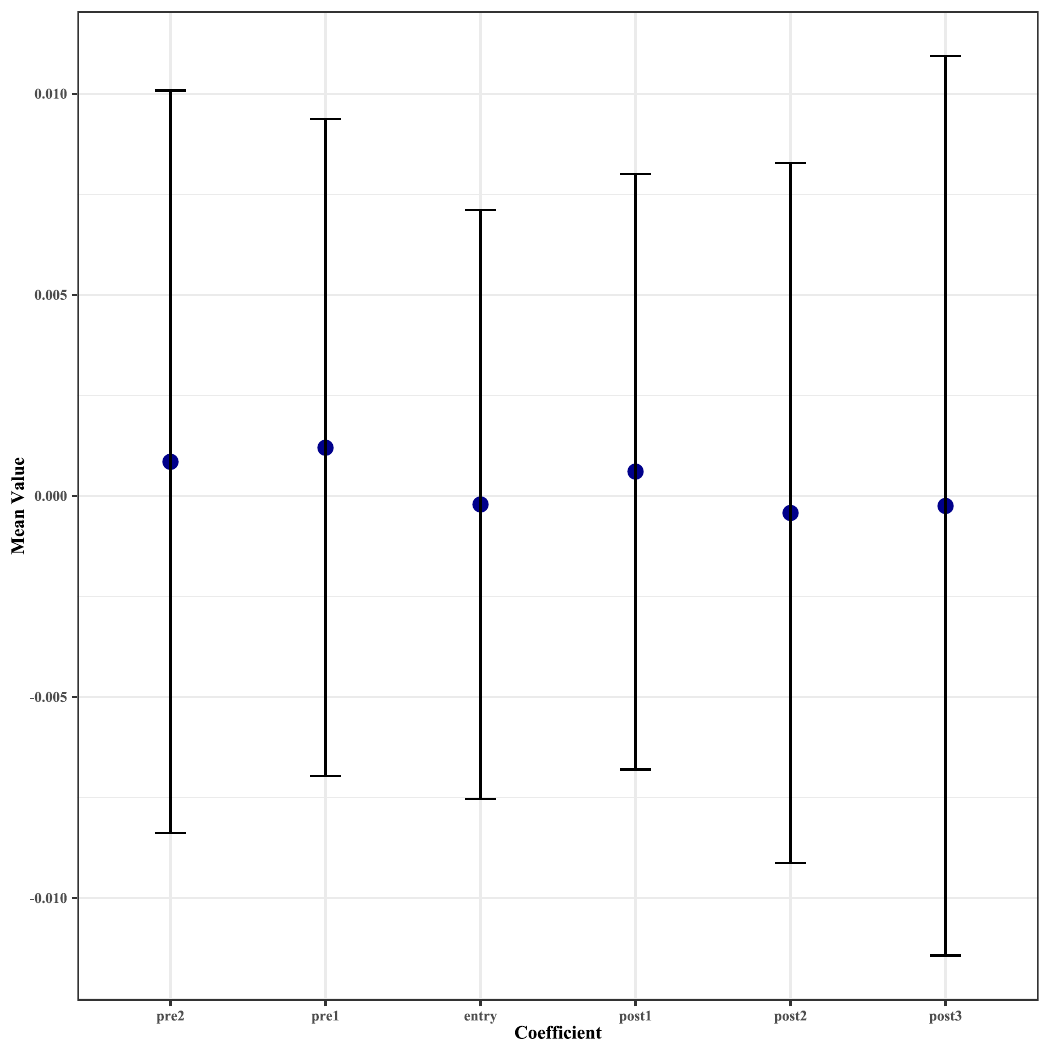}\label{fig:placebo_num_plat}}
  \subfloat[Placebo Test to number of house tours]{\includegraphics[width=0.45\textwidth]{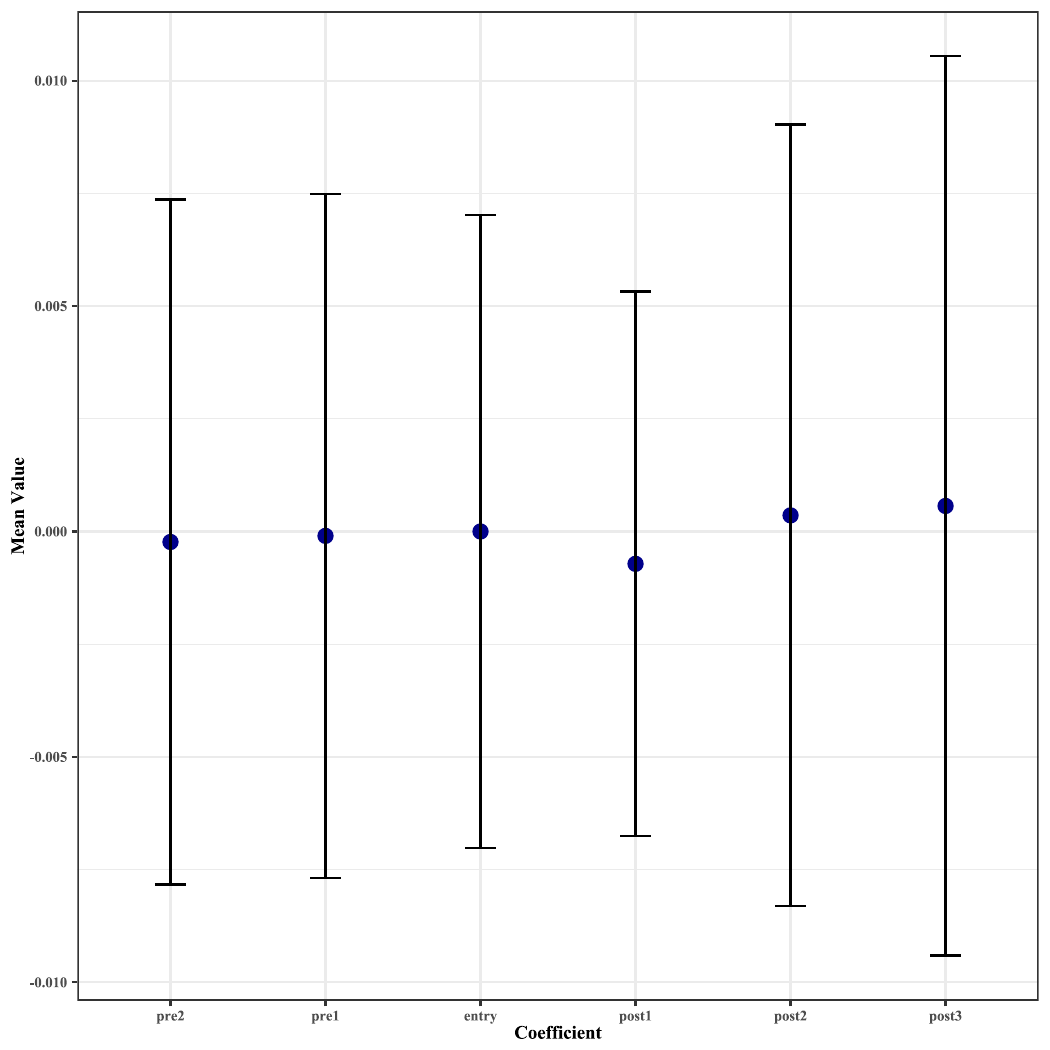}\label{fig:placebo_lead_plat}}
  \hfill % Adds horizontal space between figures
  \subfloat[Placebo Test to transaction period]{\includegraphics[width=0.45\textwidth]{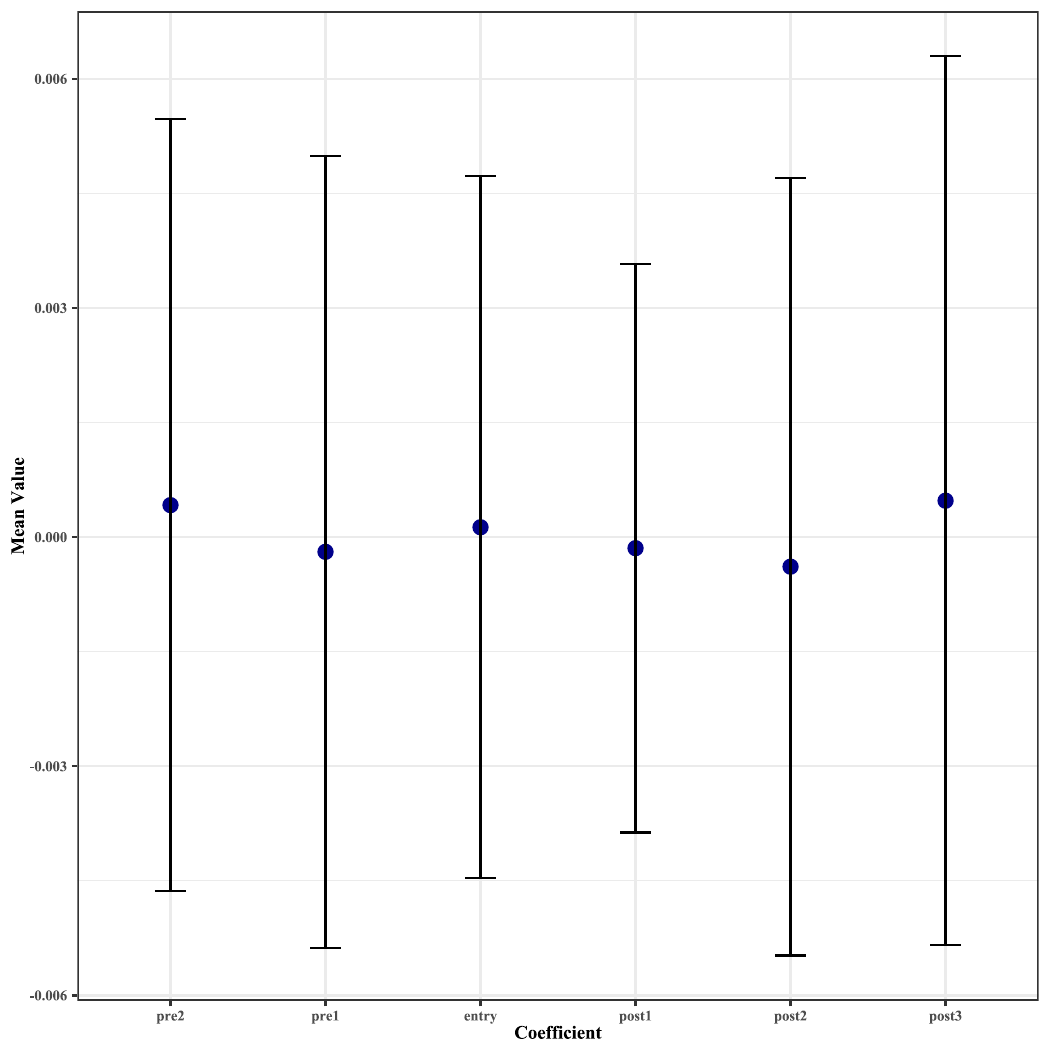}\label{fig:placebo_period_plat}}
  \subfloat[Placebo Test to price concession]{\includegraphics[width=0.45\textwidth]{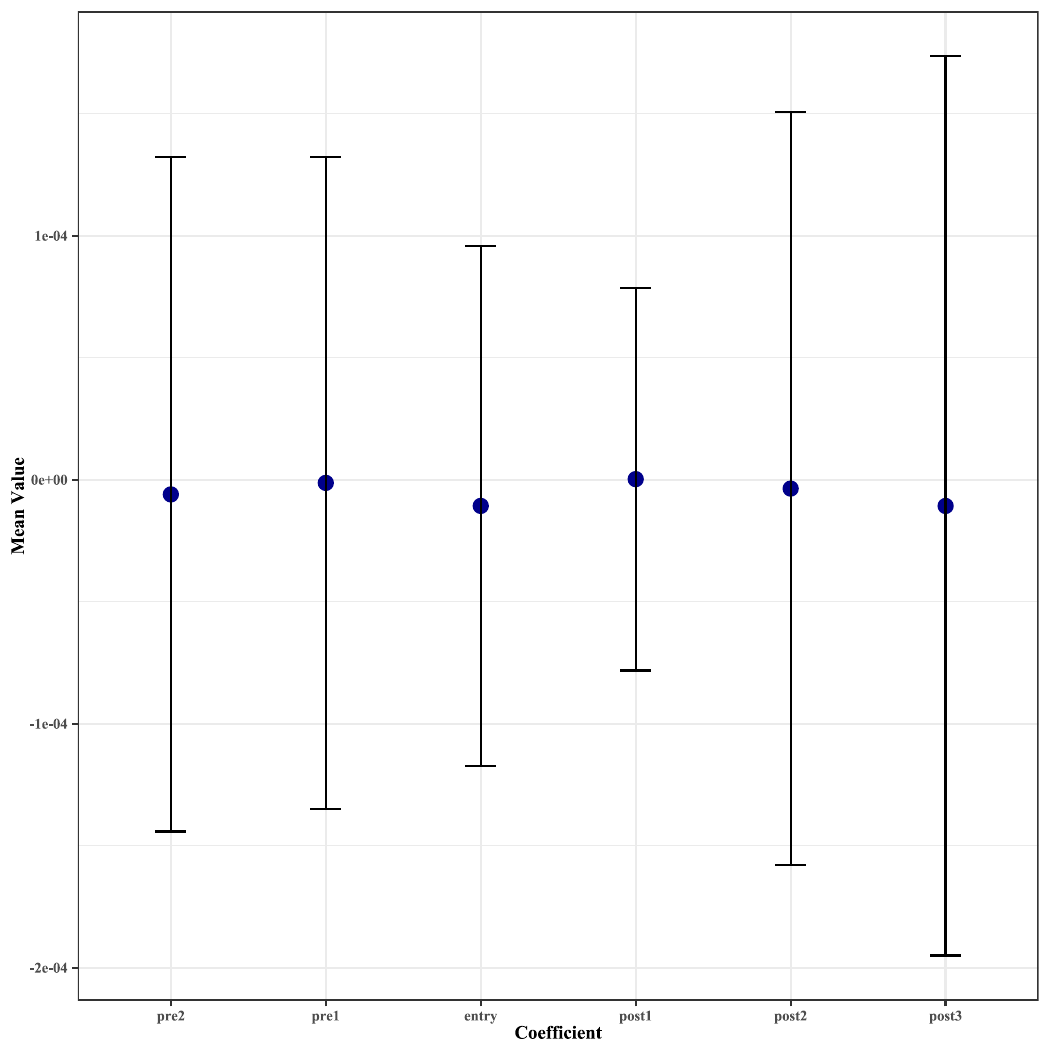}\label{fig:placebo_concession_plat}}
  \caption{Placebo Test to Platform-Mediated Consolidation Effect}
  \label{fig:placebo_test_plat}

  Note: The x-axis is the coefficient of the 

  the model is described in Equation \eqref{eq:acn_assessment}

\end{figure}

Figure \ref{fig:placebo_test_entry} and Figure \ref{fig:placebo_test_plat} show the results of the placebo test for the entry effect and the platform consolidation effect. The results show that the random treatment effect is non-significant at the 10\% level, indicating that the treatment effect is not due to other factors.

\newpage

\section{Additional Descriptions of Data} \label{sec:appendix_data}

\begin{figure}[h!]
 \centering
    \begin{minipage}{0.328\textwidth}
        \includegraphics[width=\linewidth]{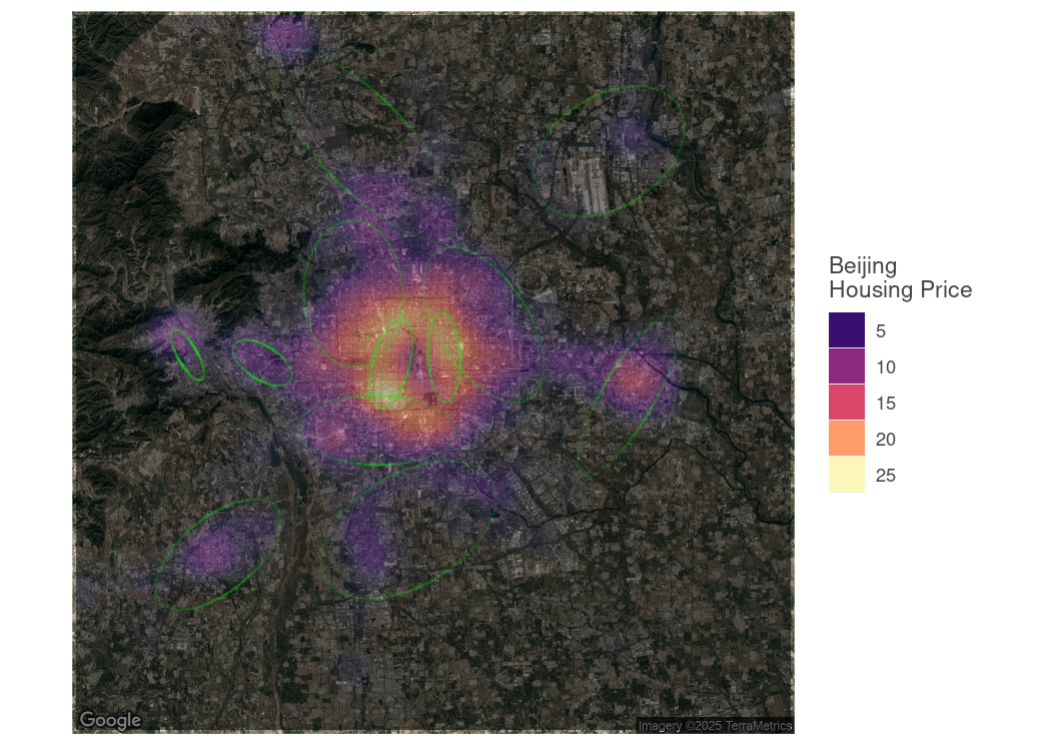}
    \end{minipage}
    \hfill
    \begin{minipage}{0.328\textwidth}
        \includegraphics[width=\linewidth]{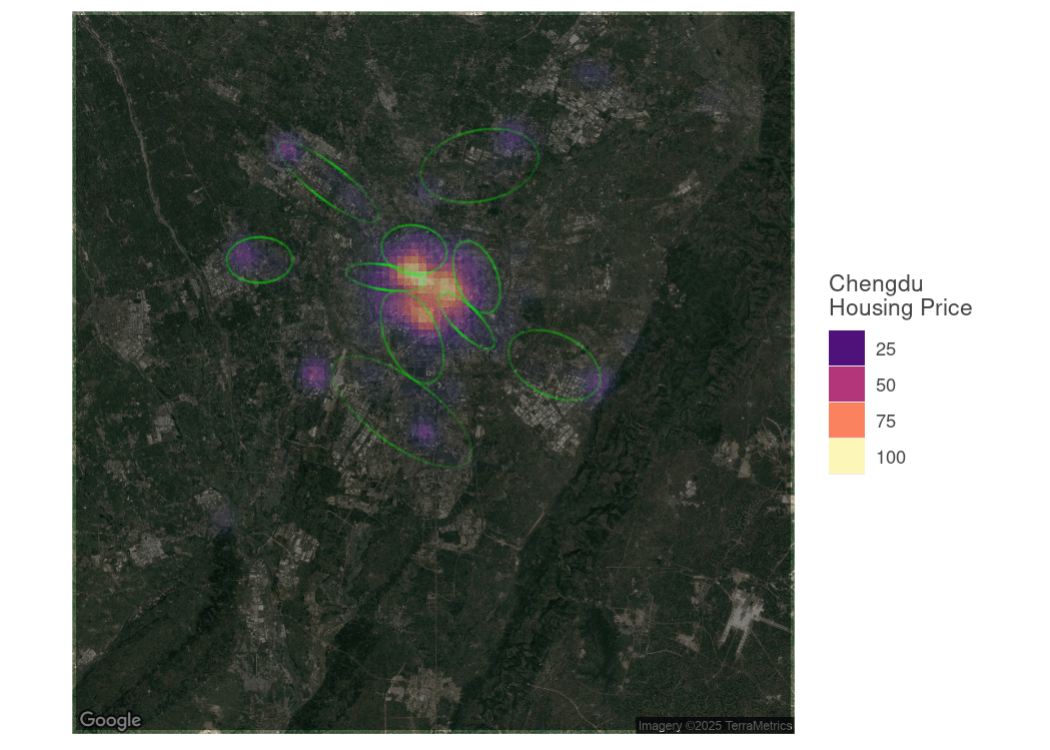}
    \end{minipage}
    \begin{minipage}{0.328\textwidth}
        \includegraphics[width=\linewidth]{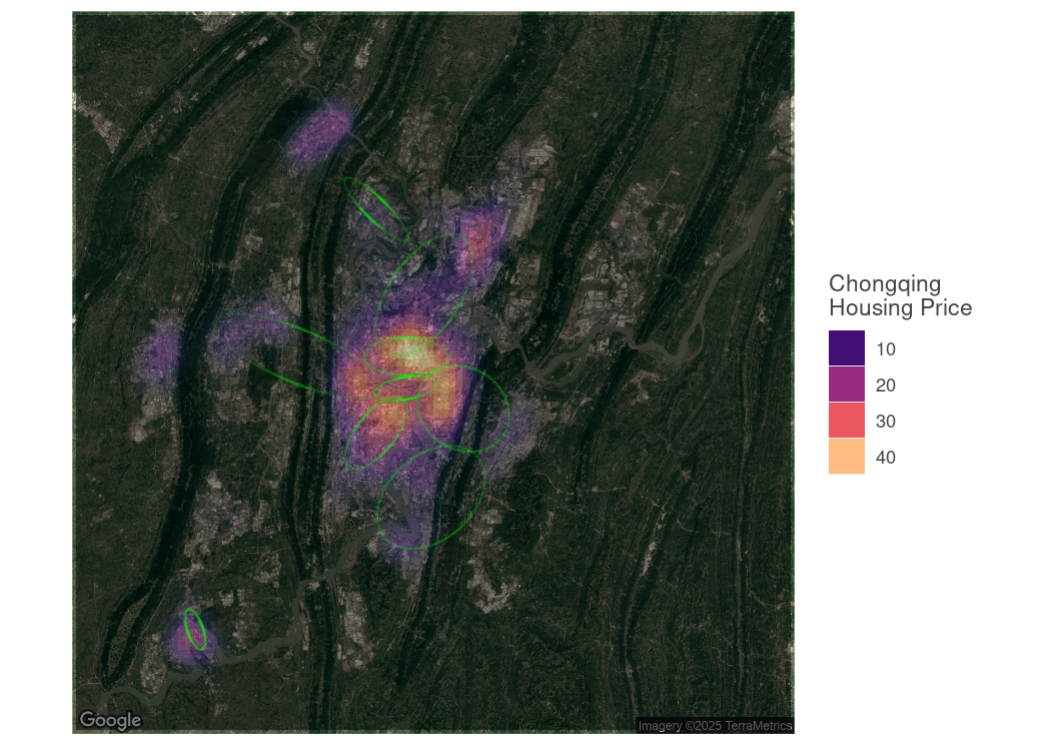}
    \end{minipage}

    \begin{minipage}{0.328\textwidth}
        \includegraphics[width=\linewidth]{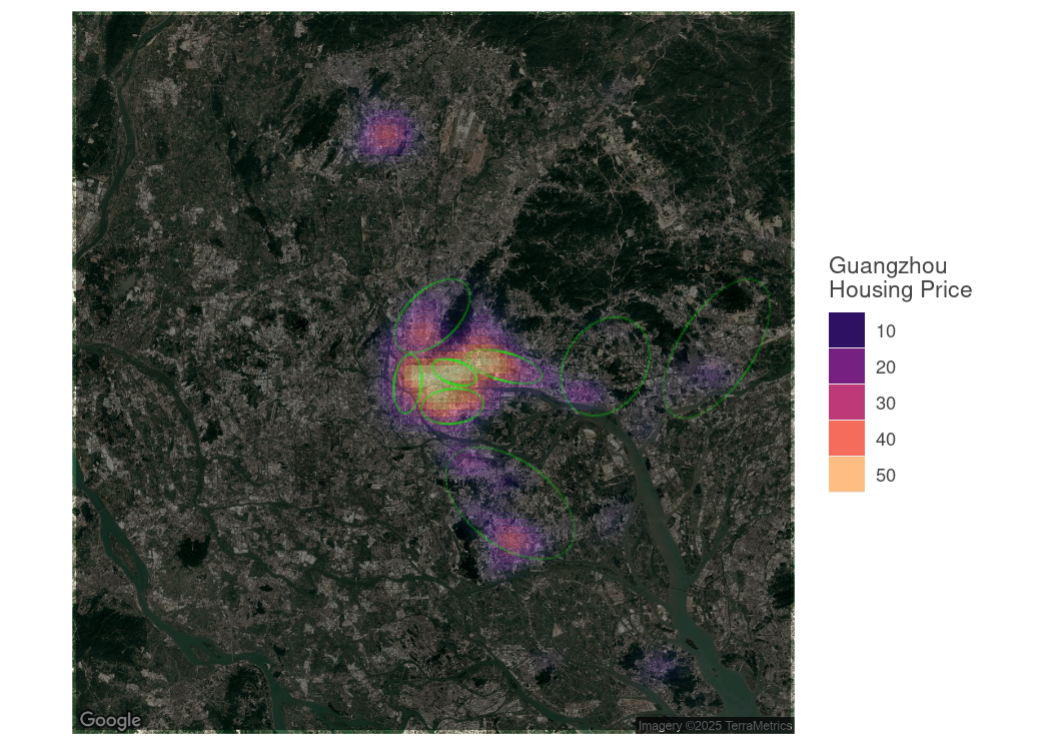}
    \end{minipage}
    \hfill
    \begin{minipage}{0.328\textwidth}
        \includegraphics[width=\linewidth]{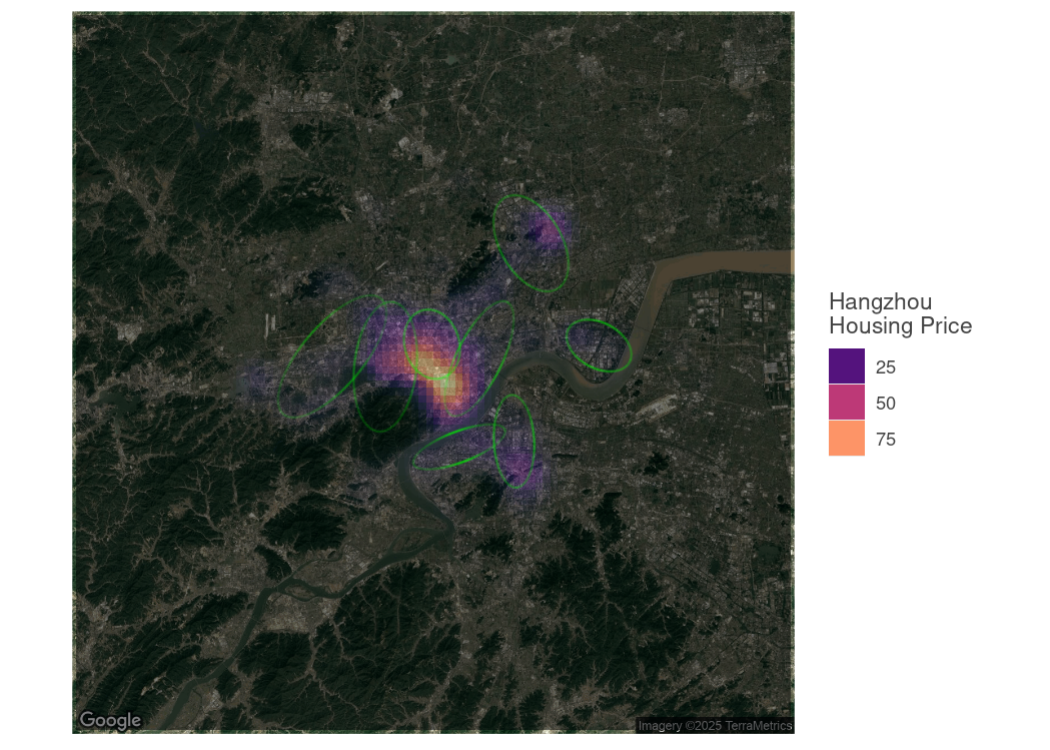}
    \end{minipage}
    \begin{minipage}{0.328\textwidth}
        \includegraphics[width=\linewidth]{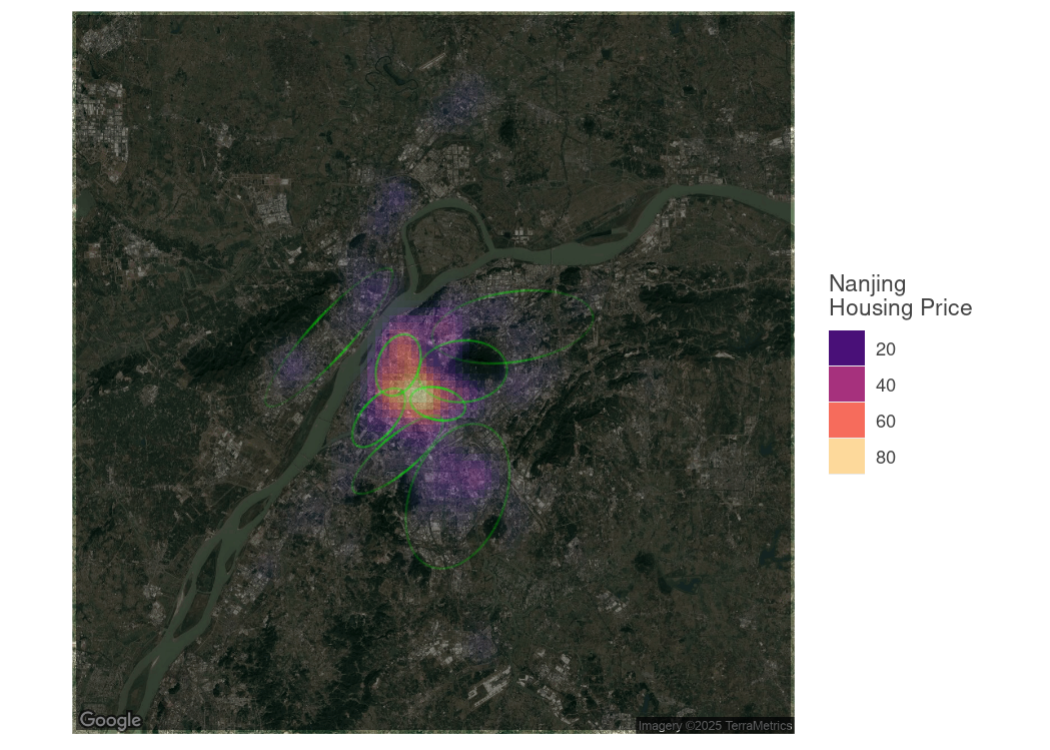}
    \end{minipage}

    \begin{minipage}{0.328\textwidth}
        \includegraphics[width=\linewidth]{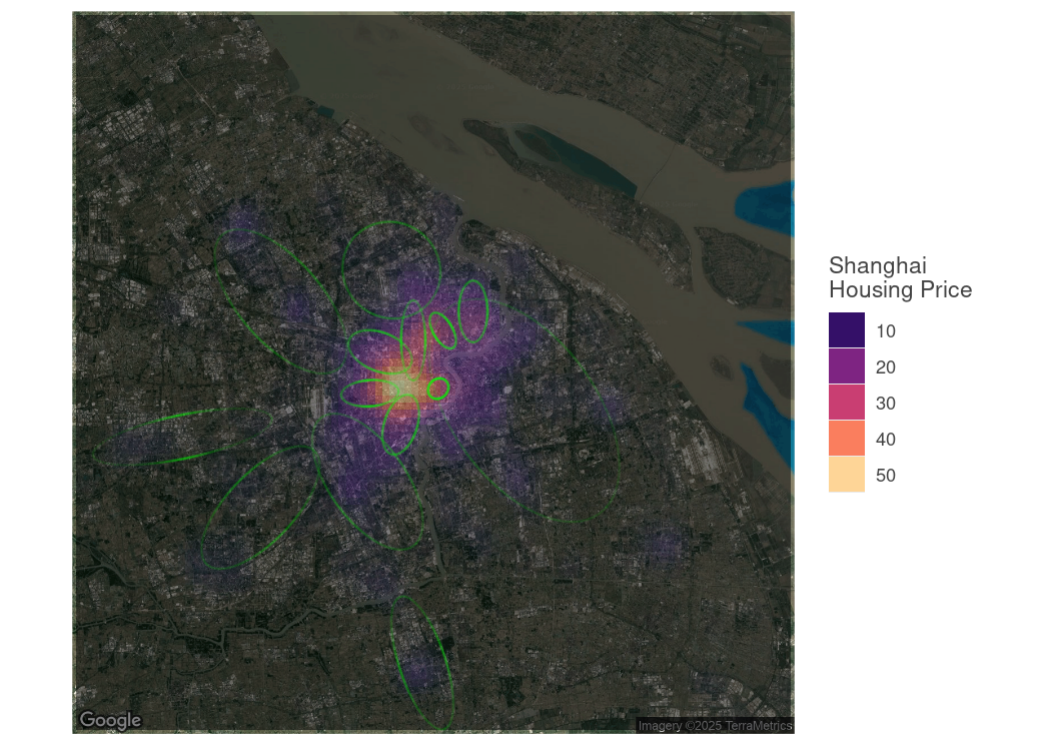}
    \end{minipage}
    \hfill
    \begin{minipage}{0.328\textwidth}
        \includegraphics[width=\linewidth]{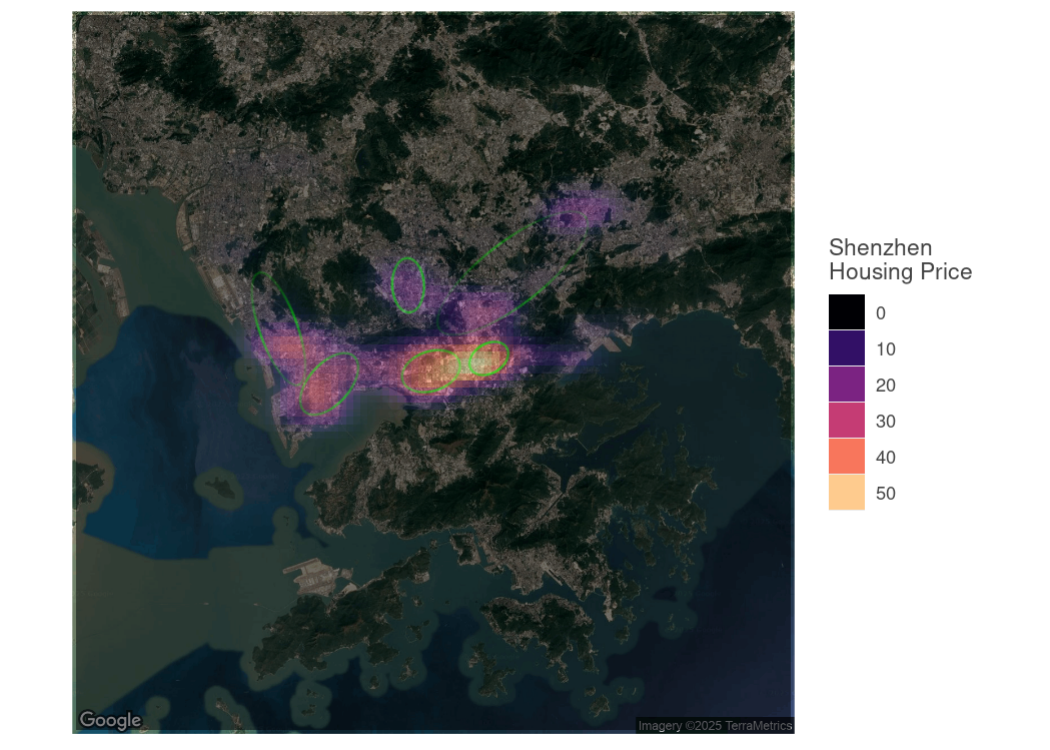}
    \end{minipage}
    \begin{minipage}{0.328\textwidth}
        \includegraphics[width=\linewidth]{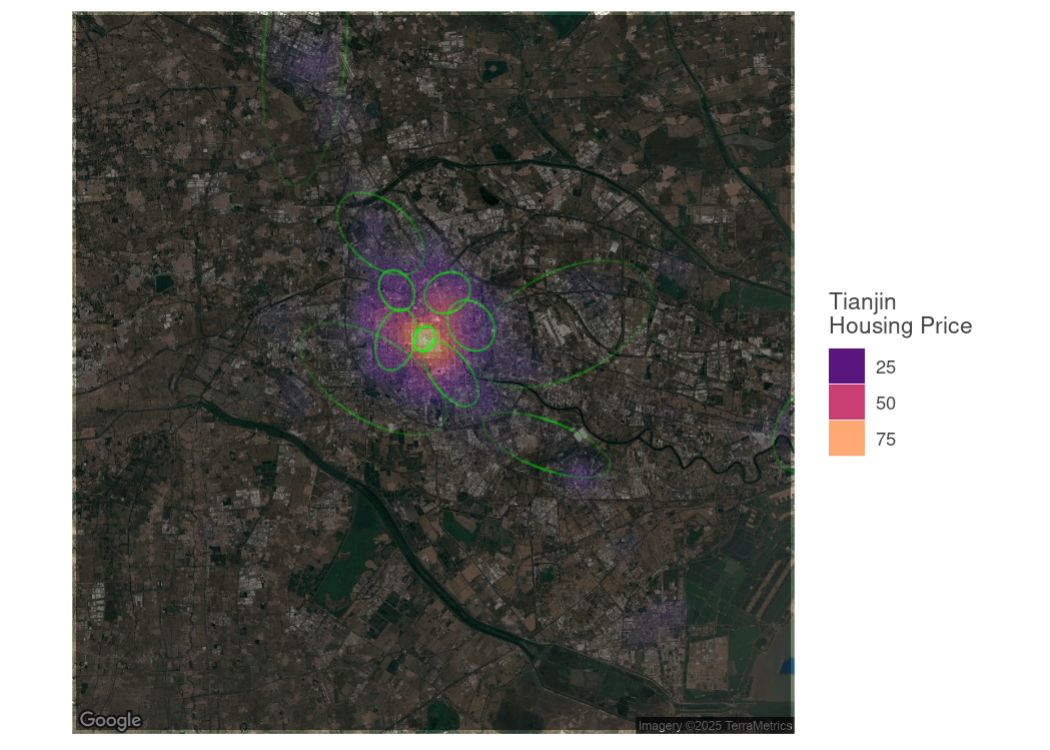}
    \end{minipage}

    \begin{minipage}{0.328\textwidth}
        \includegraphics[width=\linewidth]{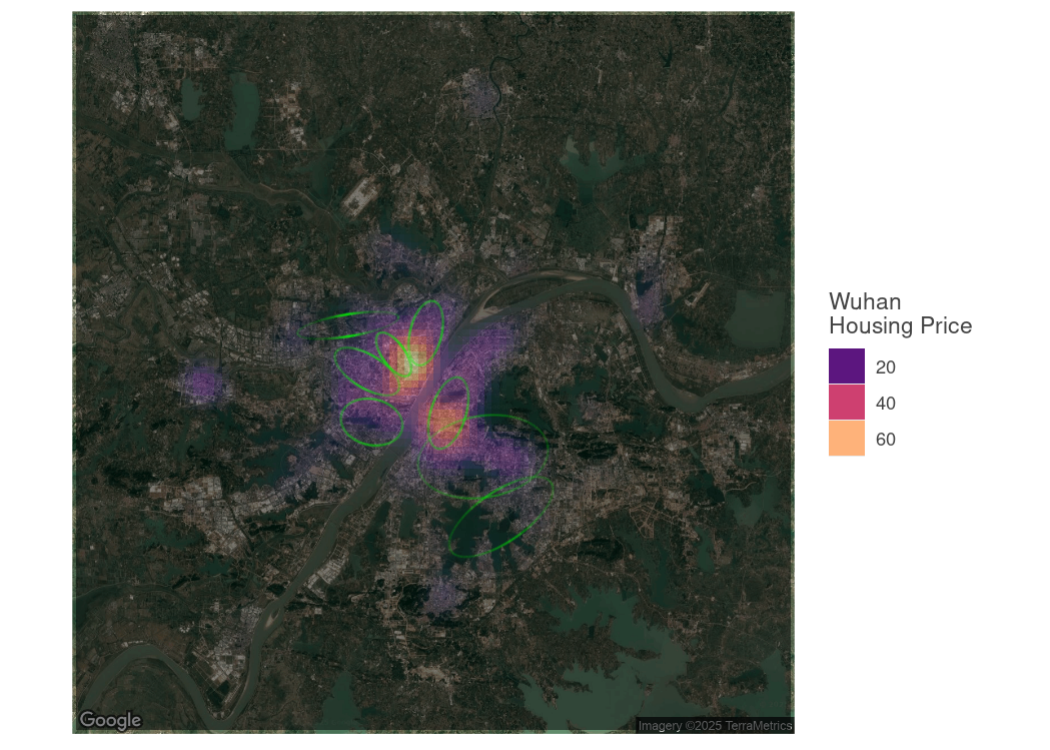}
    \end{minipage}
    \hfill

\caption{The distribution of housing prices and brokers in different cities}
\label{fig:distribution_of_housing_price_brokers_in_different_cities}

Note: The heat map represents the distribution of housing prices and the standard ellipse represents the distribution of brokers, which is calculated by the mean and standard deviation of the latitude and longitude of brokers. The larger the ellipse, the more sparse the brokerages' distribution is. The Base Map is from Google Map.
 \end{figure}

\begin{figure}[H]
  \centering
  \includegraphics[width=0.9\textwidth]{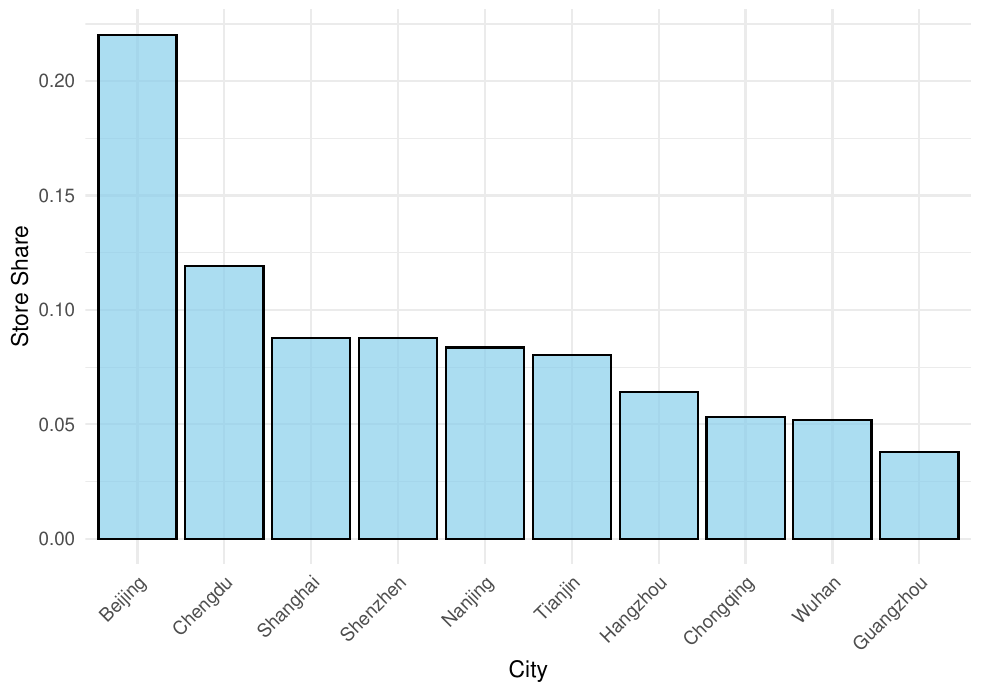}
  \caption{Distribution of Ten Cities's Brokerages Store Shares}
  \label{fig:distribution_store_shares}

  Note: the x-axis is city's name and the y-axis is the offline brokerage's stores share in the city. The data source is from AutoNavi Map.
\end{figure}

This Section Describes the presents visual representations and a description of the spatial distribution and clustering patterns of Lianjia's offline real estate brokerage stores in typical cities. Figure \ref{fig:distribution_of_housing_price_brokers_in_different_cities} shows the correlation between housing price and the distribution of stores in different cities, and the geospatial distribution of the Lianjia's offline stores. Figure \ref{fig:distribution_store_shares} presents the distribution of offline brokerage store shares in ten major Chinese cities, showing the varying market shares of Lianjia's offline stores across different cities.

\clearpage

\clearpage

\end{document}